\documentclass[a4paper,11pt,english]{article}
\pdfoutput=1
\usepackage[utf8]{inputenc}
\usepackage[T1]{fontenc}
\usepackage{lmodern, amsmath, multirow, xcolor, graphicx, cancel}
\usepackage{booktabs}
\usepackage[numbers, compress, elide]{natbib}
\usepackage[no-natbib-sort]{jheppub}

\usepackage{xstring}
\usepackage{ifthen}
\usepackage{subcaption}

\usepackage{ifpdf,feynmp}
\ifpdf\fi

\makeatletter\g@addto@macro\bfseries{\boldmath}\makeatother
\def\figureautorefname~#1\null{fig.\,#1\null}

\newcommand{\subappref}[1]{\hyperref[#1]{appendix~\ref*{#1}}}

\def\equationautorefname~#1\null{eq.\,(#1)\null}
\newcommand{\twoeqs}[2]{\hyperref[#1]{eqs.\,(\ref*{#1},}\,\hyperref[#2]{\ref*{#2})}}
\newcommand{\rangeqs}[2]{\hyperref[#1]{eqs.\,(\ref*{#1}\,--}\,\hyperref[#2]{\ref*{#2})}}

\numberwithin{equation}{section}

\usepackage{tikz}
\usetikzlibrary{positioning,calc}
\newcommand{\minitab}[2][c]{\begin{tabular}{@{}#1@{}}#2\end{tabular}}
\tikzset{every picture/.style={line width=0.75pt}} %set default line width to 0.75pt        
\usetikzlibrary{arrows,shapes}
\tikzstyle{decision} = [diamond, draw, fill=blue!20, 
text width=10em, text badly centered, node distance=3cm, inner sep=0pt]
\definecolor{c1}{rgb}{0.144759, 0.519093, 0.556572}
\definecolor{c2}{rgb}{0.134692, 0.658636, 0.517649}
\definecolor{c3}{rgb}{0.369214, 0.788888, 0.382914}
\definecolor{c4}{rgb}{0.741388, 0.873449, 0.149561}
\tikzstyle{block}         = [rectangle, draw=c1, line width=1.5pt, text centered, rounded corners, text width=3.3cm]
\tikzstyle{blockRed}      = [rectangle, draw=c2, line width=1.5pt, text centered, rounded corners, text width=3.3cm]
\tikzstyle{blockGreen}    = [rectangle, draw=c3, line width=1.5pt, text centered, rounded corners, text width=4.5cm]
\tikzstyle{blockGreenRed} = [rectangle, draw=c4, line width=1.5pt, text centered, rounded corners, text width=4.5cm]
\tikzstyle{line} = [draw, -latex']
\tikzstyle{cloud} = [draw, rectangle,fill=red!20]
\tikzset{
	position/.style args={#1:#2 from #3}{
		at=(#3.#1), anchor=#1+180, shift=(#1:#2)
	}
}

\newcommand{\kf}{\negthinspace\relax}

\newcommand{\cDipole}{c_{\text{dipole}}}
\newcommand{\cbarDipole}{c^*_{\text{dipole}}}
\newcommand{\hv}{v}
\newcommand{\fv}{V}
\newcommand{\ket}[1]{{\left|#1\right\rangle}}

\newcommand{\eg}{{\em e.g.}}
\newcommand{\ie}{{\em i.e.}}
\newcommand{\eqndot}{{\,.}}
\newcommand{\eqncomma}{{,\,}}
\newcommand{\eqCS}{{\,,\:\:}}
\renewcommand{\eqref}[1]{Eq.~(\ref{#1})}

\newcommand{\bs}[1]{{\boldsymbol{#1}}}

\newcommand{\abs}[1]{\left|#1\right|}
\newcommand{\Ket}[1]{\left|#1\right]}
\newcommand{\Bra}[1]{\left[#1\right|}
\newcommand{\tr}{{\operatorname{tr}}}

\newcommand{\parn}[1]{\left(#1\right)}

\newcommand{\sqb}[1]{[#1]}
\newcommand{\anb}[1]{\langle #1\rangle}
\newcommand{\sab}[1]{[#1\rangle}
\newcommand{\asb}[1]{\langle #1]}
\newcommand{\sqbBS}[1]{\left[\bs{#1}\right]}
\newcommand{\anbBS}[1]{\left\langle \bs{#1}\right\rangle }
\newcommand{\Sqb}[1]{[\bs{#1}]}
\newcommand{\Anb}[1]{\langle\bs{#1}\rangle}

\newcommand{\Asb}[1]{\langle\bs{#1}]}
\newcommand{\tdp}[2]{\tilde{s}_{#1#2}}
\newcommand{\ketBS}[1]{\left|\bs{#1}\right\rangle }
\newcommand{\KetBS}[1]{\left|\bs{#1}\right]}

\newcommand{\suN}[1]{\text{SU}(#1)}

\newcommand{\BraketBS}[3]{\left[\bs{#1}\bs{#2}\bs{#3}\right\rangle }
\newcommand{\braKetBS}[3]{\left\langle \bs{#1}\bs{#2}\bs{#3}\right]}
\newcommand{\ketBraBS}[2]{\ket{\bs{#1}}\Bra{\bs{#2}}}
\newcommand{\Braket}[3]{\left[#1 #2 #3\right\rangle }
\newcommand{\braKet}[3]{{\left\langle #1 #2 #3\right]}}

\newcommand{\CFfVs}[5]{{C_{#1 #2 #3 #4}^{#5}}}

\newcommand{\CudWh}[1]{{\CFfVs{}{}{}{}{#1}}}

\newcommand{\chiL}{\chi_{\text{\tiny L}}}
\newcommand{\etaR}{\eta_{\text{\tiny R}}}

\allowdisplaybreaks[4]
\newcommand{\beqa}{\begin{eqnarray}}
\newcommand{\eeqa}{\end{eqnarray}}
\newcommand{\beq}{\begin{equation}}
\newcommand{\eeq}{\end{equation}}

\newcommand{\CoeffFunc}[2]{c^{\scriptscriptstyle #1}_{#2}}
\newcommand{\suTwoTildeGen}[3]{%
\StrLen{#2}[\MyStrLenUp]% 
\StrLen{#3}[\MyStrLenDown]%
\ifthenelse{\equal{\MyStrLenUp}{0}\AND\equal{\MyStrLenDown}{0}}{
\tilde{\tau}^{#1}}{\parn{\tilde{\tau}^{#1}}^{\ #2}_{#3}}}

\newcommand{\suTwoNDelta}[2]{\delta^{\ #1}_{#2}}
\newcommand{\suTwoTgab}[2]{g^{#1 #2}}

\newcommand{\suTwoCabc}[3]{F^{#1 #2 #3}}
\def\splitstringSecond#1#2{%
    \StrCount{#1}{#2}[\nbmatch]%
    \StrCut[\nbmatch]{#1}{#2}\strfirst\strsecond
    \strsecond
}
\def\splitstringFirst#1#2{%
    \StrCount{#1}{#2}[\nbmatch]%
    \StrCut[\nbmatch]{#1}{#2}\strfirst\strsecond
    \strfirst
}

\newcommand{\ampFourPt}[4]{\mathcal{A}_4\parn{
	1^{\splitstringSecond{#1}{,}}_{\splitstringFirst{#1}{,}},
	2^{\splitstringSecond{#2}{,}}_{\splitstringFirst{#2}{,}},
	3^{\splitstringSecond{#3}{,}}_{\splitstringFirst{#3}{,}},
	4^{\splitstringSecond{#4}{,}}_{\splitstringFirst{#4}{,}}
	}}
\newcommand{\ampFivePt}[5]{\mathcal{A}_5\parn{
	1^{\splitstringSecond{#1}{,}}_{\splitstringFirst{#1}{,}},
	2^{\splitstringSecond{#2}{,}}_{\splitstringFirst{#2}{,}},
	3^{\splitstringSecond{#3}{,}}_{\splitstringFirst{#3}{,}},
	4^{\splitstringSecond{#4}{,}}_{\splitstringFirst{#4}{,}},
	5^{\splitstringSecond{#5}{,}}_{\splitstringFirst{#5}{,}}
	}}
\newcommand{\redAmpFourPt}[4]{A\parn{1^{\splitstringSecond{#1}{,}}_{\splitstringFirst{#1}{,}},\ 2^{\splitstringSecond{#2}{,}}_{\splitstringFirst{#2}{,}},\ 3^{\splitstringSecond{#3}{,}}_{\splitstringFirst{#3}{,}},\ 4^{\splitstringSecond{#4}{,}}_{\splitstringFirst{#4}{,}}}}

\newcommand{\redAmpFourPtN}[5]{A_{#1}\parn{1^{\splitstringSecond{#2}{,}}_{\splitstringFirst{#2}{,}},\ 2^{\splitstringSecond{#3}{,}}_{\splitstringFirst{#3}{,}},\ 3^{\splitstringSecond{#4}{,}}_{\splitstringFirst{#4}{,}},\ 4^{\splitstringSecond{#5}{,}}_{\splitstringFirst{#5}{,}}}}
\newcommand{\redAmpFivePtN}[6]{A_{#1}\parn{1^{\splitstringSecond{#2}{,}}_{\splitstringFirst{#2}{,}},\ 2^{\splitstringSecond{#3}{,}}_{\splitstringFirst{#3}{,}},\ 3^{\splitstringSecond{#4}{,}}_{\splitstringFirst{#4}{,}},\ 4^{\splitstringSecond{#5}{,}}_{\splitstringFirst{#5}{,}},\  5^{\splitstringSecond{#6}{,}}_{\splitstringFirst{#6}{,}}}}

\newcommand{\ampFourPtLE}[4]{\mathcal{M}\parn{\bs{1}_{#1},\ \bs{2}_{#2},\ \bs{3}_{#3},\ \bs{4}_{#4}}}

\newcommand{\nn}{\nonumber}

\author[a]{Reuven Balkin,}
\author[b]{Gauthier Durieux,}
\author[c,d]{Teppei Kitahara,}
\author[a]{Yael Shadmi,}
\author[a]{and Yaniv~Weiss}

\affiliation[a]{Physics Department, Technion---Israel Institute of Technology,\\Technion city, Haifa 3200003, Israel}
\affiliation[b]{CERN, Theoretical Physics Department, Geneva 23 CH-1211, Switzerland}
\affiliation[c]{Institute for Advanced Research, Nagoya University,\\
Furo-cho Chikusa-ku, Nagoya 464--8601, Japan}
\affiliation[d]{Kobayashi-Maskawa Institute for the Origin of Particles and the Universe, Nagoya University, Furo-cho Chikusa-ku, Nagoya 464--8602, Japan}

\title{On-shell Higgsing for EFTs}

\preprint{CERN-TH-2021-212}

\abstract{
We study the on-shell version of the Higgs mechanism in effective theories (EFTs) containing particles of different spins, focusing on contact terms as a simple starting point.
We derive the massive contact terms and their coefficients from the massless amplitudes of the EFT above the symmetry breaking scale, by covariantizing the massless contact terms under the massive little group.
In the little-group-covariant massive-spinor formalism, this notationally amounts to bolding spinor labels.
Mass-suppressed contributions to the contact-term coefficients arise from higher-point contact terms with additional soft Higgs legs.
We apply this procedure to obtain massive four-point amplitudes featuring scalars, spin 1/2 fermions and vectors, in the standard-model EFT.
The subleading helicity-flipped components of each massive contact term, which are dictated by little-group covariance, are associated with the residues of factorizable massless amplitudes.
Extra ``frozen'' Higgses emitted from each leg of a massless contact term supply the additional light-like momentum component, needed to form a massive leg of the same polarization.
As another application, we derive various components of massive three-point amplitudes from 
massless amplitudes with up to three additional Higgses, in a standard-model-like toy model.
}

\begin{document}
\begin{fmffile}{./figs/fgraph}
\fmfcmd{style_def marrow expr p = drawarrow subpath (3/4, 1/4) of p shifted 6 right withpen pencircle scaled 0.9; enddef;}
\fmfcmd{style_def marrowA expr p = drawarrow subpath (3/4, 1/4) of p shifted 6 up withpen pencircle scaled 0.4; enddef;}
\sloppy %https://tex.stackexchange.com/questions/9107/how-can-i-make-my-text-never-go-over-the-right-margin-by-always-hyphenating-or-b

% get rid of JHEP header
\makeatletter\renewcommand{\@fpheader}{\ }\makeatother

% toc depth
\setcounter{tocdepth}{2}

\maketitle

\begingroup
\lccode`\~=`\.
\lowercase{\endgroup\def~}#1{foo with #1}

%%%%%%%%%%%%%%%%%%%%%%%%%%%%%%%%%%%%%%%%%%%%%%%%%%%%%%
\section{Introduction}
%%%%%%%%%%%%%%%%%%%%%%%%%%%%%%%%%%%%%%%%%%%%%%%%%%%%%%%%
On-shell methods have proven to be very powerful in mapping effective field theory (EFT) extensions of the standard model (SM).
A systematic study of generic EFT amplitudes was first undertaken in ref.~\cite{Cohen:2010mi}, identifying in particular simple selection rules for on-shell constructibility, and conversely, for allowed contact terms, from the mass-dimensions of their couplings and the particle helicities.
In the context of the standard-model EFT (SMEFT), similar selection rules beautifully relate the non-renormalization of various operators at the one-loop level to the allowed helicity amplitudes arising in unitarity cuts~\cite{Cheung:2015aba}, and dictate the (non-)interference of tree-level SMEFT and SM amplitudes~\cite{Azatov:2016sqh}.
This approach has been recently extended to calculate the anomalous dimensions of operators in the SMEFT and other EFTs at the one-loop level~\cite{EliasMiro:2020tdv, Baratella:2020lzz, Jiang:2020mhe, AccettulliHuber:2021uoa, Baratella:2021guc}, and to relate these to the partial-wave decomposition of the amplitudes~\cite{Jiang:2020rwz, Baratella:2020dvw, Shu:2021qlr}.
Unitarity was further used to prove general non-renormalization theorems in~ref.~\cite{Bern:2019wie}, and to obtain several SMEFT anomalous dimensions at two loops~\cite{EliasMiro:2020tdv, Bern:2020ikv, Jin:2020pwh}.

The construction of EFT operator bases is facilitated in the on-shell approach where it translates to the much simpler problem of classifying bases of kinematic structures, avoiding the inherent redundancies of Lagrangians~\cite{Shadmi:2018xan,Ma:2019gtx}.
This was used to obtain EFT amplitudes with a scalar or massive vector and three gluons up to dimension~13~\cite{Shadmi:2018xan} and to derive bases for the SMEFT up to dimension~9~\cite{Ma:2019gtx,Li:2020gnx,Li:2020xlh}, 
for the low-energy effective field theory (LEFT) up to dimension~9~\cite{Li:2020tsi},
and for GRSMEFT~\cite{Ruhdorfer:2019qmk} up to dimension~8~\cite{Durieux:2019siw}.
General algorithms for finding bases of massless $n$-point contact terms were furthermore  presented in~refs.~\cite{Henning:2019enq, Falkowski:2019zdo, Durieux:2019siw, AccettulliHuber:2021uoa}, also determining the minimal dimension of EFT operators contributing to contact terms of any helicities~\cite{Durieux:2019siw}.

While dramatic, much of this recent progress boils down to results on the structure of the EFT \emph{Lagrangian,} since it is restricted to purely massless amplitudes.
To compute physical scattering processes featuring massive particles requires Higgsing, which is fully formulated only in the Lagrangian framework.
In this paper, we therefore continue the exploration of Higgsed EFTs by directly
analyzing their massive amplitudes~\cite{Shadmi:2018xan,Christensen:2018zcq,Herderschee:2019ofc,Herderschee:2019dmc,Aoude:2019tzn,Christensen:2019mch,Durieux:2019eor,Durieux:2020gip,Dong:2021yak}.
One can envision various advantages to working in a fully on-shell formulation of massive EFTs.
Bottom-up constructions of massive EFT amplitudes can be completely model-independent and thus cover the full space of possible SMEFTs.
Once the basis of independent spinor structures spanning an amplitude is known, extending the amplitude to arbitrary operator dimensions is easy.
The  spinor structure coefficients capture the full $\hv/\Lambda$ expansion, where $\hv$ is the Higgs vacuum expectation value (VEV) and $\Lambda$ is the EFT scale. The derivative expansion corresponds to an expansion in Lorentz invariants multiplying each spinor structure and is fairly straightforward to obtain.
Furthermore, the interpretation of LHC data in terms of EFT Lagrangians, which involves the modification of all SM parameters, may be simplified in an on-shell approach based on purely physical observables.

Previous studies of massive EFT amplitudes have been mostly bottom-up, with the EFT contact terms constructed from just Lorentz and locality requirements~\cite{Shadmi:2018xan,Christensen:2018zcq,Herderschee:2019ofc,Herderschee:2019dmc,Aoude:2019tzn,Christensen:2019mch,Durieux:2019eor,Durieux:2020gip,Dong:2021yak}.
In the case of the SMEFT, these requirements were augmented by imposing either just the unbroken SM symmetries, or by requiring also perturbative unitarity~\cite{Durieux:2019eor} or the full SU(2)$\times$U(1) symmetry at high-energies~\cite{Aoude:2019tzn}.
While general prescriptions for constructing bases of massive contact terms were described~\cite{Durieux:2020gip}, they are harder to implement compared to the massless case.

With this in mind, here we take instead a top-down approach, and develop an on-shell version  of the Higgs mechanism for the construction of Higgsed EFTs, in  analogy with Lagrangian formulations.
These start from the construction of operators in the massless symmetric phase.
The interactions of massive particles are then obtained by turning on the Higgs VEV.
Similarly, we will start from a contact-term basis for massless, high-energy (HE) amplitudes, and obtain from it the massive low-energy (LE) contact terms by covariantizing the massless amplitudes with respect to the massive little group.
Working with the massive spinor formalism of~ref.~\cite{Arkani-Hamed:2017jhn}, this simply amounts to bolding the massless spinor structures into massive ones.
When their leading component is forbidden by the HE gauge symmetries, the LE contact term coefficients are only generated at subleading orders in the mass expansion, by HE contact terms with additional soft Higgs legs.
The subleading helicity-flipped pieces of the LE spinor structures, which are fixed by little-group covariance, can alternatively be determined from  factorizable HE amplitudes.
Additional Higgses emitted from the legs $i$ of the LE contact term, supply the required four-vectors $i_q$ to complete one or more massless momenta into massive $\bs i=i_k + i_q$ ones.
The momenta of these additional Higgses are \emph{frozen} in a configuration such that $(i_k+i_q)^2=m_i^2$.
The HE and LE amplitudes can then be matched in the small $m_i$ expansion.

The picture of on-shell Higgsing which emerges thus combines elements of two different perspectives.
The soft Higgs limit was used to obtain massive amplitudes on the Coulomb branch of ${\cal N}=4$ supersymmetric Yang-Mills theories in ref.~\cite{Craig:2011ws}, using the massive spinors of ref.~\cite{Dittmaier:1998nn}.%
\footnote{A similar approach was applied recently in refs.~\cite{Herderschee:2019ofc,Herderschee:2019dmc} using the little-group-covariant massive-spinor formalism~\cite{Arkani-Hamed:2017jhn}.}
Alternatively, on-shell Higgsing was described as the IR unification of UV
amplitudes~\cite{Arkani-Hamed:2017jhn}, which essentially follows from the combination of the UV symmetries and Lorentz symmetry.
Part of our motivation is to elucidate the relation between these two viewpoints.

This paper is organized as follows.
After some preliminaries, \autoref{sec:general} describes how the massive LE contact terms of a Higgsed EFT arise from the bolding of massless contact terms.
The derivation of the subleading contact-term components from factorizable HE amplitudes is then detailed.
Applications to four-point contact terms are first discussed in \autoref{sec:four-point-app}, mostly focusing on the little-group covariantization, or \emph{bolding}, of their leading massless components.
Massive contact terms are enumerated in this way, and their coefficient subsequently matched to those of massless contact terms.
Various three-point contact terms are then discussed in \autoref{sec:three_point}, addressing in particular their subleading components.
Further details are collected in four appendices.

%%%%%%%%%%%%%%%%%%%%%%%%%%%%%%%%%%%%%%%%%%%%%%%%%%%%%
%%%%%%%%%%%%%%%%%%%%%%%%%%%%%%%%%%%%%%%%%%%%%%%%%%%%%
\section{Massive contact terms from on-shell Higgsing}
\label{sec:general}
%%%%%%%%%%%%%%%%%%%%%%%%%%%%%%%%%%%%%%%%%%%%%%%%%%%%%
%%%%%%%%%%%%%%%%%%%%%%%%%%%%%%%%%%%%%%%%%%%%%%%%%%%%%
To write the LE massive amplitudes, we use the little-group-covariant spinor formalism of ref.~\cite{Arkani-Hamed:2017jhn}.
Each massive particle of momentum $\bs{i}$ is described  by a pair of massless spinors, $\bs i^I]$ (or equivalently $\bs i^I\rangle$), where $I=1,2$ is a little-group index.
We use boldface to denote these spinors as well as massive momenta, to distinguish them from massless ones.
Together, the spinor pair $\bs i^I]$ parametrizes the particle momentum and the direction of its spin quantization axis.
We will often use $i_q$ and $i_k$ to denote the four-vectors $i^1\rangle[i_1$ and $i^2\rangle[i_2$, respectively.
In the high-energy limit, $i_q$ is taken to be the subleading component, scaling as $m_i^2/E_i$, while the leading $i_k$ one is unsuppressed.
The particle momentum $\bs{i}$ is given by the sum $i_q+i_k$, as $\bs i = \sum_{I=1,2} \bs i^I\rangle [ \bs i_I$, while the familiar polarizations are given by appropriate combinations of $i_k]$ and $i_q]$.
Explicit expressions for the polarizations of Dirac fermions and massive vectors can be found for example in ref.~\cite{Durieux:2019eor}.
Note that, for fixed $\bs i$, the direction $\hat{\imath}_q$ can be chosen arbitrarily,
which amounts to a choice of the spin quantization axis~\cite{Arkani-Hamed:2017jhn,Ochirov:2018uyq}.
As we will see, these light-like momenta have a clear interpretation in the on-shell Higgs mechanism.
Essentially, in the HE theory, $i_k$ is associated with the momentum of a massless particle, while $i_q$ is the momentum of an additional Higgs leg which gives mass to this particle (see, also refs.~\cite{Craig:2011ws,Kiermaier:2011cr}).
Much of our discussion here applies to three-point amplitudes as well, but because of the special three-point kinematics, we will treat these separately in \autoref{sec:three_point}.

Consider a LE tree amplitude ${\cal M}_n (\bs 1, \ldots, \bs n)$ with $n$ massive external particles.
The amplitude is given as a sum of two contributions: a factorizable part ${\cal M}_n^{\text{fac}}$, obtained by requiring correct factorization on lower-point amplitudes, 
and a contact-term part ${\cal M}_n^{\text{ct}}$, which is given as a sum over pole-free non-factorizable terms,
\beq\label{eq:basis}
{\cal M}^\text{ct}_n (\bs 1, \ldots, \bs n) = \sum_r C_{n,r} \bs{K}_{n,r}^{\{I\}}\,, 
\eeq
where $\{I\}$ collectively denotes the little-group indices of all external particles.
The kinematic $\bs{K}_{n,r}^{\{I\}}$ contains spinor structures carrying these little-group indices, multiplied by non-negative powers 
of the Lorentz invariants $\tdp{i}{j}=2 \:(\bs{i}\cdot \bs{j})$ and 
$\epsilon_{\mu\nu\rho\sigma}\bs i^\mu \bs j^\nu \bs k^\rho \bs l^\sigma$ (for $n\geq5$).
The coefficients $C_{n,r}$ are the Wilson coefficients in the LE theory. 
These constitute the ``novel'' $n$-point couplings required for calculating the full amplitude, augmenting the lower-point couplings featured in the factorizable part.
A massive spinor structure contains $2s$ factors of $\bs i^I]$ or $\bs i^I \rangle$ for each external massive particle of momentum $\bs i$ and spin $s_i$, with the little-group indices completely symmetrized.
We usually keep these indices and their symmetrization implicit.

It is useful to classify the massive spinor structures appearing in ${\cal M}^\text{ct}_n$ according to their \textit{helicity category}, namely, the helicities of the leading HE components~\cite{Durieux:2020gip}. 
The leading HE component of each spinor structure is simply obtained by naively unbolding the spinor structure: $\bs{i}]\to i]$, $\;\bs{i}\rangle\to i\rangle$, where $i$ is identified with $i_k$.
For example, for a fermion $i$, the spinor $\bs{i}]$ is in the $+1/2$ helicity category, so that, \eg, $[{\bf 12}]$ is classified as $+1/2,+1/2$ or just ++ when there is no ambiguity about the particle spins.

For simplicity, we assume that the LE theory is obtained by Higgsing the HE theory with a single Higgs field of VEV $\hv$.
Thus, all the masses are given by $\hv$ times some coupling, and the massless,
or high-energy, limit is obtained by $\hv\to0$ with all couplings kept finite.

We wish to derive the massive structures  $\bs{K}_{n,r}^{\{I\}}$ and the Wilson coefficients $C_{n,r}$ using the massless HE amplitudes as starting points.
These HE amplitudes are similarly given by factorizable plus contact-term parts
\beq
{\cal A}^\text{fac}_n( 1^{h_1}, \ldots, n^{h_n})+
 {\cal A}^\text{ct}_n( 1^{h_1}, \ldots, n^{h_n}) \,,
\eeq
where $h_i$ denote the helicities, 
and
\beq\label{eq:masslessbasis}
{\cal A}^\text{ct}_n (1^{h_1}, \ldots, n^{h_n}) = \sum_r c_{n,r} K_{n,r}^{\{h\}}\,, 
\eeq
with $\{h\}$ collectively denoting the particle helicities.
Note that we use $\cal{A}$ and $c_{n}$ for the HE amplitudes and their Wilson coefficients, and $\cal{M}$ and $C_{n}$ for the LE ones.
The kinematic $K_{n}^{\{h\}}$ is given by a massless spinor structure, namely, a product of massless spinor contractions, which carries the U(1) little-group weight of the amplitude,  multiplied by Lorentz invariants.
Since ${\cal A}^\text{ct}_n$ is pole-free, all the spinor products and the Lorentz invariants appear in non-negative powers for $n>3$.

%%%%%%%%%%%%%%%%%%%%%%%%%%%%%%%%%%%%%%%%%%%%%%%%%%%
\subsection{Constructing the EFT: bolding}\label{sec:bolding}
%%%%%%%%%%%%%%%%%%%%%%%%%%%%%%%%%%%%%%%%%%%%%%%%%%%%%%
To construct the LE EFT amplitudes ${\cal M}^\text{ct}_n$, we use the fact that the contribution it receives from each HE coupling $c_{n+n_H}$ is dictated by
(i)~the kinematic structure $K_{n+n_H}^{\{h\}}$, and 
(ii)~Lorentz symmetry and, in particular, covariance under the massive little group associated with each of the external particles.
Here $n_H\geq0$ denotes the number of \emph{additional} Higgs legs.
Consider first $n_H=0$.
To covariantize $K_{n}^{\{h\}}$, we merely need to bold the spinors and monenta it contains.
Essentially, this corresponds to matching $K_{n}^{\{h\}}$ to the leading term of $\bs K_{n}^{\{I\}}$ in the high-energy limit.
This is  particularly simple for massless particles of non-zero spins, \eg, fermions and vectors.
Each HE fermion is replaced by a massive fermion leg, and each massless vector is replaced by a massive vector in the $\pm1$ helicity category.
This identification gives the Wilson coefficient at leading order in the $\hv/\Lambda$ expansion.
Thus for example,  $c\: [12][34] \tdp13$ featured in a HE four-fermion amplitude is bolded to $(c +{\cal O}(\hv^2/\Lambda^2))\Sqb{12}\Sqb{34} \tdp13$.
The ${\cal O}(\hv^2/\Lambda^2)$ corrections are generated by higher-order HE contact terms featuring additional pairs of Higgs legs.%
\footnote{We have in mind SM-like theories.
More generally, $\mathcal{O}(\hv/\Lambda)$ may appear.}
We return to these below.
Note that the massless Mandelstam invariant $s_{ij}$ can bolded either into the massive $s_{ij}=(\bs i+\bs j)^2$, or into the massive $\tdp ij=2\:(\bs i\cdot \bs j)$.
This translates to an ambiguity in ${\mathcal O}(m^2)$ corrections to lower-dimensional contact terms, in this example $\Sqb{12}\Sqb{34}$.

Bolding the momenta and spinors associated with massless scalar degrees of freedom is more delicate.
These can be mapped either into massive radial modes or into the longitudinal helicity category of massive vectors. 
HE and LE contact terms with identical numbers of fermion, transverse vector, and scalar legs are then mapped onto each other by simply bolding the massless spinor structures,
\beq
c_n\: K_n^{\{h\}}(\{\#_f\},\{\#_\fv\},\{\#_H\}) \to C_n  {\bs K}_n^{\{I\}}(\{\#_f\},\{\#_\fv\},\{\#_h\})\,,
\label{eq:same-content-bolding}
\eeq
where $\#_f$ and $\#_\fv$ denote the numbers of fermion and vector legs,  respectively,
$\#_H$ is the number of $H$ and $H^\dagger$ legs, and $\#_h$ is the number of physical Higgs legs.
The Wilson coefficient $C_n$ is given by $C_n=c_n +{\cal O}(\hv^2/\Lambda^2)$.
From a bottom-up perspective, the bolding of~\autoref{eq:same-content-bolding} relies on matching the leading-energy components of the LE amplitudes, which are tensors in the little-group space, to the corresponding HE amplitudes, with $\bs{i}]$ and $\bs{i}\rangle$ replaced by their leading high-energy component, $i_k]$ and $i_k\rangle$ respectively, for each external particle.
It only yields LE structures in scalar, fermion, and transverse-vector helicity categories.

LE structures whose helicity categories feature longitudinal vectors require a slight generalization of the naive bolding of~\autoref{eq:same-content-bolding}.
To bold a massless scalar leg $i$ into a vector in the longitudinal helicity category, the momentum $i$ must appear in the HE contact term.
It can occur either in a Lorentz invariant $s_{ij}$, or as a momentum insertion in a spinor product, \eg, $\langle jik]$.
These HE structures are then bolded as,
\beq
s_{ij}\to \langle\bs{i}j\bs{i}],\qquad\langle jik]\to\anb{j\bs{i}}[\bs{i}k]\,.
\eeq
No prior knowledge is required about whether the scalar leg is a Goldstone mode.
This information is encoded in the amplitude: only scalar amplitudes with a momentum insertion can be bolded into longitudinal vectors.
This is consistent with Goldstones being derivatively coupled in the Lagrangian picture.
A HE contact term of this type gives rise to two different LE contact terms with equal Wilson coefficients: one corresponding to a longitudinal vector, and the second corresponding to a physical Higgs.
The latter includes a momentum insertion.

%%%%%%%%%%%%%%%%%%%%%%%%%%%%%%%%%%%%%%%%%%%%%%%%%%%%%%%%%%%%%%%%%%%
\begin{figure}[tb]
\centering\footnotesize
%%%%%%%%%%%%%%%%%%%%%%%%%%%%%%%%%%%%%%%%%%%%%%%%%%%%%%%%%%%%%%%%%%%%
\begin{align*}
&\parbox{23mm}{
	\begin{fmfgraph*}(50,35)
	\fmfleftn{i}{5} 
	\fmfrightn{o}{5}
	\fmfdotn{v}{1} 
	\fmf{plain}{i5,v1,i1}
	\fmf{phantom}{i3,v1,v2,o3}
	\fmffreeze
	\fmf{plain}{v1,v2,o3}
	\fmf{plain}{i4,v1}
	\fmflabel{$n-1$}{i1} 
	\fmflabel{$2$}{i4} 
	\fmflabel{$\vdots$}{i3} 
	\fmflabel{$1$}{i5}
	\fmfv{l=$c_n$,l.a=-60,l.d=0.2w}{v1}
	\fmfblob{0.2w}{v1}
	\fmflabel{$n$}{o3}
	\end{fmfgraph*}}
	+~~~ ~
	\parbox{23mm}{
	\begin{fmfgraph*}(50,35)
	\fmfleftn{i}{5} 
	\fmfrightn{o}{5}
	\fmfdotn{v}{1} 
	\fmf{plain}{i5,v1,i1}
	\fmf{phantom}{i3,v1,v2,o3}
	\fmffreeze
	\fmf{plain}{v1,v2,o3}
	\fmf{plain}{i4,v1}
	\fmflabel{$n-1$}{i1} 
	\fmflabel{$2$}{i4} 
	\fmflabel{$\vdots$}{i3} 
	\fmflabel{$1$}{i5}
	\fmf{dashes}{v1,o4}
	\fmfblob{0.2w}{v1}
	\fmflabel{$n$}{o3}
	\fmfv{l=$c_{n+1}$,l.a=-60,l.d=0.2w}{v1}
	\end{fmfgraph*}}
+~~~~
\parbox{23mm}{
	\begin{fmfgraph*}(50,35)
	\fmfleftn{i}{5} 
	\fmfrightn{o}{5}
	\fmfdotn{v}{1} 
	\fmf{plain}{i5,v1,i1}
	\fmf{phantom}{i3,v1,v2,o3}
	\fmffreeze
	\fmf{plain}{v1,v2,o3}
	\fmf{plain}{i4,v1}
	\fmflabel{$n-1$}{i1} 
	\fmflabel{$2$}{i4} 
	\fmflabel{$\vdots$}{i3} 
	\fmflabel{$1$}{i5}
	\fmf{dashes}{v1,o4}
	\fmf{dashes}{v1,o5}
	\fmfblob{0.2w}{v1}
	\fmflabel{$n$}{o3}
	\fmfv{l=$c_{n+2}$,l.a=-60,l.d=0.2w}{v1}
	\end{fmfgraph*}}
	 +\cdots 
	\xrightarrow[\;\;\;\;\;\;\;]{}~~~~
	\parbox{23mm}{
	\begin{fmfgraph*}(50,35)
	\fmfleftn{i}{5} 
	\fmfrightn{o}{5}
	\fmfdotn{v}{1} 
	\fmf{plain}{i5,v1,i1}
	\fmf{phantom}{i3,v1,v2,o3}
	\fmffreeze
	\fmf{plain}{i4,v1}
	\fmflabel{$n-1$}{i1} 
	\fmflabel{$2$}{i4} 
	\fmflabel{$\vdots$}{i3} 
	\fmflabel{$1$}{i5}
	\fmfv{l=$C_n$,l.a=-60,l.d=0.2w}{v1}
	\fmfblob{0.2w}{v1}
	\fmflabel{$\bs{n}$}{o3}
	\fmffreeze\fmfdraw
	\fmfpen{.8thick}
	\fmf{plain}{v1,v2,o3}
	\end{fmfgraph*}}
\end{align*}
\vskip 0.5em
\caption{\label{fig:massive_comp_origins-a}
The different terms in the $\hv/\Lambda$ expansion of the LE contact-term coefficient $C_n$ are generated by HE contact terms with additional soft Higgs legs  (some $c_n$'s may vanish because of gauge invariance).
}
\end{figure}
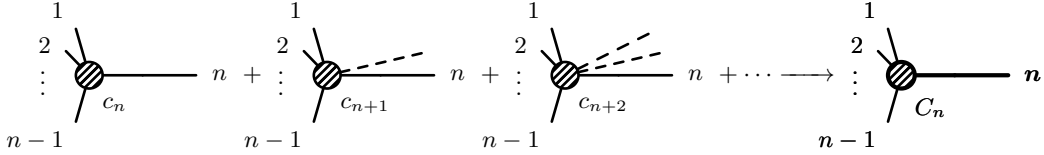

So far, we discussed the contributions of  $K_{n+n_H}^{\{h\}}$ for $n_H=0$.
These generate the full set of LE contact terms in ${\cal M}^\text{ct}_n$ that are allowed by gauge invariance in the HE theory, and give their Wilson coefficients to leading order in the $\hv/\Lambda$ expansion. 
Generically, ${\cal M}^\text{ct}_n$ contains additional terms, suppressed by powers of $\hv$.
As illustrated in~\autoref{fig:massive_comp_origins-a}, these are generated from $K_{n+n_H}^{\{h\}}$ in the limit that the $n_H$ additional Higgs momenta are soft,
\beq\label{eqn:bold2}
v^{n_H}
c_{n+n_H}\,K_{n+n_H}^{\{h\}}(1,2,..,n;0,..,0) \to v^{n_H}\, c_{n+n_H}\, \bs{K}_{n}^{\{I\}}(\bs1,\bs2,..,\bs n)\,.
\eeq
Here, the zeros on the left-hand-side stand for the $n_H$ Higgs momenta.
After setting these momenta to zero, the massless structure is bolded, just as described above for $n_H=0$.
Only contact terms with no insertion of extra Higgs momenta survive this soft limit.
The factor of $v^{n_H}$ is required on dimensional grounds, to compensate for the dimension of the $(n+n_H)$-point amplitude.
As in the Lagrangian picture, the EFT expansion of the amplitude depends on the combination $\hv+h$: each contact term featuring $h$ and no associated momentum insertion is accompanied by $\hv$ times the same contact term with $h$ removed.

Several comments are now in order.
First, a massless spinor structure can usually be bolded in several  different ways.
These are equivalent however, since two massive structures that yield the same massless structure in the HE limit are equal up to subleading mass-suppressed terms~\cite{Durieux:2020gip}.
These different choices merely correspond to different LE amplitude bases.
Second, to determine the LE contact terms, we rely on matching them to the HE contact terms.
The physical quantities are however the full amplitudes, which are the sum of the factorizable and contact-term pieces.
Thus, one could worry that the matching of contact terms is affected by factorizable pieces.
We show that this is not the case in~\autoref{app:fnf}.

%%%%%%%%%%%%%%%%%%%%%%%%%%%%%%%%%%%%%%%%%%%%%%%
%%%%%%%%%%%%%%%%%%%%%%%%%%%%%%%%%%%%%%%%%%%%%%%
\subsection{Subleading components and frozen Higgses}
\label{sec:subleading}
%%%%%%%%%%%%%%%%%%%%%%%%%%%%%%%%%%%%%%%%%%%%%%%
%%%%%%%%%%%%%%%%%%%%%%%%%%%%%%%%%%%%%%%%%%%%%%%

Above, we have seen how to determine the massive contact terms $C_{n}\bs{K}_{n}^{\{I\}}$ by identifying their leading components with the appropriate HE contact terms $c_{n} K_{n}^{\{h\}}$ such as to fix $C_{n}=c_{n}+{\cal O}(\hv)$.
We also saw that mass-suppressed contributions to $C_{n}$ arise from contact terms with additional Higgs legs.
The helicity-flipped $i_q$ components of the massive spinors structures  $\bs{K}_{n}^{\{I\}}$, which include both $i_q$ spinors and momentum insertions, 
constitute a second type of mass-suppressed contributions.
They are not required as input for the construction of the LE theory since they can be obtained from the leading $i_k$ component, by enforcing little-group covariance.
It is nevertheless instructive to see how they arise from massless HE amplitudes.
This time, higher-point \emph{factorizable} amplitudes are required, with extra Higgses emitted from the legs of an $n$-point contact term.
This splitting provides the two massless momenta needed to form a massive $\bs{i}$ momentum.
The residue of the associated pole gives rise to the desired small-mass correction, once the extra Higgs momentum is frozen to the subleading massive momentum component $i_q$ such that $\bs i^2=(i_k+i_q)^2=m_i^2$.

\begin{figure}[tb]
    \centering\footnotesize
%%%%%%%%%%%%%%%%%%%%%%%%%%%%%%%%%%%%%%%%%%%%%%%%%%%%%%%%%%%%%%%%%%%%
%%%%%%%%%%%%%%%%%%%%%%%%%%%%%%%%%%%%%%%%%%%%%%%%%%%%%%%%%%
        \begin{subfigure}[t]{1\textwidth}
        \centering
                \begin{align*}
                &
                \parbox{23mm}{
        			\begin{fmfgraph*}(50,35)
        			\fmfleftn{i}{5} 
        			\fmfrightn{o}{3}
        			\fmfdotn{v}{1} 
        			\fmf{plain}{i5,v1,i1}
        			\fmf{phantom}{i3,v1,v2,o2}
        			\fmffreeze
        			\fmf{plain}{v1,v2,o2}
        			\fmf{plain}{i1,v1} 
        			\fmf{plain}{i4,v1}
        			\fmflabel{$n-1$}{i1} 
        			\fmflabel{$2$}{i4} 
        			\fmflabel{$\vdots$}{i3} 
        			\fmflabel{$1$}{i5}
        			\fmflabel{$k$}{o2}
        			\fmfblob{0.2w}{v1}
        			\fmfv{l=$\Ket{k}$,l.a=-60,l.d=.25w}{v1}
        			\fmffreeze
				\fmfiv{l=$c_n$,l.a=60,l.d=0.2w}{vloc(__v1)}
        			\end{fmfgraph*}}
        		~~+~~~~
        		\parbox{23mm}{
        			\begin{fmfgraph*}(50,35)
        			\fmfleftn{i}{5} 
        			\fmfrightn{o}{3}
        			\fmfdotn{v}{2} 
        			\fmf{plain}{i5,v1,i1}
           		    \fmf{phantom}{i3,v1,v2,o2}
           		    \fmffreeze
        		    \fmf{plain}{v1,v2,o2}
           			\fmf{plain}{i1,v1} 
           			\fmf{plain}{i4,v1}
        			\fmf{dashes}{o3,v2} 
        			\fmflabel{$n-1$}{i1} 
        			\fmflabel{$2$}{i4} 
        			\fmflabel{$\vdots$}{i3} 
        			\fmflabel{$1$}{i5}
        			\fmflabel{$k$}{o2}
        			\fmflabel{$q$}{o3}
        			\fmfv{l=$\Ket{\eta}$,l.a=-60,l.d=.25w}{v1}
        			\fmfblob{0.2w}{v1}
        			\fmffreeze
				\fmfiv{l=$c_n$,l.a=60,l.d=0.2w}{vloc(__v1)}
        			\end{fmfgraph*}}
     +  \cdots 
        		\xrightarrow[\;\;\;\;\;\;\;]{}~~~~
        		\parbox{23mm}{
        			\begin{fmfgraph*}(50,35)
        			\fmfleftn{i}{5} 
        			\fmfrightn{o}{3}
        			\fmfdotn{v}{1} 
        			\fmf{plain}{i5,v1,i1}
        			\fmf{phantom}{i3,v1,v2,o2}
        			\fmffreeze
        			\fmf{plain}{i1,v1} 
        			\fmf{plain}{i4,v1}
        			\fmflabel{$n-1$}{i1} 
        			\fmflabel{$2$}{i4} 
        			\fmflabel{$\vdots$}{i3} 
        			\fmflabel{$1$}{i5}
        			\fmfblob{0.2w}{v1}
        			\fmfv{l=$\KetBS{p}$,l.a=-60,l.d=.25w}{v1}
				\fmfiv{l=$C_n$,l.a=60,l.d=0.2w}{vloc(__v1)}
				\fmflabel{$\bs{p}$}{o2}
        			\fmffreeze\fmfdraw
            			\fmfpen{.8thick}
        			\fmf{plain}{v1,v2,o2}
        			\end{fmfgraph*}}
                \end{align*}
    	    \vskip 0.5em
    	    \caption{Origin of the different components of a massive structure associated with an external fermion. The $\Ket{\eta}$ factor is associated with the $n$-point blob.
    	    }
    	    \label{fig:massive_comp_origins-b}
        \end{subfigure}

\vspace{8mm}

        %%%%%%%%%%%%%%%%%%%%%%%%%%%%%%%%%%%%%%%%%%%%%%%%%%%%%%%%%%%%%%%%%%%%
        \begin{subfigure}[t]{1\textwidth}
        \centering
                \begin{align*}
                &\parbox{23mm}{
            			\begin{fmfgraph*}(50,35)
            			\fmfleftn{i}{5} 
            			\fmfrightn{o}{3}
            			\fmfdotn{v}{1} 
            			\fmf{plain}{i5,v1,i1}
            			\fmf{phantom}{i3,v1,v2,o2}
            			\fmffreeze
            			\fmf{plain}{i1,v1} 
            			\fmf{plain}{i4,v1}
            			\fmf{photon}{v1,o2} 
            			\fmflabel{$n-1$}{i1} 
            			\fmflabel{$2$}{i4} 
            			\fmflabel{$\vdots$}{i3} 
            			\fmflabel{$1$}{i5}
            			\fmfblob{0.2w}{v1}
            			\fmflabel{$k$}{o2}
            			\fmfv{l=$\Ket{k}\Ket{k}$,l.a=-60,l.d=.25w}{v1}
        			\fmffreeze
				\fmfiv{l=$c_n$,l.a=60,l.d=0.2w}{vloc(__v1)}
            			\end{fmfgraph*}}
            		~~+~~~~
            		\parbox{23mm}{
            			\begin{fmfgraph*}(50,35)
            			\fmfleftn{i}{5} 
            			\fmfrightn{o}{3}
            			\fmfdotn{v}{1} 
            			\fmf{plain}{i5,v1,i1}
            			\fmf{phantom}{i3,v1,v2,o2}
            			\fmffreeze
            			\fmf{plain}{i1,v1} 
            			\fmf{plain}{i4,v1}
            			\fmf{photon}{v2,v1}
            			\fmf{dashes}{o1,v2,o3} 
            			\fmflabel{$n-1$}{i1} 
            			\fmflabel{$2$}{i4} 
            			\fmflabel{$\vdots$}{i3} 
            			\fmflabel{$1$}{i5}
            			\fmfblob{0.2w}{v1}
            			\fmfdot{v2}
            			\fmflabel{$k$}{o3}
            			\fmflabel{$q$}{o1}
            			\fmfv{l=$\Ket{\eta}\Ket{\eta}$,l.a=-67.5,l.d=.3w}{v1}
        			\fmffreeze
				\fmfiv{l=$c_n$,l.a=60,l.d=0.2w}{vloc(__v1)}
            			\end{fmfgraph*}}
            	 + \cdots
            		\xrightarrow[\;\;\;\;\;\;\;]{}~~~~
            		\parbox{23mm}{
            			\begin{fmfgraph*}(50,35)
            			\fmfleftn{i}{5} 
            			\fmfrightn{o}{3}
            			\fmfdotn{v}{1} 
            			\fmf{plain}{i5,v1,i1}
            			\fmf{phantom}{i3,v1,v2,o2}
            			\fmffreeze
            			\fmf{plain}{i1,v1} 
            			\fmf{plain}{i4,v1}
            			\fmflabel{$n-1$}{i1} 
            			\fmflabel{$2$}{i4} 
            			\fmflabel{$\vdots$}{i3} 
            			\fmflabel{$1$}{i5}
            			\fmfblob{0.2w}{v1}
            			\fmfv{l=$\KetBS{p}\KetBS{p}$,l.a=-60,l.d=.25w}{v1}
				\fmfiv{l=$C_n$,l.a=60,l.d=0.2w}{vloc(__v1)}
				\fmflabel{$\bs{p}$}{o2}
        			\fmffreeze\fmfdraw
            			\fmfpen{.8thick}
            			\fmf{photon}{v1,o2}
            			\end{fmfgraph*}}
                \end{align*}
    	    \vskip 0.5em
    	    \caption{
    	    Origin of the different components of a massive structure associated with an external vector (transverse helicity category).
The $\Ket{\eta}\Ket{\eta}$ factor is associated with the $n$-point blob.
    	    }
    	    \label{fig:massive_comp_origins-c}
        \end{subfigure}

\vspace{10mm}

       %%%%%%%%%%%%%%%%%%%%%%%%%%%%%%%%%%%%%%%%%%%%%%%%%%%%%%%%%%%%%%%%%%%%
        \begin{subfigure}[t]{1\textwidth}
        \centering
                \begin{align*}
                &	\parbox{23mm}{
            			\begin{fmfgraph*}(50,35)
            			\fmfleftn{i}{5} 
            			\fmfrightn{o}{3}
            			\fmfdotn{v}{1} 
            			\fmf{plain}{i5,v1,i1}
            			\fmf{phantom}{i3,v1,v2,o2}
            			\fmffreeze
            			\fmf{dashes}{v1,v2,o2}
            			\fmf{plain}{i4,v1}
            			\fmflabel{$n-1$}{i1} 
            			\fmflabel{$2$}{i4} 
            			\fmflabel{$\vdots$}{i3} 
            			\fmflabel{$1$}{i5}
            			\fmfblob{0.2w}{v1}
            			\fmflabel{$k$}{o2}
            			\fmfv{l=$\Ket{k}\ket{k}$,l.a=-60,l.d=.25w}{v1}
        			\fmffreeze
				\fmfiv{l=$c^\prime_n$,l.a=60,l.d=0.2w}{vloc(__v1)}
            			\end{fmfgraph*}}
            	+~~~
            		\parbox{23mm}{
            			\begin{fmfgraph*}(50,35)
            			\fmfleftn{i}{5} 
            			\fmfrightn{o}{3}
            			\fmfdotn{v}{2} 
            			\fmf{plain}{i5,v1,i1}
            			\fmf{phantom}{i3,v1,v2,o2}
            			\fmffreeze
            			\fmf{dashes}{v1,v2,o2}
            			\fmf{plain}{i4,v1}
            			\fmf{photon}{o3,v2} 
            			\fmflabel{$n-1$}{i1} 
            			\fmflabel{$2$}{i4} 
            			\fmflabel{$\vdots$}{i3} 
            			\fmflabel{$1$}{i5}
            			\fmfblob{0.2w}{v1}
            			\fmflabel{$q$}{o2}
            			\fmflabel{$k$}{o3}
            			\fmfv{l=$\Ket{\eta}\ket{\eta}$,l.a=-60,l.d=.25w}{v1}
        			\fmffreeze
				\fmfiv{l=$c^\prime_n$,l.a=60,l.d=0.2w}{vloc(__v1)}
            			\end{fmfgraph*}}
            + 	\cdots 
            		\xrightarrow[\;\;\;\;\;\;\;]{}~~~~
            		\parbox{23mm}{
            			\begin{fmfgraph*}(50,35)
            			\fmfleftn{i}{5} 
            			\fmfrightn{o}{3}
            			\fmfdotn{v}{1} 
            			\fmf{plain}{i5,v1,i1}
            			\fmf{phantom}{i3,v1,v2,o2}
            			\fmffreeze
            			\fmf{plain}{i4,v1}
            			\fmflabel{$n-1$}{i1} 
            			\fmflabel{$2$}{i4} 
            			\fmflabel{$\vdots$}{i3} 
            			\fmflabel{$1$}{i5}
            			\fmfblob{0.2w}{v1}
            			\fmfv{l=$\KetBS{p}\ketBS{p}$,l.a=-60,l.d=.25w}{v1}
        			\fmffreeze\fmfdraw\fmfpen{0.8thick}
				\fmfiv{l=$C_n$,l.a=60,l.d=0.2w}{vloc(__v1)}
            			\fmf{photon}{v1,v2,o2}
				\fmflabel{$\bs{p}$}{o2}
            			\end{fmfgraph*}}
                \end{align*}
    	    \vskip 0.5em
    	    \caption{Origin of the different components of a massive structure associated with an external vector (longitudinal helicity category). The $\Ket{\eta}\ket{\eta}$ factor is associated with the $n$-point blob. 
    	    }
    	    \label{fig:massive_comp_origins-d}
        \end{subfigure}
%%%%%%%%%%%%%%%%%%%%%%%%%%%%%%%%%%%%%%%%%%%%%%%%%%%%%%%%%%%%%%%%%%%%%%%%
\caption{High-energy origins of the leading and subleading, helicity-flipped, spinor components of massive fermions, transverse and longitudinal vectors. 
The ellipsis stands for amplitudes with additional Higgs legs.
}
\label{fig:massive_comp_origins}
\end{figure}
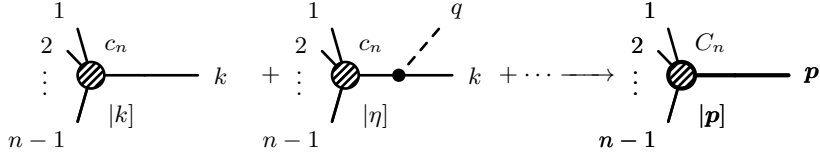
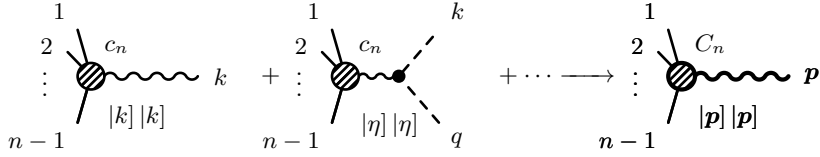
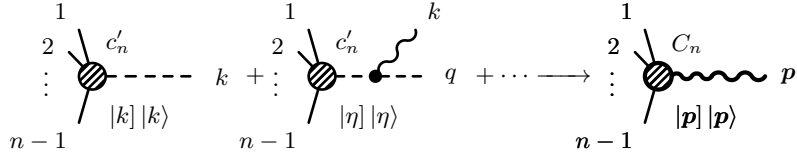

Consider for simplicity the case $n_H=1$, namely, the contribution of a factorizable HE amplitude ${\cal A}^{(c_{n})}_{n+1}(1_k,H(q);2,\ldots,n)$ with one additional Higgs of momentum $q$, which features an $n$-point contact term with coefficient $c_n$.
As we show in the following, the $1_q]$ or $1_q\rangle$ piece of the massive amplitude can be identified via the small mass limit as,
\beq
\label{eqn:lim}
{\cal M}^\text{ct}_n(\bs{1},2,\ldots,n)\Big\vert_{1_q)}
=\hv\;
\left[
\lim_{(1_k+q)^2\to 0} \frac{(1_k+q)^2}{m_1^2}\:
{\cal A}^{(c_{n})}_{n+1}(1_k,H(q);2,\ldots,n)
\right]_{q=1_q}\,.
\eeq
The $(1_k+q)^2$ residue is isolated before freezing $q$ to $1_q$ such that $(1_k+q)^2=m_1^2$.
In this process, the legs $1_k$ and $q=1_q$ are replaced by a single \emph{effective} particle of mass $m_1$ and momentum $(1_k+1_q)$. 
As mentioned above, the factor of $\hv$ is required on dimensional grounds.
Since we assume that all masses are generated from a single VEV, this is the natural---and only---scale available for the matching of the $(n+1)$- and $n$-point amplitudes in the high-energy limit.
As we will verify in the specific examples we work out, this is indeed the correct matching, with no additional numerical prefactor.
Below, we explicitly show this for a massive fermion leg, while the cases of transverse and longitudinal vectors are detailed in \autoref{app:subleading-massive-vector-spinors}.
In all cases, the contributions in~\autoref{eqn:lim} are associated with factorizations of the type,
\beq\label{eq:ntimes3}
{\cal A}_3  \times \frac1{(1_k+q)^2} \times {\cal A}^{(c_n)}_n\,,
\eeq
as illustrated in~\autoref{fig:massive_comp_origins-b}--\ref{fig:massive_comp_origins-d}.
Note that both sides of~\autoref{eqn:lim} are expansions in $1_q$, and hold for finite $\hv$.
For a scalar or fermion leg, this equation gives the full $1_q$ dependence of the contact term, including $1_k+1_q$ factors of the momenta.%
\footnote{
For a massive scalar leg, the only required modification is $1_k\to1_k+1_q$.
This originates from a (gauge invariant) pair of additional Higgses of total momentum $1_q$, with a factorization on the quartic, giving $\lambda\hv^2/(1_k+1_q)^2=1$.
}
For a vector, however, only structures with a single $1_q)$ spinor are captured by this expressions, while structures of the type $(1_q\cdots 1_q)$ require an additional Higgs leg.

In the previous subsection, we saw how ${\cal O}(\hv)$ contributions to the Wilson coefficients arise from $(n+1)$-point contact terms with an additional soft Higgs.
We can now collect the pieces and state the small-mass-limit mapping of the subleading component of the massive contact term to the HE $(n+1)$-point amplitude,
\beqa
\label{eqn:fullim}
{\cal M}^\text{ct}_n(\bs{1},\ldots,\bs{n})\Big\vert_{{\cal O}(v)}
&=& 
\lim_{q\to 0}
\, \hv \, {\cal A}^{(c_{n+1})}_{n+1} +
\sum_{i=1}^{n} \,
\lim_{(iq)=m_i\propto \hv \to 0}\, \hv \, 
{\cal A}^{(c_n)}_{n+1}(1\ldots,n;H(q))
\,,
\eeqa
where $(i q)$ denotes either $\anb{i q}$ or $\sqb{qi}$, depending on the pole to be isolated.
The left-hand side of this equation is the linear $v$ piece in ${\cal M}^\text{ct}_n$ and can also be expressed in the form  $v \Big(\frac{\partial}{\partial v} {\cal M}^\text{ct}_n\Big)_{v=0}$.
On the right-hand side, we rewrote the contribution of \autoref{eqn:lim} in the small mass limit, using the fact that $(1_k+q)^2\sim m_1^2$ near the pole.

With this form, we can further motivate the above procedure.
To determine the mass corrections, we rely on the matching  of the massive and massless amplitudes at high-energies, or equivalently for small $\hv$, with ${\cal M}^\text{ct}_n\sim \hv {\cal A}_{n+1}$.
In this limit, the $\hv$ factor picks up the $(1_k+q)^2$ pole when $q$ is frozen to $1_q$ such that $(1_k+q)^2=m_1^2$.
The Higgs is then either soft or collinear with $1_k$, such that these legs  merge into a single massive leg.
Since we are only interested in massive contact terms, no other poles can appear in ${\cal A}_{n+1}$.
The only relevant amplitudes are therefore those that feature an $n$-point contact term, namely ${\cal A}^{(c_n)}_{n+1}$.
Note that~\autoref{eqn:lim} is precisely what we expect based on the LSZ reduction: the  $(1_k+q)^2$ pole in ${\cal A}_{n+1}$ is eliminated, such that this $(n+1)$-point amplitude is converted into an $n$-point one.
Since the Higgs leg becomes part of the massive external leg, its  only effects on the amplitude are to flip the external polarization into a subleading massive component, and to replace $i_k$ momentum insertions by the massive momentum $i_k+i_q$.

While we only discussed a single massive leg here, it is easy to guess how this generalizes to higher orders.
First, one (or more) additional Higgs is required for each massive leg. The subleading components of the massive amplitude are then associated with the residue of the multi-particle pole containing all $(i_k+i_q)^2$.
For vectors, either one or two Higgs legs are required to account for all the vector polarizations.
In the latter case, the momenta of the two Higgses are automatically symmetrized over, since the radial mode $h$ can come from either one of them.
We will see one such example in~\autoref{sec:nr}.
All in all, the little-group-covariant massive amplitudes have a physical interpretation in terms of massless objects.
Momentum conservation holds at the level of both the HE and LE amplitudes.%
\footnote{The symmetrization over little-group indices in, \eg, transverse vector polarizations also emerges naturally (see~\autoref{eqn:transkq}).}

%%%%%%%%%%%%%%%%%%%%%%%%%%%%%%%%%%%%%%%%%%%%%%%%%%%%%
%%%%%%%%%%%%%%%%%%%%%%%%%%%%%%%%%%%%%%%%%%%%%%%%%%%%%
%%%%%%%%%%%%%%%%%%%%%%%%%%%%%%%%%%%%%%%%%%%%%%%%%%%%%
%%%%%%%%%%%%%%%%%%%%%%%%%%%%%%%%%%%%%%%%%%%%%%%%%%%%%

\paragraph{Massive fermion.}
\label{sec:subleading-fermion}
The $\bs{p}]$ spinor appearing in a $n$-point amplitude featuring a massive fermion line of $+1/2$ helicity category has a leading component $\bs{p}^{I=1}]=k]$, and a subleading component $-\bs{p}^{I=2}]=q]$.
If it is allowed by the gauge symmetry, the leading component comes from an $n$-point contact term in the HE theory, which is schematically of the form $\mathcal{A}_n (k^{h=+1/2},2,\ldots,n) =c_n[k\cdots)$, while the subleading component comes from an $(n+1)$-point amplitude with a negative-helicity fermion and an extra Higgs leg (see \autoref{fig:massive_comp_origins-b}).
This extra Higgs is required in a chiral theory such as the SM, to compensate for the different gauge charge of opposite-helicity fermions.
In the LE amplitude, the fermion momentum is then obtained as the sum $p=k+q$, where $k$ is the momentum of the external fermion leg, and $q$ is the momentum of the extra Higgs leg, which is  frozen such that $(k+q)^2=m_f^2$.
In the small $(k+q)^2=m_f^2$ mass limit, 
\begin{align}
\hv \, {\cal A}_{n+1}(k^{h=-1/2},2,\ldots,n;H(q))
&= \hv\,y\,\anb{k\eta} \, \frac1{(k+q)^2} \, c_n [\eta \cdots)\nonumber \\
&= - y\, \hv\, c_n\, \frac{\langle k (k+q) \cdots)}{\anb{kq}\sqb{qk}} 
=- y\, \hv\, c_n\, \frac{ [q  \cdots)}{\sqb{qk}}\,,
\label{eq:iq-fermion}
\end{align}
where $\eta = -(k+q)$.
Here we used the fact that $\hv$ isolates the pole piece in ${\cal A}_{n+1}$, and the residue is given by the product of the fermion-fermion-Higgs amplitude and the $n$-point contact $\mathcal{A}_n=c_n[\eta\cdots)$.
Using $y \hv = m_f = \anb{kq}=-[kq]$, this simply becomes
\beq
\hv \, {\cal A}_{n+1}(k^{h=-1/2},2,\ldots,n;H(q))  = -c_n \, [q \cdots) = c_n [\bs{p}^{I=2} \cdots)\,.
\eeq
Together with the leading order term $c_n [k\cdots)= c_n [\bs{p}^{I=1} \cdots)$, we get the full LE structure with a massive fermion $c_n [\bs p \cdots)$, and identify its Wilson coefficient as $C_n=c_n + \mathcal{O}(m)$.
One can also easily check that $c_n \langle {\bs p}^{I=1} \cdots )$ is obtained  as a subleading component of the LE amplitude from $\mathcal{A}_n (k^{h=-1/2},2,\ldots,n) =c_n \langle k\cdots)$ with an extra Higgs leg.

%%%%%%%%%%%%%%%%%%%%%%%%%%%%%%%%%%%%%%%%%%%%%%%
\section{Massive four-point applications}
\label{sec:four-point-app}
%%%%%%%%%%%%%%%%%%%%%%%%%%%%%%%%%%%%%%%%%%%%%%
With the understanding gained in the previous section, we can return to the construction of massive EFT contact terms.
As explained above, to obtain all the LE contact terms contributing to a given $n$-point amplitude, one needs to consider $n$- and higher-point HE contact terms with additional Higgs legs.
At some stage, the list of LE massive spinor structures is exhausted; HE contact terms with additional Higgs legs bold into massive spinor structures that were already obtained, multiplied by powers of the Lorentz invariants or $\hv/\Lambda$.
In fact, in the SM examples we study below, a single additional Higgs leg suffices to generate all the massive spinor structures.

In this section, we will apply this procedure to derive two massive four-point contact-term amplitudes in the SMEFT.
In~\autoref{sec:ffvh}, we will first derive the possible structures of a fermion-fermion-vector-scalar amplitude.
We will then consider the specific $\bar u d W h$ case and relate the LE contact-term coefficients to the HE Wilson coefficients of the SMEFT, working to dimension~8.
In~\autoref{sec:WWhh}, we will derive the $WWhh$ amplitude.

Note that the four-point contact terms we will obtain correspond to the novel couplings appearing at the four-point level, which are suppressed purely by powers of $\Lambda$.
In contrast, mass-suppressed contact terms are required in some vector amplitudes, to cancel ${\cal O}(E/m)$ terms in the factorizable parts of the amplitude.
The coefficients of these contact terms are therefore determined by the three-point couplings.
As an aside, we will also apply our top-down approach to derive one such example, namely the standard-model $WWhh$ contact term.

%%%%%%%%%%%%%%%%%%%%%%%%%%%%%%%%%%%%%%%%%%%%%%%%%%%%%
%%%%%%%%%%%%%%%%%%%%%%%%%%%%%%%%%%%%%%%%%%%%%%%%%%%%%
\subsection{\label{sec:ffvh}\texorpdfstring{$f f \fv s$}{ffVs}}
%%%%%%%%%%%%%%%%%%%%%%%%%%%%%%%%%%%%%%%%%%%%%%%%%%%%%
%%%%%%%%%%%%%%%%%%%%%%%%%%%%%%%%%%%%%%%%%%%%%%%%%%%%%
The structure of the massive $f f \fv s$ contact terms can be fully inferred by considering the four- and five-point HE amplitudes from which they originate.
For some choices of fermion chiralities and vector polarizations, the amplitude is compatible with the electroweak SU(2)$\times$U(1) gauge symmetry, and hence originates from a HE four-point amplitude.
For the remaining choices, a single additional Higgs leg (corresponding to $H$, $H^\dagger$ or $\tilde H\sim \sigma^2 H^*$) suffices to render the amplitude gauge invariant.
Thus, it is first allowed at the five-point level.
HE amplitudes with $n\geq2$ additional Higgs legs do not alter the spinor structures featured in the LE amplitude, but rather correct their prefactors by powers of $\hv^2/\Lambda^2$. 
We first discuss the derivation of the structures contributing to the general $f f \fv s$ amplitude,
and then turn to a specific example, namely $\bar u d W h$.

%%%%%%%%%%%%%%%%%%%%%%%%%%%%%%%%%%%%%%%%%%%%%%%%%%%
%%%%%%%%%%%%%%%%%%%%%%%%%%%%%%%%%%%%%%%%%%%%%%%%%%%
\subsubsection{Low-energy \texorpdfstring{$ff \fv s$}{ffvs} contact term enumeration}
\label{sec:ffvs-enumeration}
%%%%%%%%%%%%%%%%%%%%%%%%%%%%%%%%%%%%%%%%%%%%%%%%%%%
%%%%%%%%%%%%%%%%%%%%%%%%%%%%%%%%%%%%%%%%%%%%%%%%%%%
The relevant massless HE amplitudes are  $\mathcal{A}_4(f,f,V,s)$, $\mathcal{A}_4(f,f,s,s)$, $\mathcal{A}_5(f,f,\fv,s,s)$ and $\mathcal{A}_5(f,f,s,s,s)$,
where we use $\fv,f,s$ to denote particles of spin $1,1/2$ and 0 respectively.
The SM gauge symmetry requires same-helicity fermion pairs to only arise with another such pair or with an odd number of Higgses.
Conversely, odd numbers of Higgses only arise together with an odd number of same-helicity fermion pairs~\cite{Durieux:2019siw}.
The helicity amplitudes allowed in SMEFT are thus:
\beqa\label{eq:spin45}\label{eq:spin45last}
\mathcal{A}_4(f^+,f^+,\fv^\pm,s)		:& [12]\langle3123\rangle\,,~[13][23]\,,\nonumber\\
\mathcal{A}_4(f^+,f^-,s,s)		:& [132\rangle\,,\nonumber\\
\mathcal{A}_5(f^+,f^-,\fv^+,s,s)		:& [13]\langle243]\,,~[13]\langle253]\,,~ [1243]\langle243]\,,\\
\mathcal{A}_5(f^+,f^+,s,s,s)		:& [12]\,,~[1342]\,,\nonumber
\eeqa
as well as those obtained from label exchanges and parity conjugation which swaps angle and square spinors.
For each helicity amplitude, we listed the independent spinor structures needed to span the corresponding contact terms (forming bases of \emph{stripped contact terms}~\cite{Durieux:2020gip}).
The most general amplitude with these helicities is given as a linear combination of these structures,  
multiplied by coefficient functions which are power series in the Lorentz invariants, $s_{ij}/\Lambda^2$.
Furthermore, any other spinor structures with the correct little-group weight can be written as linear combinations of the structures in~\autoref{eq:spin45last} multiplied by positive powers of the invariants.
The full amplitudes, including SU(2) factors, must be symmetric over the exchange of identical Higgs legs.

\begin{figure}[t]
\begin{tikzpicture}
\node (uv) [inner sep=0mm] {
	\fmfframe(25,-5)(25,0){\begin{fmfgraph*}(50,50)
	\fmfleft{l0,l1,l2,l3,l4}
	\fmfright{r0,r1,r2,r3}
	\fmf{vanilla}{l1,v,l3}
	\fmf{vanilla}{l2,v}
	\fmfv{lab=ct, d.shape=circle, d.size=8mm, d.fill=0, l.dist=0mm}{v}
	\fmf{dashes,tens=1.5}{v,r1}
	\fmf{vanilla,tens=1.5}{v,r2}
	\fmfv{lab=$1$, l.dist=2mm, l.angl=180}{l3}
	\fmfv{lab=\raisebox{2mm}{$\vdots$}, l.dist=2.25mm, l.angl=180}{l1}
	\fmfv{lab=$n$, l.dist=2mm, l.angl=0}{r2}
	\fmfv{lab=$n\!+\!1$, l.dist=1mm, l.angl=0}{r1}
	\end{fmfgraph*}}
};
\node (p) [below=1cm of uv, xshift=0cm, inner sep=0mm]{
	\fmfframe(25,0)(25,0){\begin{fmfgraph*}(50,50)
	\fmfleft{l0,l1,l2,l3,l4}
	\fmfright{r0,r1,r2,r3}
	\fmf{vanilla}{l1,v,l3}
	\fmf{vanilla}{l2,v}
	\fmfv{lab=ct, d.shape=circle, d.size=8mm, d.fill=0, l.dist=0mm}{v}
	\fmf{phantom,tens=1.5}{v,r1}
	\fmf{vanilla,tens=1.5}{v,r2}
	\fmfv{lab=$1$, l.dist=2mm, l.angl=180}{l3}
	\fmfv{lab=\raisebox{2mm}{$\bs{\vdots}$}, l.dist=2.25mm, l.angl=180}{l1}
	\fmfv{lab=$\bs{n}$, l.dist=2mm, l.angl=0}{r2}
	\fmfv{lab=$\hv$, l.dist=-1mm, l.angl=0}{r1}
	\end{fmfgraph*}}
};
\node [above=8mm of uv, anchor=north]{\minitab{no $p_{n+1}$}};
\draw[->,c3, line width=2] (uv.south) to[out=-90,in=90]
	node[right]{\hyperref[item:ii]{\textcolor{c2}{(ii)}}}
	(p.north);
\node [below=2mm of p, anchor=north]{\minitab{leading HE\\breaks gauge inv.}};
\node (uv0) [inner sep=0mm, left=-1mm of uv] {
	\fmfframe(25,-5)(25,0){\begin{fmfgraph*}(50,50)
	\fmfleft{l0,l1,l2,l3,l4}
	\fmfright{r1}
	\fmf{vanilla}{l1,v,l3}
	\fmf{vanilla}{l2,v}
	\fmfv{lab=ct, d.shape=circle, d.size=8mm, d.fill=0, l.dist=0mm}{v}
	\fmf{vanilla,tens=3}{v,r1}
	\fmfv{lab=$1$, l.dist=2mm, l.angl=180}{l3}
	\fmfv{lab=\raisebox{2mm}{$\vdots$}, l.dist=2.25mm, l.angl=180}{l1}
	\fmfv{lab=$n$, l.dist=2mm, l.angl=0}{r1}
	\end{fmfgraph*}}
};
\node (nop) [below=1cm of uv0, xshift=0cm, inner sep=0mm]{
	\fmfframe(25,0)(25,0){\begin{fmfgraph*}(50,50)
	\fmfleft{l0,l1,l2,l3,l4}
	\fmfright{r1}
	\fmf{vanilla}{l1,v,l3}
	\fmf{vanilla}{l2,v}
	\fmfv{lab=ct, d.shape=circle, d.size=8mm, d.fill=0, l.dist=0mm}{v}
	\fmf{vanilla,tens=3}{v,r1}
	\fmfv{lab=$1$, l.dist=2mm, l.angl=180}{l3}
	\fmfv{lab=\raisebox{2mm}{$\bs{\vdots}$}, l.dist=2.25mm, l.angl=180}{l1}
	\fmfv{lab=$\bs{n}$, l.dist=2mm, l.angl=0}{r1}
	\end{fmfgraph*}}
};
\draw[->,c1, line width=2] (uv0.south) to[out=-90,in=90]
	node[right]{\hyperref[item:i]{\textcolor{c1}{(i)}}}
	(nop.north);
\node [below=2mm of nop, anchor=north]{\minitab{same spin content\\at HE and LE}};
\node (uv1) [inner sep=0mm, right=0mm of uv] {
	\fmfframe(25,-5)(25,0){\begin{fmfgraph*}(50,50)
	\fmfleft{l0,l1,l2,l3,l4}
	\fmfright{r1}
	\fmf{vanilla}{l1,v,l3}
	\fmf{vanilla}{l2,v}
	\fmfv{lab=ct, d.shape=circle, d.size=8mm, d.fill=0, l.dist=0mm}{v}
	\fmf{dashes,tens=3}{v,r1}
	\fmfv{lab=$1$, l.dist=2mm, l.angl=180}{l3}
	\fmfv{lab=\raisebox{2mm}{$\bs{\vdots}$}, l.dist=2.25mm, l.angl=180}{l1}
	\fmfv{lab=$p_n$, l.dist=2mm, l.angl=0}{r1}
	\end{fmfgraph*}}
};
\node (p1) [below=1cm of uv1, xshift=0cm, inner sep=0mm]{
	\fmfframe(25,0)(25,0){\begin{fmfgraph*}(50,50)
	\fmfleft{l0,l1,l2,l3,l4}
	\fmfright{r1}
	\fmf{vanilla}{l1,v,l3}
	\fmf{vanilla}{l2,v}
	\fmfv{lab=ct, d.shape=circle, d.size=8mm, d.fill=0, l.dist=0mm}{v}
	\fmf{dashes,tens=3}{v,r1}
	\fmfv{lab=$\bs{1}$, l.dist=2mm, l.angl=180}{l3}
	\fmfv{lab=\raisebox{2mm}{$\bs{\vdots}$}, l.dist=2.25mm, l.angl=180}{l1}
	\fmfv{lab=$\bs{p_n}$, l.dist=2mm, l.angl=0}{r1}
	\end{fmfgraph*}}
};
\node (p12) [below=1mm of p1, xshift=0cm, inner sep=0mm]{
	\fmfframe(25,0)(25,0){\begin{fmfgraph*}(50,50)
	\fmfleft{l0,l1,l2,l3,l4}
	\fmfright{r1}
	\fmf{vanilla}{l1,v,l3}
	\fmf{vanilla}{l2,v}
	\fmfv{lab=ct, d.shape=circle, d.size=8mm, d.fill=0, l.dist=0mm}{v}
	\fmf{photon,tens=3}{v,r1}
	\fmfv{lab=$\bs{1}$, l.dist=2mm, l.angl=180}{l3}
	\fmfv{lab=\raisebox{2mm}{$\bs{\vdots}$}, l.dist=2.25mm, l.angl=180}{l1}
	\fmfv{lab=$\bs{n}\rangle[\bs{n}$, l.dist=1mm, l.angl=0}{r1}
	\end{fmfgraph*}}
};
\node [above=8mm of uv1, anchor=north]{\minitab{$p_n$ insertion}};
\draw[->,c2, line width=2] (uv1.south)  to[out=-90,in=90]
	node[right]{\hyperref[item:iii]{\textcolor{c3}{(iii)}}}
	(p1.north);
\node (uv2) [inner sep=0mm, right=0mm of uv1] {
	\fmfframe(25,-5)(25,0){\begin{fmfgraph*}(50,50)
	\fmfleft{l0,l1,l2,l3,l4}
	\fmfright{r0,r2,r1,r3}
	\fmf{vanilla}{l1,v,l3}
	\fmf{vanilla}{l2,v}
	\fmfv{lab=ct, d.shape=circle, d.size=8mm, d.fill=0, l.dist=0mm}{v}
	\fmf{dashes,tens=1.5}{v,r1}
	\fmf{dashes,tens=1.5}{v,r2}
	\fmfv{lab=$1$, l.dist=2mm, l.angl=180}{l3}
	\fmfv{lab=\raisebox{2mm}{$\bs{\vdots}$}, l.dist=2.25mm, l.angl=180}{l1}
	\fmfv{lab=$p_n$, l.dist=2mm, l.angl=0}{r1}
	\fmfv{lab=$n\!+\!1$, l.dist=1mm, l.angl=0}{r2}
	\end{fmfgraph*}}
};
\node (p2) [below=1cm of uv2, xshift=0cm, inner sep=0mm]{
	\fmfframe(25,0)(25,0){\begin{fmfgraph*}(50,50)
	\fmfleft{l0,l1,l2,l3,l4}
	\fmfright{r0,r2,r1,r3}
	\fmf{vanilla}{l1,v,l3}
	\fmf{vanilla}{l2,v}
	\fmfv{lab=ct, d.shape=circle, d.size=8mm, d.fill=0, l.dist=0mm}{v}
	\fmf{dashes,tens=1.5}{v,r1}
	\fmf{phantom,tens=1.5}{v,r2}
	\fmfv{lab=$\bs{1}$, l.dist=2mm, l.angl=180}{l3}
	\fmfv{lab=\raisebox{2mm}{$\bs{\vdots}$}, l.dist=2.25mm, l.angl=180}{l1}
	\fmfv{lab=$\bs{p_n}$, l.dist=1mm, l.angl=0}{r1}
	\fmfv{lab=$\hv$, l.dist=-1mm, l.angl=0}{r2}
	\end{fmfgraph*}}
};
\node (p22) [below=1mm of p2, xshift=0cm, inner sep=0mm]{
	\fmfframe(25,0)(25,0){\begin{fmfgraph*}(50,50)
	\fmfleft{l0,l1,l2,l3,l4}
	\fmfright{r0,r2,r1,r3}
	\fmf{vanilla}{l1,v,l3}
	\fmf{vanilla}{l2,v}
	\fmfv{lab=ct, d.shape=circle, d.size=8mm, d.fill=0, l.dist=0mm}{v}
	\fmf{photon,tens=1.5}{v,r1}
	\fmf{phantom,tens=1.5}{v,r2}
	\fmfv{lab=$\bs{1}$, l.dist=2mm, l.angl=180}{l3}
	\fmfv{lab=\raisebox{2mm}{$\bs{\vdots}$}, l.dist=2.25mm, l.angl=180}{l1}
	\fmfv{lab=$\bs{n}\rangle[\bs{n}$, l.dist=1mm, l.angl=0}{r1}
	\fmfv{lab=$\hv$, l.dist=-1mm, l.angl=0}{r2}
	\end{fmfgraph*}}
};
\node [above=8mm of uv2, anchor=north]{\minitab{$p_n$ insertion, no $p_{n+1}$}};
\draw[->,c4, line width=2] (uv2.south) to[out=-90,in=90]
	node[right]{\hyperref[item:iv]{\textcolor{c4}{(iv)}}}
	(p2.north);
\draw[draw=none] ($(p)+(0cm,-3cm)$) -- node[yshift=-1mm]{only transverse vectors}  ($(nop)+(0cm,-3cm)$);
\draw[draw=none] ($(p1)+(0cm,-3cm)$) -- node[yshift=-1mm]{also longitudinal vectors}  ($(p2)+(0cm,-3cm)$);
\node (irlab) [left=0mm of nop, yshift=2mm] {LE:};
\node (uvlab) [above=2cm of irlab] {HE:};
\end{tikzpicture}
\caption{Massive low-energy (LE) contact terms are obtained from the little-group covariantization of the massless high-energy (HE) contact terms corresponding to their leading components.
In the simplest case, this simply amounts to \emph{bolding} massless spinors into massive ones (\hyperref[item:i]{\textcolor{c1}{(i)}}).
An additional \emph{soft} Higgs leg is required to produce massive LE contact terms whose leading high-energy component is a massless HE amplitude that breaks gauge invariance (\hyperref[item:ii]{\textcolor{c2}{(ii)}} and \hyperref[item:iv]{\textcolor{c4}{(iv)}}).
Massless HE amplitudes featuring a scalar momentum insertion are required to generate massive longitudinal vectors at LE (\hyperref[item:iii]{\textcolor{c3}{(iii)}} and \hyperref[item:iv]{\textcolor{c4}{(iv)}}).}
\label{fig:bolding-contact-terms}
\end{figure}
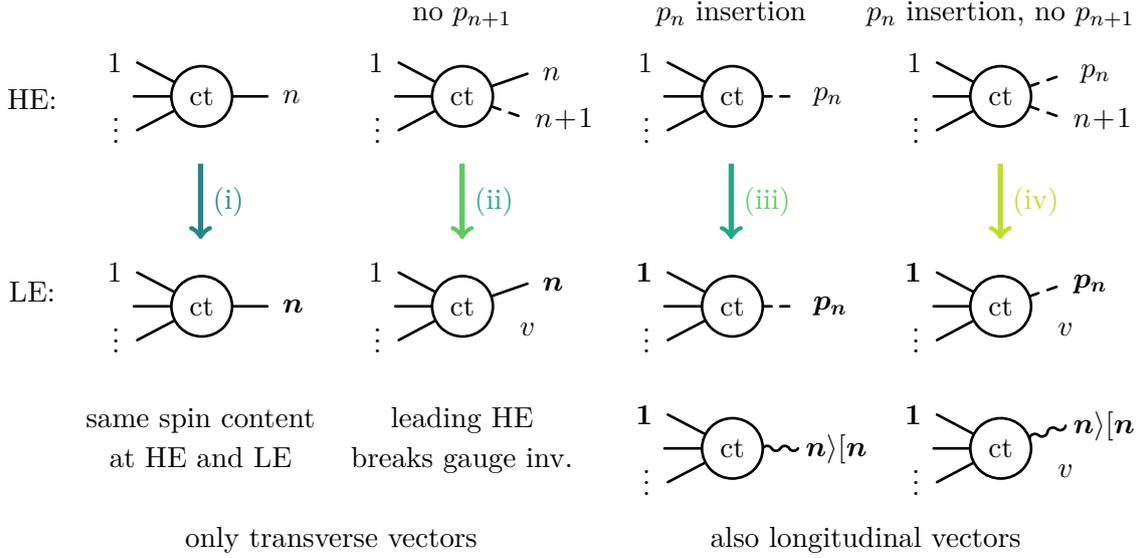

We can now obtain the LE contact terms.
There are four types of qualitatively different patterns that emerge (see \autoref{fig:bolding-contact-terms}), and which would hold quite generally:
\begin{enumerate}
\renewcommand{\labelenumi}{(\roman{enumi})}%
\item \label{item:i} \emph{LE and HE amplitudes of same spin content:}
These cases are the easiest to treat.
The HE amplitudes feature the highest-weight states of the massive representations, and 
should merely be covariantized with respect to the full SU(2) little group(s), which simply amounts to bolding.
From the massless $ff\fv s$ structures on the first line of \autoref{eq:spin45}, we can thus directly get massive ones,
\beq
\Sqb{12} \Anb{3123}\,,~\Sqb{13}\Sqb{23}\,.
\eeq
\item \label{item:ii} \emph{LE amplitude $\leftrightarrow$ HE amplitude of same spin  content plus additional scalar leg(s) required by gauge invariance:} 
Here one Higgs momentum, say 5 (or $4+5$) is first set to zero.
The remaining structures can then be straightforwardly bolded as in \hyperref[item:i]{(i)}.
Thus, $\sqb{13}\braKet 2 5 3$ is eliminated and $\sqb{13}\braKet 2 4 3$ gives
$\Sqb{13}\Asb{243}$.%
\footnote{Note that if we freeze 5 instead, identifying it for instance as $3_q$, $\sqb{13}\braKet 2 5 3\to \sqb{13}\braKet 2 {3_q} 3$ which is bolded to $m_3\Sqb{13}\Anb{23}$, giving a subleading mass correction to one of the structures found below.
}
\item \label{item:iii} \emph{LE longitudinal vector $\leftrightarrow$ HE scalar:}
The remaining four-point amplitude is $[132\rangle$.
With its $ffss$ spin content, it cannot be directly bolded to a $ff\fv s$ structure.
To get the correct little-group weight, one can however write the scalar momentum $p_3$ as $3\rangle [3$ and then bold,
\begin{align}
    \braKet231 &= \anb{23}\sqb{31}  \to \Sqb{13}\Anb{23}\,.
\end{align}
\item \label{item:iv} \emph{LE longitudinal vector $\leftrightarrow$ HE scalar plus additional Higgs leg(s) required by gauge invariance:}
Here we have \hyperref[item:ii]{(ii)} and \hyperref[item:iii]{(iii)} combined. As in \hyperref[item:ii]{(ii)}, one Higgs momentum is
set to zero. We then proceed as in \hyperref[item:iii]{(iii)}. The relevant scalar momentum, $p_3$ in this case, must be replaced by $3\rangle [3$ which can then be bolded. 
In some cases, the scalar momentum $p_3$ does not appear to leading order in the EFT expansion. To get the correct little-group structure one must go to higher orders to obtain factors of $p_3\cdot p_i/\Lambda^2$.
The five-point $\mathcal{A}_5(f^+,f^+,s,s,s)$ amplitude features examples of both types:
\begin{enumerate}
\item The structure $[12]$ does not contain an insertion of $p_3$.
However, terms that are of higher order in the EFT expansion do contain factors of $p_3$ through the Lorentz invariants.
The independent lowest-order terms are
\begin{align}
    \sqb{12} \tdp13\eqCS
    \sqb{12} \tdp23\eqCS
\end{align}
which give
\begin{align}\label{eqn:longv}
    \sqbBS{12}\braKetBS{3}{1}{3}\eqCS
    \sqbBS{12}\braKetBS{3}{2}{3}\eqCS
\end{align}
Note that $\sqb{12} \tdp34$ is not an independent structure once we set $p_5=0$.
\item Setting $p_5=0$, and using momentum conservation, the structure $[1342]=-[1312]$ merely reduces to structures we encountered before.
\end{enumerate}
\end{enumerate}
Collecting the results above, the massive amplitude can be written as a linear combination (with coefficients that are polynomials in $s_{ij}/\Lambda^2$) of the following spinor structures,
\beq
\Sqb{13}\Sqb{23}\,, \Sqb{13}\Anb{23}\,, \Sqb{12}\Anb{3123}\,, \Sqb{13}\Asb{213}\,, \Sqb{12}\Asb{313}\,,
\eeq
plus the ones obtained by swapping square and angle brackets and $1\leftrightarrow2$.
Comparing with the bottom-up derivation of ref.~\cite{Durieux:2019eor}, we note that the top-down procedure above indeed generates all the LE $ff V s$ structures.
The full set of structures, and their  HE origins are summarized in~\autoref{fig:f1f2Wh_origins} (for the specific $\bar udWh$ particle content).
We can further derive the coefficients of the LE amplitude in terms of the HE ones to the desired order in the EFT expansion.
We do this in the next subsection for the $\bar udWh$ amplitude.

In this $ff\fv s$ case, we derived the full set of independent spinor structures from four- and five-point amplitudes.
For massive amplitudes with different spin content, even higher-point massless amplitudes could be needed. At some stage, bolding these massless amplitudes would not produce any
new massive structures; rather, they would lead to spinor structures already obtained from lower point amplitudes multiplied by Lorentz invariants. It will be interesting to study this systematically, but we leave this for future work.

Note that $\Sqb{12}\Anb{3123}$ can be traded for other massive structure via the identity~\cite{Durieux:2019eor},
\beq
	\sqbBS{12}\anbBS{3123}=
	\frac{\sqbBS{12}}{m_V}
	\left(
		\tdp{2}{3}\, \braKetBS{3}1{3} -
		\tdp{1}{3}\, \braKetBS{3}{2}{3}
	\right)
	-\tdp 12\sqbBS{13}\sqbBS{23} 
	-m_{f^\prime}\BraketBS 321\sqbBS{23}-m_f\BraketBS 312\sqbBS{13}\,,
\label{eqn:f1f2Wh_identity}
\eeq
which has a smooth massless limit, and holds for any three four-momenta.
However, in the SMEFT, or in other Higgsed theories, this structure originates from the massless structure
 $[12]\langle3123\rangle$ and its Wilson coefficient is given, to leading order in the $\hv/\Lambda$ expansion, by the HE Wilson coefficient of $[12]\langle3123\rangle$ which is of dimension~8.
 Since we are interested in the list of distinct contact terms corresponding to independent Wilson coefficients, $\Sqb{12}\Anb{3123}$ is kept as an independent structure.

It is also worth commenting on the $1/m_V$ appearing on the right-hand side of \autoref{eqn:f1f2Wh_identity}.
The structures $\Sqb{12}\braKetBS{3}i{3}$, with $i=1,2$, are in the list of independent four-point contact terms above
(see~\autoref{eqn:longv}).
In a bottom-up construction, the factor $1/m_V$ gives the appropriate normalization of structures featuring longitudinal vectors~\cite{Durieux:2020gip}.
Therefore, $\Sqb{12}\braKetBS{3}i{3}/m_V$ first arises at dimension~6.
Specifically, it contributes to the amplitude as
$c_6\Sqb{12}\braKetBS{3}i{3}/{m_V\Lambda^2}$, where $c_6$ is dimensionless.
Good high-energy behavior of the full amplitude, or equivalently, gauge invariance of the HE theory,
implies that any $E/m$ terms cancel between the factorizable and contact-term parts of the LE amplitude in the high-energy limit.
Therefore, $c_6$ is not an independent coupling: its ${\cal O}(m^0)$ piece is determined by the three-point LE couplings.
Thus, $\Sqb{12}\braKetBS{3}i{3}$ should not be included in  the list of independent four-point contact terms at dimension~6.
Instead, an independent coefficient only arises for this structure at dimension~8, corresponding to the $\hv^2/\Lambda^2$ correction to $c_6$, which is not related to any of the lower-point couplings.
We will see the matching of these various coefficients in the next section in an explicit example, namely the $\bar u d Wh$ amplitude.

%%%%%%%%%%%%%%%%%%%%%%%%%%%%%%%%%%%%%%%%%%%%%%%%%%%
%%%%%%%%%%%%%%%%%%%%%%%%%%%%%%%%%%%%%%%%%%%%%%%%%%%
\subsubsection{Low-energy \texorpdfstring{$\bar{u}dWh$}{udWh} contact term matching}
%%%%%%%%%%%%%%%%%%%%%%%%%%%%%%%%%%%%%%%%%%%%%%%%%%%

%%%%%%%%%%%%%%%%%%%%%%%%%%%%%%%%%%%%%%%%%%%%%%%%%%%
%%%%%%%%%%%%%%%%%%%%%%%%%%%%%%%%%%%%%%%%%%%%%%%%%%%
\begin{figure}[tb]
	\centering
	\begin{subfigure}{1\textwidth}
		\centering
			\begin{tikzpicture}[auto]
\node [block] (RRRAtLE) {\footnotesize{}$\begin{aligned}
	&\ampFourPt{Q^c}{D}{W}{H}\\
	&{\color{c1}\Downarrow} \text{ bolding}\\
	&\sqbBS{13}\sqbBS{23}\CudWh{RRR}
\end{aligned} $};
\node [block,right=2mm of RRRAtLE] (LLLAtLE) {\footnotesize{}$\begin{aligned}
	&\ampFourPt{U^c}{Q}{W}{H}\\
	&{\color{c1}\Downarrow} \text{ bolding}\\
	&\anbBS{13}\anbBS{23}\CudWh{LLL}
	\end{aligned} $};
\node [block,right=2mm of LLLAtLE] (LLRAtLE) {\footnotesize{}$\begin{aligned}
	&\ampFourPt{U^c}{Q}{W}{H^\dagger}\\
	&{\color{c1}\Downarrow} \text{ bolding}\\
	&\anbBS{12}\sqbBS{3123}\CudWh{LLR}
	\end{aligned} $};
\node [block,right=2mm of LLRAtLE] (RRLAtLE) {\footnotesize{}$\begin{aligned}
	&\ampFourPt{Q^c}{D}{W}{H^\dagger}\\
	&{\color{c1}\Downarrow} \text{ bolding}\\
	&\sqbBS{12}\anbBS{3123}\CudWh{RRL}
	\end{aligned} $};
\node [blockRed,below=2mm of LLLAtLE] (RLZAtLE) {\footnotesize{}$\begin{aligned}
	&\ampFourPt{Q^c}{Q}{H}{H^\dagger}\\
	&{\color{c2}\Downarrow}\: p_3 \to \ketBraBS{3}{3}\\
	&\sqbBS{13}\anbBS{23}\CudWh{RL0}
	\end{aligned} $};
\node [blockRed,right=2mm of RLZAtLE] (LRZAtLE) {\footnotesize{}$\begin{aligned}
	& \ampFourPt{U^c}{D}{H}{H}\\
	&{\color{c2}\Downarrow}\: p_3 \to \ketBraBS{3}{3}\\
	&\anbBS{13}\sqbBS{23}\CudWh{LR0}
	\end{aligned} $};
			\end{tikzpicture}
			\caption{%
Four-point high-energy origin of massive low-energy $\bar{u} d W h$ spinor structures.
}
	\end{subfigure}\vskip 1em
	\begin{subfigure}{1\textwidth}
		\begin{tikzpicture}[auto]
\node [blockGreen] (RLRAtLE) {\footnotesize{}$\begin{aligned}
	&\ampFivePt{Q^c}{Q}{W}{H^\dagger}{H}\\
	&{\color{c3}\Downarrow} \text{ soft Higgs}\times v \\
	&\sqbBS{13}\BraketBS{3}{1}{2}\CudWh{RLR}
	\end{aligned} $};
\node [blockGreen,right=2mm of RLRAtLE] (LRRAtLE) {\footnotesize{}$\begin{aligned}
	& \ampFivePt{U^c}{D}{W}{H}{H}\\
	&{\color{c3}\Downarrow} \text{ soft Higgs}\times v \\
	&\sqbBS{23}\BraketBS{3}{2}{1}\CudWh{LRR}
	\end{aligned} $};
\node [blockGreenRed,right=2mm of LRRAtLE] (RRZAAtLE) {\footnotesize{}$\begin{aligned}
	&\ampFivePt{Q^{c}}{D}{H}{H}{H^{\dagger}}\\
	&{\color{c4}\Downarrow} \text{ soft Higgs}\times v;\ p_3 \to \ketBraBS{3}{3} \\
	&\sqbBS{12}\BraketBS{3}{\parn{1 \pm 2}}{3}\CudWh{RR0_{A,S}}
	\end{aligned} $};
\node [blockGreen,below=2mm of RLRAtLE] (LRLAtLE) {\footnotesize{}$\begin{aligned}
	&\ampFivePt{U^{c}}{D}{W}{H}{H}\\
	&{\color{c3}\Downarrow} \text{ soft Higgs}\times v \\
	&\anbBS{13}\BraketBS{2}{1}{3}\CudWh{LRL}
	\end{aligned} $}; 
\node [blockGreen,right=2mm of LRLAtLE] (RLLAtLE) {\footnotesize{}$\begin{aligned}
	&\ampFivePt{Q^c}{Q}{W}{H^\dagger}{H}\\
	&{\color{c3}\Downarrow} \text{ soft Higgs}\times v \\
	&\anbBS{23}\BraketBS{1}{2}{3}\CudWh{RLL}
	\end{aligned} $};
\node [blockGreenRed,right=2mm of RLLAtLE] (LLZAAtLE) {\footnotesize{}$\begin{aligned}
	&\ampFivePt{U^{c}}{Q}{H}{H}{H^\dagger}\\
	&{\color{c4}\Downarrow} \text{ soft Higgs}\times v;\ p_3 \to \ketBraBS{3}{3} \\
	&\anbBS{12}\BraketBS{3}{\parn{1 \pm 2}}{3}\CudWh{LL0_{A,S}}
	\end{aligned} $};
		\end{tikzpicture}
		\caption{
Five-point high-energy origins of the massive low-energy $\bar{u} d W h$ spinor structures.
}
	\end{subfigure}
	\caption{\label{fig:f1f2Wh_origins}High-energy origins of the massive low-energy $\bar{u} d W h$ spinor structures.
$\CudWh{}$ are the Wilson coefficients in the LE theory, and their superscripts denote fermion chiralities and the $W$ polarization (for details see text).
The different colors categorize the different Higgsing processes
(\hyperref[item:i]{\textcolor{c1}{(i)}},
\hyperref[item:ii]{\textcolor{c2}{(ii)}},
\hyperref[item:iii]{\textcolor{c3}{(iii)}},
\hyperref[item:iv]{\textcolor{c4}{(iv)}})
taking massless SMEFT contact terms to massive ones.
	} 
\end{figure}
%%%%%%%%%%%%%%%%%%%%%%%%%%%%%%%%%%%%%%%%%%%%%%%%%%%
%%%%%%%%%%%%%%%%%%%%%%%%%%%%%%%%%%%%%%%%%%%%%%%%%%%
We take all particles to be incoming. 
Thus, a massless right-handed (RH) (left-handed (LH)) chiral fermion, with momentum $p$, is associated with a $\Ket{p}$ ($\ket{p}$) helicity spinor,
while a massless RH (LH) chiral anti-fermion, with momentum $p$, is associated with a $\ket{p}$ ($\Ket{p}$) helicity spinor.
In the following, we ignore SU(3)$_C$, which can trivially be restored with a Kronecker delta contracting the quark and anti-quark color indices.
The quarks in the  $\bar u d W^+ h$ amplitude originate in the HE from the LH doublet $Q$, and RH singlets $U,D$, with,
\beq
\ket{2} \leftrightarrow Q_2=d\,,~~~ \Ket{2} \leftrightarrow D\,,~~~
\Ket{1} \leftrightarrow Q^{c,1}=u\,,~~~ \ket{1} \leftrightarrow U^c\,,
\eeq
where $c$ denotes charge conjugation and $1,2$ are weak $\suN{2}$ isospin indices.
As discussed above, the massive four-point contact terms with helicity categories allowed by gauge invariance can be matched directly to four-point HE contact terms.
The massive contact terms whose helicity categories are forbidden by SU(2)$\times$U(1) in the HE theory are generated at the five-point level with the addition of a single Higgs leg.

We will need the HE four-point amplitudes corresponding to the first two lines of \autoref{eq:spin45},
\begin{subequations}\label{eqn:f1f2Wh_SM_HE_4pt_and_5ptbis}
	\begin{align}
	&\label{eqn:f1f2Wh_SM_HE_4ptbis} \begin{aligned}
			& \ampFourPt{Q^{c,i}}{D}{W^{a},h_3}{H_j}  \,,\    
			  \ampFourPt{U^c}{Q_i}{W^{a},h_3}{H_{j}} \,,\   \\
			  &\ampFourPt{Q^{c, j}}{Q_i}{H_k}{H^{\dagger l}} \,,\ 
			  \ampFourPt{U^{c}}{D}{H_i}{H_j}\,,
	\end{aligned}
	\end{align}
	and the five-point amplitudes corresponding to the last two,
	\begin{equation}
	\begin{gathered}
		\ampFivePt{Q^{c, i}}{Q_j}{W^{a},h_3}{H_k}{H^{\dagger l}}\,,\
		\ampFivePt{U^{c}}{D}{W^{a},h_3}{H_i}{H_j}\,,\\
		\ampFivePt{Q^{c,i}}{D}{H_j}{H_k}{H^{\dagger l}}\,,\
		\ampFivePt{U^{c}}{Q_i}{H_j}{H_k}{H^{\dagger l}} 
		 \eqndot
	\end{gathered}
	\label{eqn:f1f2Wh_SM_HE_5ptbis}
	\end{equation}
\end{subequations}
Here $i,j,k,l$ are (anti-)fundamental weak SU(2) indices, $h_3$ is the $W$ helicity, and $a$ is an adjoint $\suN{2}$ index. 

Generically, each HE amplitude ${\cal A}$ features several independent SU(2) structures.
It is convenient to define \emph{reduced amplitudes} (denoted by $A$'s), which multiply these different SU(2) factors.
To reduce clutter, the full expressions for the amplitudes, including the SU(2) factors, are given in~\autoref{app:udWh}. Here we will only display the reduced amplitudes. These feature spinor structures from~\autoref{eq:spin45}.
At four-point, one gets:
\begin{subequations}\label{eqn:f1f2Wh_HE_4pt}
	\begin{align}
	    \label{eqn:udWh_QQHH_reduced}\redAmpFourPtN{r}{Q^{c}}{Q}{H}{H^\dagger} &=  [132\rangle \, \CoeffFunc{Q^{c} QH H^\dagger}{r}  \quad \text{for~}r=1,2 \,, \\
	    \redAmpFourPtN{-}{U^{c}}{D}{H}{H} &= \braKet{1}{\parn{3-4}}{2}\, 
	    \CoeffFunc{U^{c} D H H}{-} \eqncomma \\
	    \redAmpFourPt{Q^{c}}{D}{W,-}{H} &= \sqb{12}\anb{3123} \,\CoeffFunc{Q^{c} D W^{-} H}{}     \eqncomma \\
    	    \redAmpFourPt{U^c}{Q}{W,-}{H} &= \anb{12}\sqb{3123} \, \CoeffFunc{U^{c} Q W^{+} H}{}    \eqncomma \\
	    \redAmpFourPt{Q^{c}}{D}{W,+}{H} &= \sqb{13}\sqb{23}\, \CoeffFunc{Q^{c} D W^{+} H}{}    \eqncomma 	    \label{eqn:redFourPt_cf_Qc_D_Wp_H} \\
    	    \redAmpFourPt{U^c}{Q}{W,+}{H} &= \anb{13}\anb{23} \, \CoeffFunc{U^{c} Q W^{-} H}{}    \eqncomma
	\end{align}
\end{subequations}
where the coefficient functions $\CoeffFunc{}{}$ are power series in $s_{ij}/\Lambda^2$.
In the first line above, and throughout this section, $r$ runs over the allowed SU(2) factors. 
In the second line, and throughout this section, $\pm$ subscripts denote (anti)symmetry with respect to identical Higgs legs,
such that the full expression, including the SU(2) factor is symmetric.

At five-point, one gets: 
\begin{subequations}\label{eqn:f1f2Wh_HE_5pt}
		\begin{align}
    		\redAmpFivePtN{r}{Q^c}{Q}{W,+}{H}{H^\dagger} =& \sqb{13}\parn{\braKet 2{p^-_{45}}3 \, 
    		  \CoeffFunc{Q^{c} Q  W^+ H H^\dagger}{1, r} + \braKet 2{p^+_{45}}3 \, \CoeffFunc{Q^{c} Q  W^+ H H^\dagger}{2, r}} \nn\\
    		&+ \sqb{1243}\braKet{2}{4}{3} \, \CoeffFunc{Q^{c } Q W^+ H H^\dagger}{r}
    		 \quad 
    		 \text{for~} r=1,2,3\,,\\
    		\redAmpFivePtN{r}{Q^c}{Q}{W,-}{H}{H^\dagger} =& \anb{23}\parn{\braKet 3{p^-_{45}}1 \, \CoeffFunc{Q^{c} Q  W^- H H^\dagger}{1,r} +\braKet 3{p^+_{45}}1 \, \CoeffFunc{Q^{c} Q  W^- H H^\dagger}{2,r} } \nn\\
    		&+\anb{2143}\braKet{3}{4}{1} \, \CoeffFunc{Q^{c } Q  W^- H H^\dagger}{r}
    		\quad 
    		 \text{for~} r=1,2,3\,, \\
    		\redAmpFivePtN{+}{U^c}{D}{W,+}{H}{H}  =&\sqb{23}\parn{\braKet 1{p^-_{45}}3 \,\CoeffFunc{U^{c } D  W^+ H^2}{1,-} +\braKet 1{p^+_{45}}3 \, \CoeffFunc{U^{c } D  W^+ H^2}{2,+}}\nn\\ & +
    		\sqb{21p^{-}_{45}3}\braKet{1}{p^{-}_{45}}{3} \, \CoeffFunc{U^{c } D  W^+ H^2}{3,+} 
    		\eqncomma \\
    		\redAmpFivePtN{+}{U^c}{D}{W,-}{H}{H}  =& \anb{13}\parn{\braKet 3{p^-_{45}}2 \, \CoeffFunc{U^{c } D  W^- H^2}{1,-} +\braKet 3{p^+_{45}}2 \, \CoeffFunc{U^{c } D  W^- H^2}{2,+}} \nn\\ &+
    		\anb{12p^{-}_{45}3}\braKet{3}{p^{-}_{45}}{2} \, \CoeffFunc{U^{c } D  W^- H^2}{3,+}
    		\eqncomma \\
    		\label{eqn:udWh_QDHHH_reduced}\redAmpFivePtN{\pm}{Q^c}{D}{H}{H}{H^\dagger}  =& \sqb{12} \, \CoeffFunc{Q^{c } D H^2 H^\dagger}{1,\pm}+\parn{\sqb{1342}-\sqb{1432}} \, \CoeffFunc{Q^{c } D H^2 H^\dagger}{2,\mp}\eqncomma\\
    		\redAmpFivePtN{\pm}{U^c}{Q}{H}{H}{H^\dagger}  =&\anb{12} \,  \CoeffFunc{U^{c } Q H^2 H^\dagger}{1,\pm}+\parn{\anb{1342}-\anb{1432}}\, \CoeffFunc{U^{c } Q H^2 H^\dagger}{2,\mp}\eqncomma
	\end{align}%
\end{subequations}%
where $p^\pm_{45}\equiv p_4 \pm p_5$.
The five-point amplitude coefficient functions, $\CoeffFunc{}{}$, are  power series in $s_{ij}/\Lambda^2$, and  $\epsilon^{\mu\nu\rho\sigma}p_{1,\mu}p_{2,\nu}p_{3,\rho}p_{4,\sigma}/\Lambda^4$.

The different spinor and Lorentz structures appearing in~\twoeqs{eqn:f1f2Wh_HE_4pt}{eqn:f1f2Wh_HE_5pt} can be bolded into massive structures as described in the previous \autoref{sec:ffvs-enumeration}, to give the full LE $\bar{u}dWh$ amplitude,
\begin{align}\label{eqn:f1f2Wh_nf}
\ampFourPtLE{\bar{u}}{d}{W}{h} =&
\sqbBS{13}\sqbBS{23}\CudWh{RRR}
+\sqbBS{12}\parn{\braKet \bs3{\parn{\bs1+\bs2}\bs3} \CudWh{RR0_{A}} 
+\braKet \bs3{\parn{\bs1-\bs2}\bs3} \CudWh{RR0_{S}} }\nonumber \\& 
+\sqbBS{13}\anbBS{23} \CudWh{RL0}
+\BraketBS 312\sqbBS{13}\CudWh{RLR}
+\braKetBS 321\anbBS{23}\CudWh{RLL}\nonumber \\
 & +\anbBS{13}\sqbBS{23}\CudWh{LR0}
+\BraketBS 321\sqbBS{23}\CudWh{LRR}
+\braKetBS 312\anbBS{13}\CudWh{LRL}\nonumber \\
 & +\anbBS{13}\anbBS{23}\CudWh{LLL}
+\anbBS{12}\parn{\braKet \bs3{\parn{\bs1+\bs2}\bs3} \CudWh{LL0_{A}}
+\braKet \bs3{\parn{\bs1-\bs2}\bs3} \CudWh{LL0_{S}}}\nonumber \\
 & +\anbBS{12}\sqbBS{3123} \CudWh{LLR}
 +\sqbBS{12}\anbBS{3123} \CudWh{RRL}\,.
\end{align}
The coefficients 
$\CudWh{}$'s  are determined in terms of the $\CoeffFunc{}{}$'s of~\twoeqs{eqn:f1f2Wh_HE_4pt}{eqn:f1f2Wh_HE_5pt}
up to $\hv^2/\Lambda^2$ corrections.
We now discuss a few representative examples of this matching in detail. 
The massless $[13][23]$ structure featured in the $Q^cDWH$ amplitude simply bolds to $\Sqb{13}\Sqb{23}$, so
\beq
\CudWh{RRR} = \frac{1}{2} \, \CoeffFunc{Q^{c} D W^{+} H}{} +{\cal O}(\hv^2/\Lambda^2)\,.
\eeq
Here, the factor of $1/\sqrt{2}$ comes from $H \supset h/\sqrt{2}$.
Explicitly we have,
\beq
\CoeffFunc{Q^{c} D W^{+} H}{}= c_6/\Lambda^2 + c_8\, s_{12}/\Lambda^4 + 
c^\prime_8\, s_{13}/\Lambda^4 +\cdots \,,
\eeq
where the $c$'s are SMEFT Wilson coefficients.
Then
\beq
\CudWh{RRR}= \frac{1}{2}\left( c_6/\Lambda^2 + c_8\,\tilde{s}_{12}/\Lambda^4 + 
c^\prime_8\,\tilde{s}_{13}/\Lambda^4 \right)+ {\cal O}(\hv^2/\Lambda^2)+\cdots \,.
\eeq
This matching gives $\CudWh{RRR}$ to leading order in $\hv$ only, because we only used
the four-point HE amplitude contributing to this coefficient.
There are two sorts of ${\cal O}(m^2)$ corrections.
First, higher-order corrections, scaling as $\hv^2/\Lambda^2$ come from six-point (and higher) amplitudes with an extra $H^\dagger$ and $H$.
Second, there is a ${\cal O}(m^2)$ discrepancy between
the massive $\tilde{s}_{ij}=2 \bs{p}_i\cdot \bs{p}_j$ and the massless $s_{ij}$ by $\sim m^2$.
Thus, this second effect is a higher-order correction that we can consistently neglect.
Note that the $Q^c D W  H$ amplitude also contributes to  other LE amplitudes, namely $\bar{d} d Z h$, and $\bar{d} d \gamma h$, so the same HE coefficients will appear in these amplitudes (up to numerical factors involving the weak mixing angle).

As another example, consider the massless four-point amplitude of~\autoref{eqn:udWh_QQHH_reduced}, where the bolding is less trivial. 
The spinor structure $[132\rangle $ is bolded into $- \Sqb{13}\Anb{23}$ and 
 we have,
\begin{align}
    \CudWh{RL0} = -\frac{1}{2} \, \CoeffFunc{Q^{c} QH H^\dagger}{2} + \mathcal{O}\parn{\hv^2 / \Lambda^2}\eqndot
\end{align}
Next consider the five-point amplitude $\redAmpFivePtN{\pm}{Q^c}{D}{H}{H}{H^\dagger}$ of~\autoref{eqn:udWh_QDHHH_reduced}.\footnote{$\epsilon^{\mu\nu\rho\sigma} p_{1\mu}p_{2\nu}p_{3\rho}p_{4\sigma}$ is ${\cal O}\parn{m_3^2}$ 
in the limit that 5 is frozen, so we can omit its contribution in the coefficient functions.}
The first term on the right-hand side of~\autoref{eqn:udWh_QDHHH_reduced} is
\begin{equation*}
\sqb{12}\, \CoeffFunc{}{1,\pm} \eqndot
\end{equation*}
In order to get the correct little-group factor for the massive $W$,
we need one factor of $p_3$ from $\CoeffFunc{}{1,\pm}$, just as expected for a Goldstone amplitude.
In fact, it is simpler in this case to rewrite the five-point amplitude in terms of the Goldstone and physical Higgs.
Working to up dimension~8, this amplitude can be written as,
\beq\label{eqn:ghh}
A(Q^c_1,D,G^+,h,h)= \frac{b_8}{\Lambda^4}\, [12] \left(s_{31}+s_{32}\right) + 
\frac{b^\prime_8}{\Lambda^4}\,[12] \left(s_{31}-s_{32}\right)\,,
\eeq
where we used the fact that the amplitude features $p_3$ and is symmetric under $4\leftrightarrow 5$ exchange.% 
\footnote{The coefficients $b_8$ and $b^\prime_8$ can be written in terms of those appearing in the expansion~\autoref{eqn:udWh_QDHHH_reduced}.
The amplitude $A(Q^cDHHH^\dagger)$ contains six independent coefficients at dimension~8: The
polynomials $\CoeffFunc{}{1\pm}$, $\CoeffFunc{}{2\pm}$ of~\autoref{eqn:udWh_QDHHH_reduced} can be written at this dimension as
$\CoeffFunc{}{1+}=c_1 s_{34} + c_2(s_{13}+s_{14})+c_3(s_{23}+s_{24})$,
$\CoeffFunc{}{1-}=c_4(s_{13}-s_{14})+c_5(s_{23}-s_{24})$,
$\CoeffFunc{}{2+}=c_6$,
$\CoeffFunc{}{2-}=0$, where $c_i$ are order $1/\Lambda^4$.
Thus $b_8$ and $b^\prime_8$ of~\autoref{eqn:ghh} are given by $b_8/\Lambda^4=c_2+c_3$, $b^\prime_8/\Lambda^4=c_2-c_3$.
}
Bolding this and dropping mass-suppressed terms, we get
\beq\label{eqn:ghhbold}
 \frac{b_8}{\Lambda^4} \,\Sqb{12} \asb{\bs3(\bs1+\bs2)\bs3} +   \frac{b^\prime_8}{\Lambda^4}\, \Sqb{12}\asb{\bs3(\bs1-\bs2)\bs3}\,,
 \eeq
so that
\begin{align}\label{eqn:finalbold}
    \CudWh{RR0_A} &= \hv \frac{b_8}{\Lambda^{4}}+\cdots
     \,,\\
    \CudWh{RR0_S} &= \hv 
    \frac{b^\prime_8}{\Lambda^{4}}+\cdots
     \,.
\end{align}
The result can be compared with a standard Feynman diagram calculation. 
It is easy to check that, at dimension~8, there are  six independent coefficients in the $Q^cDHHH^\dagger$ amplitude.
Indeed, there are six independent dimension-8 operators that contribute to this amplitude~\cite{Li:2020gnx}.
Only two of these contribute to HE Goldstone amplitude, or to  $\bar u d W^+ h$ as we see in~\autoref{eqn:ghh}, in agreement with the result~\autoref{eqn:ghhbold}.

%%%%%%%%%%%%%%%%%%%%%%%%%%%%%%%%%%%%%%%%%%%%%%%%%%%
%%%%%%%%%%%%%%%%%%%%%%%%%%%%%%%%%%%%%%%%%%%%%%%%%%%
\subsection{\texorpdfstring{$W^{+}W^{-}hh$}{WWhh}}
\label{sec:WWhh}
%%%%%%%%%%%%%%%%%%%%%%%%%%%%%%%%%%%%%%%%%%%%%%%%%%%
%%%%%%%%%%%%%%%%%%%%%%%%%%%%%%%%%%%%%%%%%%%%%%%%%%%
Another example of interest is the massive $W^+ W^- hh$ amplitude. 
In the following, we will determine both the renormalizable and non-renormalizable contact terms in the LE theory, from the HE amplitudes.
%%%%%%%%%%%%%%%%%%%%%%%%%%%%%%%%%
%%%%%%%%%%%%%%%%%%%%%%%%%%
\subsubsection{EFT contact terms}
The following four- and five-point contact terms contribute to the massive $W^{+}W^{-}hh$ LE  amplitude,
\begin{equation}
\begin{gathered}
\ampFourPt{W^{a},h_1}{W^{b},h_2}{H^{\dagger i}}{H_j}\eqncomma \quad
\ampFourPt{H^{\dagger i}}{H^{\dagger j}}{H_{k}}{H_{l}}\eqncomma\\
\ampFivePt{W^{a},h_1}{H^{\dagger i}}{H^{\dagger j}}{H_{k}}{H_{l}}\eqndot
\end{gathered}
\end{equation}
As we will see, these amplitudes, when bolded, saturate the full list of massive spinor structures of the massive $WWhh$ amplitude given in ref.~\cite{Durieux:2020gip}.
Higher-point amplitudes, therefore, merely correct  the Wilson coefficients by $\mathcal{O}(\hv^2/\Lambda^2)$ terms. 
As in the previous section, we first strip off the $\suN{2}$ group theory factors.
Since we are working with the mass eigenstates $W^\pm$, it is convenient to work with corresponding SU(2) generators, instead of the usual Pauli matrices.
We define the various group theory factors needed in~\hyperref[sec:su2_basis]{appendix\,\ref*{sec:su2_basis}}.
The amplitudes are then given by,
\begin{align}
\ampFourPt{W^{a},h}{W^{b},h}{H^{\dagger i}}{H_j} &= 
\suTwoTgab{a}{b} \suTwoNDelta{j}{i} \, A_{+}\parn{\dots} + \suTwoCabc{a}{b}{c}\suTwoTildeGen{c}{j}{i} \, A_{-}\parn{\dots}\eqncomma\\    
\ampFourPt{W^{a},h}{W^{b},-h}{H^{\dagger i}}{H_j} &= 
\suTwoTgab{a}{b}\suTwoNDelta{j}{i} \, A_{1}\parn{\dots} + \suTwoCabc{a}{b}{c}\suTwoTildeGen{c}{j}{i} \, A_{2}\parn{\dots}\eqncomma\\    
\label{eqn:HHHH_HE}\ampFourPt{H^{\dagger i}}{H^{\dagger j}}{H_{k}}{H_{l}} &= \parn{\suTwoNDelta{k}{i}\suTwoNDelta{l}{j} + \suTwoNDelta{l}{i}\suTwoNDelta{k}{j}} \, A_{+}\parn{\dots}+
\epsilon_{ij}\epsilon^{kl} \, A_{-}\parn{\dots}\eqncomma\\ 
\ampFivePt{W^{a},h_1}{H^{\dagger i}}{H^{\dagger j}}{H_{k}}{H_{l}} &= \parn{\suTwoTildeGen{a}{k}{i} \suTwoNDelta{l}{j} \pm \suTwoTildeGen{a}{l}{i}\suTwoNDelta{k}{j}} \, A_{\pm}\parn{\dots}\eqncomma
\end{align}
where, to avoid clutter, the arguments of the reduced amplitudes have been omitted. 
Symmetry under exchange of identical bosons is denoted by a $\pm$ subscript. 
In the first two amplitudes, to make Bose symmetry manifest, we distinguished the case of same- and opposite-helicity bosons.
At the four-point level, each reduced contact term is spanned by a single spinor structure, dictated by the external helicities, 
\begin{subequations}\label{eqn:WWHH_HE}
    \begin{align}
    \redAmpFourPtN{\pm}{W,+}{W,+}{H^{\dagger}}{H} &= 
        \sqb{12}^2 \, \CoeffFunc{W^+ W^+ H^\dagger H}{\pm}\eqncomma \\
    \redAmpFourPtN{i}{W,+}{W,-}{H^{\dagger}}{H} &= 
        \braKet{2}{\parn{3-4}}{1}^{2} \, \CoeffFunc{W^+ W^- H^\dagger H}{r}\eqndot
    \end{align}
\end{subequations}
Here $r=1,2$, the $\CoeffFunc{}{}$ coefficient functions are  power series in the Lorentz invariants, and the $\pm$ subscript denotes (anti)symmetrization under $1\leftrightarrow2$ exchange.
The five-point contact terms require several spinor structures,
{\small%
\begin{align}
\label{eqn:WHHHH_5pt}
\redAmpFivePtN{\pm}{W,+}{H^{\dagger}}{H^{\dagger}}{H}{H} =& 
    \parn{\sqb{1231} - \sqb{1321}} \, \CoeffFunc{W^+ H^\dagger H^\dagger HH}{\mp,\pm}
    \nn\\ & +
    \parn{\sqb{1451} - \sqb{1541}} \, \CoeffFunc{W^+ H^\dagger H^\dagger HH}{\pm,\mp}\\ &+
    \parn{\sqb{1241} - \sqb{1251} + \sqb{1351} - \sqb{1341}} \, \CoeffFunc{W^+ H^\dagger H^\dagger HH}{\mp,\mp}\eqncomma\nn
\end{align}}%
where again the $\CoeffFunc{}{}$ coefficient functions are power series in the Lorentz invariants, and the first and second $\pm$ subscripts denote (anti)symmetrization with respect to $2\leftrightarrow3$ exchange and $4\leftrightarrow5$ exchange  respectively.
The parity conjugates of the contact terms in~\twoeqs{eqn:WWHH_HE}{eqn:WHHHH_5pt} are obtained by exchanging square and angle brackets.

We now derive the LE $WWhh$ amplitude as in~\autoref{sec:ffvh}.
The four-point HE amplitudes of~\autoref{eqn:WWHH_HE} are easily covariantized with respect to the massive little group, giving four distinct helicity categories: $++$, $+-$ and their parity conjugates. 
The $00$ helicity category is obtained from $\redAmpFourPtN{\pm}{H^{\dagger}}{H^{\dagger}}{H}{H}$ by bolding the appropriate  Lorentz invariants into spinor structures containing 1 and 2 momenta. 
The leading order terms that can be bolded into the appropriate structures are, sacrificing explicit Bose symmetry,
\begin{subequations}\label{eqn:HHHH_HE2}
\begin{align}
    \redAmpFourPtN{+}{H^{\dagger}}{H^{\dagger}}{H}{H}  &= \frac{c_{6+}}{\Lambda^{2}}\,\parn{\tdp12+\tdp34}
    + \frac{c_{8+}}{\Lambda^{4}}\, \tdp13\tdp23 + \dots \eqncomma \\ 
    \redAmpFourPtN{-}{H^{\dagger}}{H^{\dagger}}{H}{H} &= \frac{
    c_{8-}}{\Lambda^{4}} \,\left(\tdp13-\tdp23\right)\tdp12  + \dots \eqndot
    \end{align}
\end{subequations}
where the $c$ coefficients are pure numbers, and the ellipses stand for higher-order terms.
Bolding these terms, we get 
\begin{gather}
\anbBS{12}\sqbBS{12}\left(
\frac{c_{6+}}{\Lambda^{2}}+\cdots\right) +
\braKetBS 131\braKetBS 232\left(
\frac{c_{8+}}{\Lambda^{4}} + \cdots\right)
\eqncomma\\
\anbBS{12}\sqbBS{12}\,\left( 
\frac{c_{8-}}{\Lambda^4} \left(\tilde{s}_{13}-\tilde{s}_{23}\right) + \cdots \right)\eqndot
\end{gather}

Finally, the five-point amplitudes of~\autoref{eqn:WHHHH_5pt} 
give additional contributions at ${\cal O}(\hv)$.
To get these, we first take one of the Higgses, say 5, to be soft 
and bold the remaining structures such that 2 gets the appropriate little-group weight
and multiply the resulting expressions by $\hv$.
Thus for example,
\begin{equation}
\sqb{1231}\to \#  \hv \sqbBS{12}\braKetBS 231\eqndot
\end{equation}
Assembling these results, we obtain the contact-term $WWhh$
four-point 
\begin{align}\label{eqn:WWhh_massive_full}
\mathcal{M}_4^{\text{ct}}\parn{\bs 1_{W^{+}}, \bs 2_{W^{-}}, \bs 3_{h}, \bs 4_{h}}  =&\anbBS{12}\sqbBS{12}{C}_{4,+}+\braKetBS 1{\parn{3-4}}1\braKetBS 2{\parn{3-4}}2C_{4,+}^{\prime}+\nonumber \\
 & +\sqbBS{12}^{2}C_{4}^{W_{1}^{+}W_{2}^{+}}+\anbBS{12}^{2}C_{4}^{W_{1}^{-}W_{2}^{-}}\nonumber \\
 & +\braKetBS 1{\parn{3-4}}2^{2}C_{4}^{W_{1}^{-}W_{2}^{+}}+\braKetBS 2{\parn{3-4}}1^{2}C_{4}^{W_{1}^{+}W_{2}^{-}}\nonumber \\
 & +\hv\sqbBS{12}\braKetBS 2{\parn{3-4}}1{C}_{5,-}^{W_{1}^{+}}+\hv\anbBS{12}\braKetBS 1{\parn{3-4}}2{C}_{5,-}^{W_{1}^{-}}\nonumber \\
 & +\hv\sqbBS{12}\braKetBS 1{\parn{3-4}}2{C}_{5,-}^{W_{2}^{+}}+\hv\anbBS{12}\braKetBS 2{\parn{3-4}}1{C}_{5,-}^{W_{2}^{-}}\eqncomma
\end{align}
where each ${C}$ is some combination of the $\CoeffFunc{}{}$'s in 
\hyperref[eqn:WWHH_HE]{eqs.\,(\ref*{eqn:WWHH_HE})}--\hyperref[eqn:HHHH_HE2]{(\ref*{eqn:HHHH_HE2})}.

%%%%%%%%%%%%%%%%%%%%%%%%%%%%%%%%%%%%%%%%%%%%%%%%%%
%%%%%%%%%%%%%%%%%%%%%%%%%%%%%%%%%%%%%%%%%%%%%%%%%%
\subsubsection{Standard-model contact term}
\label{sec:wwhh_sm}
%%%%%%%%%%%%%%%%%%%%%%%%%%%%%%%%%%%%%%%%%%%%%%%%%%
%%%%%%%%%%%%%%%%%%%%%%%%%%%%%%%%%%%%%%%%%%%%%%%%%%
The standard-model $W^+W^-hh$ amplitude also features an $m_W^2$-suppressed contact term.
This term can be determined by calculating the factorizable four-point massive amplitude, and requiring good high-energy behavior, as done
for example in ref.~\cite{Bachu:2019ehv}.
It is interesting to ask however whether we can get this contact term directly, along the lines of our top-down derivation.
The HE standard-model $W^+W^-H^\dagger H$ amplitude is given by,
{\small\begin{align}
   \ampFourPt{W^{a},+}{W^{b},-}{H^{\dagger i}}{H_j} &=2 g^2_2\,
    \frac{\sab{13 2}}{\sab{2 3 1}}\,\bigg[ 
    \frac{1}{4}\suTwoTgab{a}{b} \suTwoNDelta{j}{i} -\frac{i}{2}\suTwoCabc{a}{b}{c}\suTwoTildeGen{c}{j}{i}\frac{s_{13}-s_{23} }{s_{12}} 
    \bigg]\,.
\end{align}}%
Setting $a=+, b=-, i=j=2$, and summing the contributions with $3_{H^\dagger}4_H$ and $3_H4_{H^\dagger}$, the $WWhh$ 
component of the HE amplitude is
\begin{align}\label{eqn:HE_WWHH_smcontact_term}
    \mathcal{A}_4\parn{1^+_{W^+},2^-_{W^-},3_h,4_h} &= \frac{g^2_2}2\,
    \frac{[1_k 3_k 2_k \rangle}{[2_k 3_k 1_k \rangle}\,.
\end{align}
A factor $(1/\sqrt{2})^2$ is included, to account for the fact that $(H^\dag)^{i=2} \supset h/\sqrt{2}$ and $H_{i=2} \supset h/\sqrt{2}$.
Here and hereafter, for future convenience, we  replace the spinor labels $i$ by $i_k$ in the HE amplitudes.
This expression should match the high-energy limit of the LE $\mathcal{M}_4\parn{\bs 1_{W^{+}}, \bs 2_{W^{-}}, \bs 3_{h}, \bs 4_{h}}$ which features the spinors $1_q$, $2_q$, etc.
In the high-energy limit, these $q$ spinors map to the reference spinors appearing in polarization vectors and can be replaced by any of the $k$ spinors in the problem, thanks to gauge invariance.
Reversing this mapping, one can attempt to recast~\autoref{eqn:HE_WWHH_smcontact_term} in the form of a local LE contact term, by rewriting some $k$'s in terms of the LE $q$'s.
For $1_q$ and $2_q$ both aligned with $3_k$, we have 
\beq
3_k\rangle =\frac{\anb{1_k 3_k}}{m_1} 1_q\rangle \,, \quad 
3_k]=\frac{[3_k 2_k]}{m_2} 2_q]\,,
\eeq 
so that a local contact term of the form $\frac{g_2^2}{2 m_W^2}\,\sqb{1_k2_q}\anb{1_q2_k}$ is indeed obtained from \autoref{eqn:HE_WWHH_smcontact_term}.
Bolding this to $-\,\frac{g_2^2}{2 m_W^2}\sqb{{\bf 1}^{I=1} {\bf 2}^{K=2}}\anb{{\bf 1}^{J=1} {\bf 2}^{L=2}} $, we get the LE contact term,
\beq
\mathcal{M}_4\parn{\bs 1_{W^{+}}, \bs 2_{W^{-}}, \bs 3_{h}, \bs 4_{h}} = - \frac{g_2^2}{2m_W^2}\, \Sqb{12}\Anb{12}\,.
\label{eqn:LE_WWhh_factrizable}
\eeq

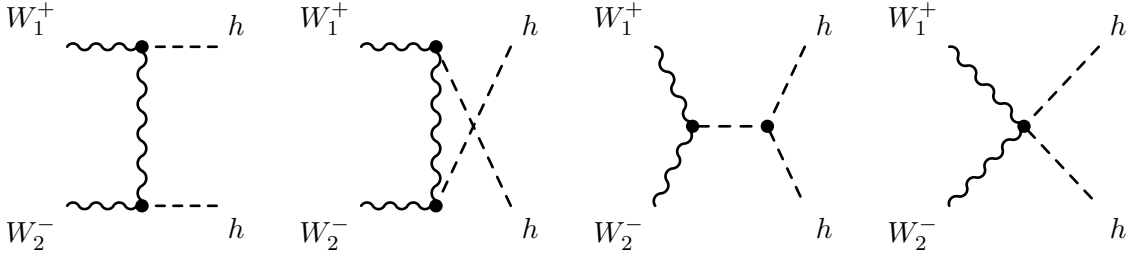
\begin{figure}[t]
\fmfframe(15,15)(15,15){\begin{fmfgraph*}(70,60)         
	\fmfleftn{i}{2} 
	\fmfrightn{o}{2}
	\fmfdotn{v}{2} 
	\fmf{phantom}{i1,v1,o1}
	\fmf{phantom}{i2,v2,o2}
	\fmffreeze
	\fmf{dashes}{o2,v2}
	\fmf{dashes}{o1,v1}
	\fmf{photon}{i1,v1,v2,i2}
	\fmflabel{$h$}{o2}
	\fmflabel{$h$}{o1}
	\fmflabel{$W^+_1$}{i2}
	\fmflabel{$W^-_2$}{i1}
\end{fmfgraph*}}\hfill
\fmfframe(15,15)(15,15){\begin{fmfgraph*}(70,60)         
\fmfleftn{i}{2} 
	\fmfrightn{o}{2}
	\fmfdotn{v}{2} 
	\fmf{phantom}{i1,v1,o1}
	\fmf{phantom}{i2,v2,o2}
	\fmffreeze
	\fmf{dashes}{o1,v2}
	\fmf{dashes}{o2,v1}
	\fmf{photon}{i1,v1,v2,i2}
	\fmflabel{$h$}{o2}
	\fmflabel{$h$}{o1}
	\fmflabel{$W^+_1$}{i2}
	\fmflabel{$W^-_2$}{i1}
\end{fmfgraph*}}\hfill
\fmfframe(15,15)(15,15){\begin{fmfgraph*}(70,60)         
\fmfleftn{i}{2} 
	\fmfrightn{o}{2}
	\fmfdotn{v}{2} 
	\fmf{photon}{i1,v1,i2}
	\fmf{dashes}{v1,v2}
	\fmf{dashes}{o2,v2,o1}
	\fmflabel{$h$}{o2}
	\fmflabel{$h$}{o1}
	\fmflabel{$W^+_1$}{i2}
	\fmflabel{$W^-_2$}{i1}
\end{fmfgraph*}}\hfill
\fmfframe(15,15)(15,15){\begin{fmfgraph*}(70,60)         
	\fmfleftn{i}{2} 
	\fmfrightn{o}{2}
	\fmfdotn{v}{1} 
	\fmf{dashes}{o2,v1,o1}
	\fmf{photon}{i2,v1,i1}
	\fmffreeze
	\fmflabel{$h$}{o2}
	\fmflabel{$h$}{o1}
	\fmflabel{$W^+_1$}{i2}
	\fmflabel{$W^-_2$}{i1}
\end{fmfgraph*}}
\caption{\label{fig:LE_WWHH_4pt_diagrams} Factorizable and contact-term contributions to the LE $W^+ W^- hh$ amplitude.
}
\end{figure}

To see why this guess works, consider the LE standard-model amplitude
$\mathcal{M}_4\parn{\bs 1_{W^{+}}, \bs 2_{W^{-}}, \bs 3_{h}, \bs 4_{h}}$
(see~\autoref{fig:LE_WWHH_4pt_diagrams}),
\begin{align}
&\mathcal{M}_4\parn{\bs 1_{W^{+}}, \bs 2_{W^{-}}, \bs 3_{h}, \bs 4_{h}} =\frac{g^{2}_2}{s_{13}-m_{W}^{2}}\parn{\frac{\BraketBS 131\BraketBS 242}{2m_{W}^{2}}-\anbBS{12}\sqbBS{12}}\nonumber \\
  & \quad +\frac{g^{2}_2}{s_{14}-m_{W}^{2}}\parn{\frac{\BraketBS 141\BraketBS 232}{2m_{W}^{2}}-
  \anbBS{12}\sqbBS{12}}
  - \frac{i g_2c_{hhh}}{m_W}\,\frac{\anbBS{12}\sqbBS{12}}{s_{12}-m_h^2}
  -\frac{g^2_2}{2m_W^2}\anbBS{12}\sqbBS{12}\,,
  \label{eqn:LE_WWhh_amplitude}
\end{align}
where the last term is the contact term, and the $c_{hhh}$ term originates from the Higgs self-coupling.
For opposite-helicity vectors, the high-energy limit of this expression is
\begin{align}
&\mathcal{M}_4\parn{{\bs 1}^{11}_{W^{+}}, {\bs 2}^{22}_{W^{-}}, \bs 3_{h}, \bs 4_{h}}
\xrightarrow{\text{HE limit}} \nonumber \\
&\frac{g_{2}^{2}}{2 m_W^2}\,\parn{\anb{1_{q}2_{k}}\sqb{1_{k}2_{q}}-\frac{\braKet{2_{k}}4{2_{q}}\Braket{1_{k}}3{1_{q}}}{\tilde s_{13}}-\frac{\braKet{2_{k}}3{2_{q}}\Braket{1_{k}}4{1_{q}}}{\tilde s_{14}}}+\mathcal{O}\parn{m_{W}^2}\eqndot
\label{eqn:LE_WWhh_pm_HE_limit_amplitude1}
\end{align}
In full generality, the spinors $1_q\rangle$ and $2_q]$ can be decomposed along two other spinors:
\begin{equation}
1_q\rangle =
  \frac{\anb{1_k 1_q}}{\anb{1_k\xi_1}} \xi_1\rangle
+ \frac{\anb{\xi_1 1_q}}{\anb{\xi_1 1_k}} 1_k\rangle
\eqncomma\qquad
2_q] =
  \frac{\sqb{2_k 2_q}}{\sqb{2_k\xi_2}} \xi_2]
+ \frac{\sqb{\xi_2 2_q}}{\sqb{\xi_2 2_k}} 2_k]
\eqndot
\label{eq:xi_decomposition}
\end{equation}%
Choosing $\xi_1\rangle$ such that $\anb{1_k\xi_1}\ne0$ and $\anb{1_q\xi_1}=0$, and similarly for $\xi_2]$, one gets:
\begin{align}
&\mathcal{M}_4\parn{{\bs 1}^{11}_{W^{+}}, {\bs 2}^{22}_{W^{-}}, \bs 3_{h}, \bs 4_{h}}
\xrightarrow{\text{HE limit}} 
- \frac{g_2^2}{2}\frac{
	\anb{\xi_12_{k}}\sqb{1_{k}\xi_2}
	-{\asb{2_{k}4\xi_2}\sab{1_{k}3\xi_1}}/{\tilde s_{13}}
	-{\asb{2_{k}3\xi_2}\sab{1_{k}4\xi_1}}/{\tilde s_{14}}
}{
	\anb{1_k\xi_1}\sqb{2_k\xi_2}
}
\eqndot
\label{eqn:LE_WWhh_pm_HE_limit_amplitude}
\end{align}%
Remarkably, this expression is independent of the choice of $\xi_{1,2}$ which is obviously a consequence of gauge invariance.
One can for instance take $\xi_1\rangle[\xi_2=2_k\rangle[1_k$, or $3_k\rangle[3_k$, $4_k\rangle[4_k$.
Either way, the expression~\autoref{eqn:LE_WWhh_pm_HE_limit_amplitude} reduces to the high-energy result~\autoref{eqn:HE_WWHH_smcontact_term}.
It is then no accident that we got the LE contact term by choosing $1_q,2_q\propto 3_k$: with this choice, only the LE contact term contribution in~\autoref{eqn:LE_WWhh_amplitude} survives at high-energies.
In contrast, for $\xi_1\rangle[\xi_2=2_k\rangle[1_k$, only the factorizable part of~\autoref{eqn:LE_WWhh_amplitude} contributes.
It is also instructive to consider the Feynman diagram computation of the massless amplitude, written in terms of arbitrary reference momenta.
It is easy to check that $1_q$ and $2_q$ directly map to these reference momenta, and that the choice $1_q, 2_q\propto 3_k$ isolates the four-point vertex diagram.

%%%%%%%%%%%%%%%%%%%%%%%%%%%%%%%%%%%%%%%%%%%%%%%%%%
%%%%%%%%%%%%%%%%%%%%%%%%%%%%%%%%%%%%%%%%%%%%%%%%%%
%%%%%%%%%%%%%%%%%%%%%%%%%%%%%%%%%%%%%%%%%%%%%%%%%%
%%%%%%%%%%%%%%%%%%%%%%%%%%%%%%%%%%%%%%%%%%%%%%%%%%

This example clearly demonstrates the equivalence of  perturbative unitarity and the UV gauge symmetry. 
For transverse $W$ polarizations, the factorizable LE amplitude and the  contact term of~\autoref{eqn:HE_WWHH_smcontact_term} are both finite in the high-energy limit, and no contact term is required to restore gauge invariance.
In contrast, for longitudinal $W$'s, the LE factorizable amplitude and the contact term both scale as $s_{12}/m_W^2$, and the contact term is required for this contribution to cancel.

%%%%%%%%%%%%%%%%%%%%%%%%%%%%%%%%%%%%%%%%%%%%%%%%%%
%%%%%%%%%%%%%%%%%%%%%%%%%%%%%%%%%%%%%%%%%%%%%%%%%%
\section{Massive three-point applications}
%%%%%%%%%%%%%%%%%%%%%%%%%%%%%%%%%%%%%%%%%%%%%%%%%%%
%%%%%%%%%%%%%%%%%%%%%%%%%%%%%%%%%%%%%%%%%%%%%%%%%
\label{sec:three_point}
Massive three-point amplitudes are the simplest type of massive contact terms, 
so it is natural to extend our top-down approach to derive them.
These amplitudes were studied extensively in the literature~\cite{Arkani-Hamed:2017jhn,Christensen:2018zcq,Durieux:2019eor,Bachu:2019ehv},
mainly using a bottom-up construction, and matching the \emph{leading-order} behavior of the amplitudes to the HE theory, which corresponds to (un)bolding.
In this case too, Higgsing can be viewed as a unification of different HE amplitudes~\cite{Arkani-Hamed:2017jhn}.
Our focus here will be on the \emph{sub-leading} parts of the amplitudes, mostly of the type discussed in~\autoref{sec:subleading}.
For simplicity, we consider a Higgsed U(1) toy model, with chiral fermions which obtain mass via the Higgs mechanism, and include both renormalizable and non-renormalizable interactions. This toy model captures all the relevant ingredients of the SM tree amplitudes. 
As above, we start by simply bolding the allowed massless three-point amplitudes into massive ones.
For the renormalizable three-point couplings, this essentially parallels the analysis of~\cite{Bachu:2019ehv} which derived the SM renormalizable couplings by matching the massive and massless amplitudes.
We then apply the methods of \autoref{sec:general} to show how mass-suppressed contributions to the amplitudes are generated
from four-, five-, or six-point amplitudes, by freezing some Higgs legs.
Some details of the transition from four- to three-point kinematics are discussed in~\subappref{app:4to3}.

%%%%%%%%%%%%%%%%%%%%%%%%%%%%%%%%
\subsection{HE theory}
%%%%%%%%%%%%%%%%%%%%%%%%%%%%%%%%
We consider a chiral U(1) gauge theory toy model.
The HE (massless) theory consists of two fermions, the left-handed $\chiL$ (helicity $-1/2$) 
and the right-handed $\etaR$ (helicity $+1/2$), a vector $\gamma$, 
and a charged complex scalar $\phi$.
The analogy with the Lagrangian picture is most transparent if we decompose the Higgs field in terms of the Goldstone $\theta$  and the physical Higgs $h$. 
We implicitly assume that additional fermions exist, on top of the ones considered here, such that the theory is anomaly free.
Since, however, we are only interested in tree-amplitudes, these will not affect the analysis.

The three-point amplitudes involving the 
positive-helicity vector are given by,
\begin{gather}
{\cal A}_3(1^+_{\chiL^c}, 2^{-}_{\chiL}, 3_\gamma^+) = \sqrt{2}g Q_\chi \frac{[13]^2}{[21]}\,, 
\quad {\cal A}_3(1^-_{\etaR^c}, 2^{+}_{\etaR}, 3_\gamma^+) = \sqrt{2}g Q_\eta \frac{[23]^2}{[12]}\,,\nonumber \\
{\cal A}_3(1_{\theta}, 2_{h}, 3_\gamma^+) = \sqrt{2}ig Q_\phi \frac{[13][23]}{[12]}\,,
\label{eq::gauge_3pt_amps}
\end{gather}
where 
$g$ is the U(1) gauge coupling,\kf\footnote{
With the factors of 2 and $\sqrt2$ appearing in the HE three-point amplitudes, the various couplings correspond to the usual Lagrangian conventions for the gauge, Yukawa and dipole couplings, \eg,  $g$ appears in the
covariant derivative  as $D_{\mu} \psi = \left( \partial_{\mu} + i Q_{\psi} g A_{\mu}\right) \psi$. 
} 
and $Q_\chi, Q_\eta$, $Q_\phi$ are the U(1) charges of  $\chiL$, $\etaR$ and $\phi$ respectively.
If $Q_\phi = Q_\eta-Q_\chi$,
the following Yukawa interactions are allowed,
\begin{align}
{\cal A}_3(1^-_{\etaR^c}, 2^{-}_{\chiL}, 3_h) &=-\frac{y_\phi}{\sqrt{2}} \langle 12\rangle\,, \;\;\; {\cal A}_3(1^+_{\chiL^c}, 2^{+}_{\etaR}, 3_{h}) =-\frac{y^*_\phi }{\sqrt{2}}[12] \,, \nonumber
\\
{\cal A}_3(1^-_{\etaR^c}, 2^{-}_{\chiL}, 3_\theta) &=-i\frac{y_\phi}{\sqrt{2}} \langle 12\rangle\,, \;\;\; {\cal A}_3(1^+_{\chiL^c}, 2^{+}_{\etaR}, 3_{\theta}) =i\frac{y^*_\phi}{\sqrt{2}} [12] \,.
\end{align}
Note that we can take $y_\phi$ to be real by rescaling $1\rangle$ or $2\rangle$.
In a vector-like theory, the following dimension-five, three-point amplitudes are allowed as well
\begin{align}\label{eqn:fordip}
{\cal A}_3(1^-_{\etaR^c}, 2^{-}_{\chiL}, 3_\gamma^-)  \sim 
\langle13\rangle\langle23\rangle\,, \;\;\; {\cal A}_3(1^+_{\chiL^c}, 2^{+}_{\etaR}, 3_\gamma^+)  \sim 
[13][23] \;\;\;\;\;\; \text{if}\;\;\; Q_\eta=Q_\chi\,.
\end{align}
However, since we are interested in Higgsed, SM-like, chiral theories, we will take $Q_\phi=Q_\eta- Q_\chi \neq 0$.
The amplitudes~\autoref{eqn:fordip} are then forbidden.
Instead, the following four-point  amplitudes are allowed and will prove to be relevant for our discussion,
\begin{align}
{\cal A}^{\text{ct}}_4(1^-_{\etaR^c}, 2^{-}_{\chiL}, 3_\gamma^-, 4_h) = \frac{2 \cDipole}{\Lambda^2}\langle13\rangle\langle23\rangle \,, \;\;\;
{\cal A}^{\text{ct}}_4(1^+_{\chiL^c}, 2^{+}_{\etaR}, 3_\gamma^+, 4_{h})  =\frac{2\cbarDipole}{\Lambda^2}[13][23] \,, \nonumber
\\
{\cal A}^{\text{ct}}_4(1^-_{\etaR^c}, 2^{-}_{\chiL}, 3_\gamma^-, 4_\theta) = \frac{2i\cDipole}{\Lambda^2}\langle13\rangle\langle23\rangle \,, \;\;\;
{\cal A}^{\text{ct}}_4(1^+_{\chiL^c}, 2^{+}_{\etaR}, 3_\gamma^+, 4_{\theta})  =-\frac{2i\cbarDipole}{\Lambda^2}[13][23] \,, 
\label{eq::non_renorm_ints}
\end{align}
where $\Lambda$ is the cutoff scale of the HE theory and $\cDipole$ is dimensionless.
These expressions give the full tree amplitudes, since their factorizable parts vanish due to helicity selection rules~\cite{Cheung:2015aba}. 
Recall that $\Lambda$ is assumed to be hierarchically bigger than the Higgs VEV, \ie, $\hv\ll\Lambda$.

\subsection{LE theory}

The LE (massive) theory contains a massive fermion $\psi$, a massive vector $Z$ and a massive scalar $h$.
Several three-point amplitudes are allowed, namely
\begin{align}
{\cal M}_3({\bf {1}}_{\psi^c},{\bf {2}}_{\psi},{\bf {3}}_{Z})\,, \quad
{\cal M}_3({\bf {1}}_{h},{\bf {2}}_{Z},{\bf {3}}_{Z})\,, \quad
{\cal M}_3({\bf {1}}_{\psi^c},{\bf {2}}_{\psi},{\bf {3}}_{h})\,.
\end{align}
Let us investigate how these amplitudes are constructed from the HE theory.
For concreteness, we focus on ${\cal M}_3({\bf {1}}_{\psi^c},{\bf {2}}_{\psi},{\bf {3}}_{Z})$ for the remainder of this section.
To facilitate the matching of the massive and massless amplitudes, we display the mapping of the massive and massless external states,
\begin{equation}
\begin{gathered}
{\bs {1}}_{\psi^c}^{I=1}] \equiv {1_k}] 
\leftrightarrow
1_{\chiL^c}] \,, 
\quad~~ 
{\bs {1}}^{I=2}_{\psi^c} \rangle 
\equiv {1_k}\rangle 
\leftrightarrow
1_{\etaR^c}\rangle\,,\\
{\bs {2}}^{I=1}_{\psi}]
\equiv
2_k] 
\leftrightarrow
2_{\etaR}] \,, 
\quad~~ 
{\bs {2}}^{I=2}_{\psi}\rangle
\equiv 
2_k\rangle 
\leftrightarrow 2_{\chiL}\rangle \,, \\
{\bs {3}}_{Z}^{I=1}] [ {\bs {3}}_{Z}^{J=1} 
\equiv  3_k] [3_k
\leftrightarrow  3_{\gamma}] [3_{\gamma}\,, 
\quad 
{\bs {3}}_{Z}^{I=2}\rangle \langle{\bs {3}}_{Z}^{J=2} 
\equiv  3_k\rangle \langle 3_k
\leftrightarrow  3_{\gamma}\rangle \langle 3_{\gamma}\,,\\
{\bs {3}}_{Z}\rangle [ {\bs {3}}_{Z} ^{\{IJ\}=\{12\}}
\equiv \frac{1}{\sqrt{2}}\left(  {\bs {3}}_{Z}^{I=1}\rangle [ {\bs {3}}_{Z}^{I=2} + 
{\bs {3}}_{Z}^{I=2}\rangle [ {\bs {3}}_{Z}^{I=1} \right) 
\equiv \frac{1}{\sqrt{2}}\Bigl( 3_k \rangle [ 3_k - 3_q\rangle [ 3_q 
 \Bigr) 
 \\ 
{\bs {p}}_{h,\alpha\dot\alpha} \equiv 3_k\rangle [3_k + 3_q \rangle [3_q
\,.
\end{gathered}
\end{equation}

%%%%%%%%%%%%%%%%%%%%%%%%%%%%%%%%%%%%%%%%%%%%%%%%%%%%%%%%%%%%%%%%%%%%%%%%%%%%%%%%%%
\subsection{\texorpdfstring{$\mathcal{O}(m^0)$}{O(m 0)}: massive amplitudes arising from HE three-point amplitudes}\label{sec:m0vector}
%%%%%%%%%%%%%%%%%%%%%%%%%%%%%%%%%%%%%%%%%%%%%%%%%%%%%%%%%%%%%%%%%%%%%%%%%%%%
We begin the construction of the massive amplitude by bolding the three-point  
fermion-fermion-photon HE amplitude.
The  bolding is somewhat subtle in this case, owing to the non-local nature of the three-point HE amplitude.
It can however be achieved by introducing an arbitrary spinor $r\rangle$.
Starting with ${\cal A}_3(1^+_{\chiL^c}, 2^{-}_{\chiL}, 3_\gamma^+)$ of~\autoref{eq::gauge_3pt_amps}, multiplied and divided by $\anb{3 r}$, one gets,
\begin{align}
\label{eqn:3rewrite}
{\cal A}_3(1^+_{\chiL^c}, 2^{-}_{\chiL}, 3_\gamma^+) =
\sqrt{2}g Q_\chi \frac{[13]^2}{[21]}=
\sqrt{2}g Q_\chi \frac{[ 13] [ 1 3 r \rangle}{[21]\langle3 r  \rangle}=
\sqrt{2}g Q_\chi \frac{[ 13] [1 2  r \rangle}{[12]\langle3 r \rangle}
=\sqrt{2}g Q_\chi \frac{[ 13] \langle 2 r \rangle}{\anb{3r}}\,,
\end{align}
where in the second step we used momentum conservation.
Identifying furthermore $r\rangle=3_q\rangle$ (or alternatively taking $3_q\rangle\propto r\rangle $)
with $\anb{33_q}=m_Z$, this expression gives the $O(m^0)$ piece of the massive amplitude,
\beq
\sqrt{2}g Q_\chi \frac{[ 1_k3_k] \langle 2_k 3_q \rangle}{m_Z}\,,
\eeq
which can be readily bolded to
\begin{align}
\label{eqn:ffzbold}
\sqrt{2}g Q_\chi\frac{[{\bf{1 3 }}]\langle  {\bf{2 3}} \rangle}{m_Z}\,.
\end{align}
Clearly, the form of the final expression in~\autoref{eqn:3rewrite} is related to the gauge symmetry, and in fact $r$ is nothing but the reference momentum for $3$. By introducing it,
the physical non-locality of the amplitude is traded for a seemingly unphysical singularity.
Furthermore,  $3_q$ of the massive amplitude is mapped into the arbitrary reference momentum of the HE amplitude, and the singularity again has a clear physical meaning: it is the inverse 
mass.\footnote{Alternatively, the massive amplitude~\autoref{eqn:ffzbold} can be obtained via one of its subleading components, as  in~\autoref{sec:subleading}, by starting from a four-point amplitude with an extra Higgs leg, as discussed in the next \autoref{sec:3pt-order-m1}.}
Repeating this procedure for ${\cal A}_3(1^-_{\etaR^c}, 2^{+}_{\etaR}, 3_\gamma^+)$ one gets the full massive amplitude,
\begin{align}
\label{eqn:mffz}
{\cal M}_3({\bf {1}}_{\psi^c}, {\bf {2}}_{\psi}, {\bf {3}}_{Z}) = \sqrt{2}g Q_\chi\frac{[{\bf{1 3 }}]\langle  {\bf{2 3}} \rangle}{m_Z}+\sqrt{2}g Q_\eta\frac{\langle{\bf{1 3 }}\rangle[  {\bf{2 3}} ]}{m_Z}\,.
\end{align}
Additional structures may appear as a result of non-renormalizable interactions, in particular those of~\autoref{eq::non_renorm_ints}.
However, at this point, for simplicity, we take $\Lambda \to \infty$  and consider only renormalizable interactions.  
One can check that the two terms above reproduce, by construction, the correct high energy amplitudes for the little-group indices $(1,2,11),(2,1,11)$, and similarly for  $(1\leftrightarrow2)$.

With the LE amplitude~\autoref{eqn:mffz} in hand, we can examine the remaining choices of little-group indices that are allowed in the HE theory,
namely the (would-be) Yukawa interaction $(1,1,\{12\})$ and $(2,2,\{12\})$.
After some manipulations, the former takes the form
\begin{align}
\label{eqn:massamp}
{\cal M}_3({\bf {1}}^{1}_{\psi^c}, {\bf {2}}^{1}_{\psi},{\bf {3}}^{\{12\}}_{Z})
=
 -g Q_\phi \frac{m_\psi}{m_Z}[1_k 2_k] + gQ_\phi\frac{m_\psi}{m_Z}\langle 1_q  2_q\rangle
 +2 g Q_\chi \frac{[1_k 3_q  2_q\rangle}{m_Z} + 2 g Q_\eta \frac{[2_k 3_q   1_q\rangle}{m_Z}\,,
\end{align}
where we used $\anb{i_k i_q}= \sqb{i_q i_k} = m_\psi$ for $i=1,2$ (which holds for real $y_\phi$), and $Q_\phi = Q_{\eta}-Q_{\chi}$.
Matching this to the appropriate high-energy amplitude,
\begin{align}
{\cal M}_3({\bf {1}}^{1}_{\psi^c},{\bf {2}}^{1}_{\psi},{\bf {3}}^{\{12\}}_{Z})\Big\vert_{{\cal O}(m^0)} = i {\cal A}_3(1^+_{\chiL^c}, 2^{+}_{\etaR}, 3_{\theta})\,,
\end{align}
we recover the relation between the fermion and vector masses,
\begin{align}
\frac{m_{\psi}}{m_Z} 
=  \frac{y_\phi /\sqrt{2}}{g Q_\phi}\,.
\label{eqn::fermion_vector_masses}
\end{align}

%%%%%%%%%%%%%%%%%%%%%%%%%%%%
\subsection{\texorpdfstring{$\mathcal{O}(m)$}{O(m 1)}: massive amplitudes arising from HE four-point amplitudes}
\label{sec:3pt-order-m1}
%%%%%%%%%%%%%%%%%%%%%%%%%%%%%%%%%%%%%%%%%%%%%%%%%%%%%%%%%%%%%%%%%%%%%
%%%%%%%%%%%%%%%%%%%%%%%%%%%%%%%%%%%%%%%%%%%%%%%%%%%%%%%%%%%%%%%%%%%%%
For the remaining six choices of little-group indices, namely $(1,2,\{12\}), (2,2,11), (2,2,22)$ (plus $1\leftrightarrow2$), the amplitudes vanish at $\mathcal{O}(m^0)$: the global charges of the massless particles involved forbid these amplitudes from appearing in the high-energy theory.
These components  arise from amplitudes with additional Higgs legs.

Let us start with the longitudinal-vector amplitude,  ${\mathcal M}_3({\bf {1}}^{1}_{\psi^c},{\bf {2}}^{2}_{\psi},{\bf {3}}^{\{12\}}_{Z})$.
Expanding~\autoref{eqn:mffz} for this choice
of little-group indices we have,
\begin{align}\label{eqn:massive_1212}
&{\mathcal M}_3({\bf {1}}^{1}_{\psi^c},{\bf {2}}^{2}_{\psi},{\bf {3}}^{\{12\}}_{Z})
=
\frac{y_\phi}{\sqrt{2}}\langle 1_q 2_k \rangle
+\frac{y_\phi}{\sqrt{2}}[1_k 2_q]
+2g Q_\chi\frac{[1_k 3_q 2_k  \rangle}{m_Z}
-2g Q_\eta\frac{\langle1_q 3_q 2_q  ]}{m_Z} 
\\
&=\frac{y_\phi}{\sqrt{2}}m_1 \left(\frac{\langle 1_q 2_k \rangle}{\langle 1_k 1_q \rangle}
\right)+\frac{y_\phi}{\sqrt{2}}m_2\left(\frac{[1_k 2_q]}{[2_q 2_k]}\right)
+2g Q_\chi m_Z\left(\frac{[1_k 3_q 2_k  \rangle}{\langle 3_k 3_q \rangle [3_q 3_k]}\right)
+{\cal O}(m^3)\,,\nonumber
\end{align}
where  in the second step the mass dependence is made manifest, and we distinguish $m_1$, $m_2$ for clarity, although $m_1=m_2=m_\psi$. 
The first three  terms of~\autoref{eqn:massive_1212}  
can be obtained from a four-point amplitude with one additional Higgs leg. The last term, on the other hand, is
${\cal O}(m^3)$ and originates from a six-point HE amplitude.
\\ \\
Our starting point is the massless amplitude ${\mathcal A}_4(1^+_{\chiL^c}, 2^{-}_{\chiL}, 3_{\theta}, 4_{h})$ given in~\autoref{eq:massles_4pt_pmGBh}.
As discussed in~\autoref{sec:subleading}, each of the ${\cal O}(m)$ terms of the massive amplitude is obtained by freezing the momentum 4, identifying it with either $1_q$, $2_q$, or $3_q$, and multiplying the result by $\hv$.
Thus, for example, the first term of~\autoref{eqn:massive_1212}, which is proportional to $m_1$, originates from the $\tdp14$ pole, 
\beq
\lim_{\anb{14} 
\to0}
\, \hv
    i {\mathcal A}_4(1^+_{\chiL^c}, 2^{-}_{\chiL}, 3_{\theta}, 4_{h})
= \lim_{\anb{14}
\to0}
-\left(\frac{y_\phi\, }{\sqrt{2}} \right)^2  \frac{\hv\,\langle 24 \rangle}{\langle 14 \rangle}=
\frac{y_\phi}{\sqrt{2}}\,\anb{ 1_q 2_k}\,,
\label{eq:1212_matching_aa}
\eeq
where at the last step we identified $4$ and $2$ with $1_q$ and $2_k$, respectively. 
We used the fact that $\anb{14}=m_1\propto \hv$;
in the small mass limit, multiplying by $\hv$ isolates the pole in the four-point amplitude. The second and third terms in~\autoref{eqn:massive_1212} are similarly obtained by taking
$\anb{24}$ and $\anb{34}$ to zero,
\begin{align}
\lim_{\sqb{24}
\to0}
\, \hv
    i {\mathcal A}_4(1^+_{\chiL^c}, 2^{-}_{\chiL}, 3_{\theta}, 4_{h})
&= \lim_{\sqb{24}
\to0}
\left(\frac{y_\phi\, }{\sqrt{2}} \right)^2  \frac{\hv\,[14]}{[ 42]}=
\frac{y_\phi}{\sqrt{2}}\,\sqb{1_k 2_q}\,,
\label{eq:1212_matching_bb}\\
%%%
 \lim_{\anb{34} 
\to0}
\, \hv
    i {\mathcal A}_4(1^+_{\chiL^c}, 2^{-}_{\chiL}, 3_{\theta}, 4_{h})
&= \lim_{\anb{34}\to0}
2(g Q_\phi) (g Q_\chi) \frac{\hv [1 4 2\rangle}{s_{34}}
=
2 g Q_\chi \frac{[1_k 3_q 2_k \rangle}{m_Z}\,.
\label{eq:1212_matching_cc}
\end{align}
Note that the matching in \hyperref[eq:1212_matching_aa]{eqs.\,(\ref*{eq:1212_matching_aa})}, \hyperref[eq:1212_matching_bb]{(\ref*{eq:1212_matching_bb})}, and \hyperref[eq:1212_matching_cc]{(\ref*{eq:1212_matching_cc})} required the expected relations between the fermion and vector masses and the scale $\hv$;
\begin{align}
    m_\psi = \frac{y_\phi }{\sqrt{2}} \hv\,, \;\;\; m_Z = g\, Q_{\phi} \hv \,.
\end{align}
It is also instructive to express these results using the notation introduced in \autoref{eqn:fullim} as, 
\begin{subequations}
\begin{align}
\mathcal{M}_3({\bf {1}}^{1}_{\psi^c},{\bf {2}}^{2}_{\psi},{\bf {3}}^{\{12\}}_{Z})\Big\vert_{{\cal O}(m_1)} = i\lim_{\anb{14}\to0} vA_4(1^+_{\chiL^c},2^{-}_{\chiL},3_{\theta},4_{h}) \,, 
 \label{eq:1212_matching_a}
 \\
\mathcal{M}_3({\bf {1}}^{1}_{\psi^c},{\bf {2}}^{2}_{\psi},{\bf {3}}^{\{12\}}_{Z})\Big\vert_{{\cal O}(m_2)} = i\lim_{\anb{24}\to0} vA_4(1^+_{\chiL^c},2^{-}_{\chiL},3_{\theta},4_{h})  \,,
 \label{eq:1212_matching_b}
 \\
 \mathcal{M}_3({\bf {1}}^{1}_{\psi^c},{\bf {2}}^{2}_{\psi},{\bf {3}}^{\{12\}}_{Z})\Big\vert_{{\cal O}(m_3)} = i\lim_{\anb{34}\to0} \hv A_4(1^+_{\chiL^c},2^{-}_{\chiL},3_{\theta},4_{h})  \,.
  \label{eq:1212_matching_c}
\end{align}
\end{subequations}

Next consider ${\mathcal M}_3({\bf {1}}^{2}_{\psi^c}, {\bf {2}}^{2}_{\psi},{\bf {3}}^{11}_{Z})$, which features a transverse vector. Expanding the massive amplitude we have
\begin{align}
    {\mathcal M}_3({\bf {1}}^{2}_{\psi^c}, {\bf {2}}^{2}_{\psi},{\bf {3}}^{11}_{Z})
&=
- \sqrt{2}g Q_\chi\frac{[1_q 3_k ]\langle  2_k 3_q \rangle}{m_Z}-\sqrt{2}g Q_\eta\frac{\langle1_k 3_q \rangle[  2_q 3_k ]}{m_Z} \,.
\label{eq:4pt_2211}
\end{align}
This expression features two $q$'s, so naively it seems to originate from massless  amplitudes with at least two additional Higgs legs.   Just as in~\autoref{sec:m0vector} however, the naive counting does not hold when a transverse vector is involved.
The massless amplitude ${\mathcal A}_4(1^-_{\etaR^c}, 2^{-}_{\chiL}, 3^+_{\gamma}, 4_{h})$ is given by (see~\autoref{eq:massles_4pt_mmph})
\beq
    {\mathcal A}_4(1^-_{\etaR^c}2^{-}_{\chiL}3^+_{\gamma}4_{h})
    =
     -\left( \frac{y_\phi}{\sqrt{2}} \right)\left[
    \sqrt{2}g Q_\chi   \frac{[34]\langle 24  \rangle}{[14]\langle 34\rangle} 
    +
     \sqrt{2}gQ_{\eta} \frac{[34] \langle 14\rangle }{[24]\langle 34\rangle}
    \right]\,.
\eeq
For $[14]\to0$,
\beq
{\mathcal A}_4(1^-_{\etaR^c}, 2^{-}_{\chiL}, 3^+_{\gamma}, 4_{h})\sim
-\sqrt{2} g Q_\chi\,\left( \frac{y_\phi}{\sqrt{2}} \right) \frac{[34]}{[14]}\, \frac{\anb{24}}{\anb{34}}
\sim
-\sqrt{2} g Q_\chi\,\left( \frac{y_\phi}{\sqrt{2}} \right) \frac{[34]}{[14]}\, \frac{\anb{2r}}{\anb{3r}}\,,
\eeq
where $r\rangle$ is an arbitrary spinor. In the last step, we used the three-point kinematics obtained for $[14]\to0$ (see the derivation in~\subappref{app:4to3}).
Identifying $r=3_q$ and $4=1_q$, and multiplying by $\hv$, we find
\beq
\label{eqn:2211_matching}
\lim_{[14]\to0} \hv\, {\mathcal A}_4(1^-_{\etaR^c}, 2^{-}_{\chiL}, 3^+_{\gamma}, 4_{h})=
-\sqrt{2} g Q_\chi\left( \frac{y_\phi \hv}{\sqrt{2}} \right) \, \frac{[31_q]}{[11_q]}\, \frac{\anb{23_q}}{\anb{33_q}}
= 
-\sqrt{2} g Q_\chi \frac{[1_q3]\anb{23_q}}{m_Z}
\,,
\eeq
where again we used the fact that $[1_q 1] = m_\psi = y_{\phi}\hv/\sqrt{2}$, reproducing the first term in~\autoref{eq:4pt_2211}.
In fact, we can repeat this derivation without relying on the three-point kinematics, 
by introducing the arbitrary spinor $r$ right away, just as we did in~\autoref{sec:m0vector}. Multiplying and dividing ${\mathcal A}_4(1^-_{\etaR^c}, 2^{-}_{\chiL}, 3^+_{\gamma}, 4_{h})$ by $\anb{3r}$, and using the Schouten identity,
\beq
\frac{\anb{24}}{\langle34 \rangle [41]}= \frac{\anb{2r}}{[14]\anb{r3}} -\frac{\anb{r4}}{[12]\anb{r3}}\,,
\eeq
and then picking the $[14]$ pole as above. 
As in~\autoref{sec:m0vector}, $r$ corresponds to the reference momentum of the massless amplitude. When expressed in terms of this reference momentum, the pole associated with  3 is replaced by a $1/\anb{3r}$, and the physical poles are separated, with each term containing just a single pole, one with $1/[12]$ and the other with $1/[14]$. The second $\mathcal{O}(m_2)$ term can be similarly found by taking $[24]\to 0$. Thus we find
\begin{subequations}
\begin{align}
\mathcal{M}_3({\bf {1}}^{2}_{\psi^c},{\bf {2}}^{2}_{\psi},{\bf {3}}^{11}_{Z})\Big\vert_{{\cal O}(m_1)} = \lim_{[14]\to0} \hv\, {\mathcal A}_4(1^-_{\etaR^c}, 2^{-}_{\chiL}, 3^+_{\gamma}, 4_{h}) \,, 
 \label{eq:2211_matching_a}
 \\
\mathcal{M}_3({\bf {1}}^{2}_{\psi^c},{\bf {2}}^{2}_{\psi},{\bf {3}}^{11}_{Z})\Big\vert_{{\cal O}(m_2)} = \lim_{[24]\to0} \hv\, {\mathcal A}_4(1^-_{\etaR^c}, 2^{-}_{\chiL}, 3^+_{\gamma}, 4_{h}) \,.
 \label{eq:2211_matching_b}
\end{align}
\end{subequations}

Finally, the $(2,2,22)$ component,
\beq\label{eqn:azero}
     {\mathcal M}_3({\bf {1}}^{2}_{\psi^c}, {\bf {2}}^{2}_{\psi},{\bf {3}}^{22}_{Z})=  \sqrt{2}g Q_\chi\frac{[1_q 3_q ]\langle  2_k 3_k \rangle}{m_Z}+\sqrt{2}g Q_\eta\frac{\langle1_k 3_k \rangle[  2_q 3_q ]}{m_Z}\,,
\eeq
corresponds to the HE amplitude ${\mathcal A}_4(1^-_{\etaR^c}, 2^{-}_{\chiL}, 3^-_{\gamma}, 4_{h})$,
which  vanishes in a renormalizable theory due to helicity selection rules~\cite{Cheung:2015aba}.
However, our discussion of the transverse amplitude ${\mathcal A}_4(1^-_{\etaR^c}, 2^{-}_{\chiL}, 3^+_{\gamma}, 4_{h})$ above
suggests that, for transverse vector amplitudes, 
it may be useful to express the massless amplitude in terms of reference momenta in order to bold it into a massive expression.
Writing ${\mathcal A}_4(1^-_{\etaR^c}, 2^{-}_{\chiL}, 3^-_{\gamma}, 4_{h})$ in this way 
using Feynman diagrams,   
the different individual contributions give rise to each of the terms of~\autoref{eqn:azero}, although their sum vanishes. For details, see \autoref{app::massless_amplitudes}.

\subsection{\texorpdfstring{$\mathcal{O}(m^2)$}{O(m 2)} corrections}
So far, we derived the different components of the massive amplitude 
based on their \emph{leading} energy behavior, whether it is ${\cal O}(m^0)$ or 
${\cal O}(m)$. It is also interesting to see how the subleading components of these amplitudes arise from higher-point HE amplitudes.
The only example we examine here is ${\cal M}_3({\bf {1}}^{1}_{\psi^c}, {\bf {2}}^{1}_{\psi},{\bf {3}}^{\{12\}}_{Z})$, corresponding to the (would-be) Yukawa interaction (the transverse vector case is treated in~\autoref{app::massless_amplitudes} using reference momenta).
Rewriting the amplitude in~\autoref{eqn:massamp} using~\autoref{eqn::fermion_vector_masses},
such that the mass expansion is manifest, we have
\begin{align}
&{\cal M}_3({\bf {1}}^{1}_{\psi^c}, {\bf {2}}^{1}_{\psi},{\bf {3}}^{\{12\}}_{Z}) =
 -\frac{y_\phi}{\sqrt{2}}[1_k 2_k] + \frac{y_\phi}{\sqrt{2}}\langle 1_q  2_q\rangle
 +2 g Q_\chi \frac{[1_k 3_q] \langle 3_q   2_q\rangle}{m_Z} + 2 g Q_\eta \frac{[2_k 3_q] \langle 3_q   1_q\rangle}{m_Z} \nonumber\\
 &=
  -\frac{y_\phi}{\sqrt{2}}[1_k 2_k] 
  + \frac{y_\phi}{\sqrt{2}}m_1 m_2 \left(\frac{\langle 1_q  2_q\rangle}{\langle 1_k 1_q \rangle \langle 2_k 2_q \rangle}\right)
  + 2 g Q_\eta m_Z m_1 \left(\frac{[2_k 3_q   1_q\rangle}{[3_q 3_k]\langle 3_k 3_q \rangle\langle 1_k 1_q \rangle}\right)\nonumber
 \\
 & \quad +2 g Q_\chi m_Z m_2 \left(\frac{[1_k 3_q   2_q\rangle}{[3_q 3_k]\langle 3_k 3_q \rangle\langle 2_k 2_q \rangle} \right)\,.
\end{align}
The $\mathcal{O}(m^2)$ terms arise from a five-point amplitude, with each Higgs leg supplying a single $q$. The lengthy expression for the relevant five-point amplitude ${\mathcal A}_5(1^+_{\chiL^c}, 2^{+}_{\etaR}, 3_{\theta}, 4_{h}, 5_{h})$ can be found in \autoref{eq:massles_5pt_ppGBhh}.
Importantly for our goal, it contains the following terms
\begin{align}
 i {\mathcal A}_5(1^+_{\chiL^c}, 2^{+}_{\etaR}, 3_{\theta}, 4_{h}, 5_{h}) =&
 \left(\frac{y_\phi}{\sqrt{2}}\right)^3\left( \frac{\langle 45\rangle}{\langle 14 \rangle \langle 25 \rangle}\right)
+\left(2g^2 Q_\eta Q_\phi\right)\left(\frac{y_\phi}{\sqrt{2}}\right)\left(\frac{[254 \rangle }{s_{35}\langle 14 \rangle  }\right)
 \nonumber
\\&
+ \left(2g^2 Q_{\chi} Q_\phi\right) \left(\frac{y_\phi}{\sqrt{2}} \right)\left(\frac{[154\rangle }{ s_{35}\langle 24 \rangle}\right)
+...\,.
\label{eq:massless_5pt_ppGBhh}
\end{align}
The Higgs momenta 4 and 5 are associated with two of the three $q$'s.
We then find
\begin{subequations}
\begin{align}
     \mathcal{M}_3({\bf {1}}^{1}_{\psi^c},{\bf {2}}^{1}_{\psi},{\bf {3}}^{\{12\}}_{Z})\Big\vert_{{\cal O}(m_1m_2)}
     &=
     \lim_{\anb{14}, \anb{25}\to0}\hv^2 i {\mathcal A}_5(1^+_{\chiL^c}, 2^{+}_{\etaR}, 3_{\theta}, 4_{h}, 5_{h})  \,,
    \\
     \mathcal{M}_3({\bf {1}}^{1}_{\psi^c},{\bf {2}}^{1}_{\psi},{\bf {3}}^{\{12\}}_{Z})\Big\vert_{{\cal O}(m_1m_Z)}
     &=
     \lim_{\anb{14}, s_{35}\to0}\, 
     \hv^2 i \mathcal{A}_5(1^+_{\chiL^c}, 2^{+}_{\etaR}, 3_{\theta}, 4_{h}, 5_{h})  \,,
         \\
     \mathcal{M}_3({\bf {1}}^{1}_{\psi^c},{\bf {2}}^{1}_{\psi},{\bf {3}}^{\{12\}}_{Z})\Big\vert_{{\cal O}(m_2m_Z)}
     &=
    \lim_{\anb{24}, s_{35}\to0}\, 
    \hv^2 i \mathcal{A}_5(1^+_{\chiL^c}, 2^{+}_{\etaR}, 3_{\theta}, 4_{h}, 5_{h})  \,,
\end{align}
\end{subequations}
where in the first line, $\anb{14},\anb{25}\propto \hv\to 0$ and similarly for the other terms.
Note that we multiply by two powers of $\hv$ since we are going from a five-point to a three-point amplitude. 
This isolates the singularities of interest, with all the other terms (which are not shown in~\autoref{eq:massless_5pt_ppGBhh})  tending to zero.\footnote{Since the original amplitude is symmetric under $4 \leftrightarrow 5$ exchange, it does not matter which limit we take as long as we consider the three possible pairings of $\{4,5\}$ with $\{1,2,3\}$.}

%%%%%%%%%%%%%%%%%%%%%%%%%%%%%%%%%%%%%%%%%%
\subsection{Including non-renormalizable interactions}\label{sec:nr}
%%%%%%%%%%%%%%%%%%%%%%%%%%%%%%%%%%%%%%%%%%%
As discussed above, at the renormalizable level, the massive amplitude starts at $\mathcal{O}(m)$ for all-plus or all-minus helicities, and the four-point amplitude ${\mathcal A}_4(1^-_{\etaR^c}, 2^{-}_{\chiL}, 3^-_{\gamma}, 4_{h})$ vanishes. Let us now include also non-renormalizable interactions, with a cutoff scale $\Lambda$. 
The four-point amplitude now receives an additional contribution from the dipole contact term
\begin{align}
{\mathcal A}^{\text{ct}}_4(1^-_{\etaR^c}, 2^{-}_{\chiL}, 3_\gamma^-, 4_h) = \frac{2\cDipole}{\Lambda^2}\langle13\rangle\langle23\rangle\,.
\end{align}
Taking the soft Higgs limit $4\to0$, as discussed in~\autoref{sec:bolding} (see~\autoref{eqn:bold2}), is trivial, since the Higgs momentum does not appear in this expression.
Thus, we find
\begin{align}
\lim_{4\to 0}\,{\hv \mathcal A}^{\text{ct}}_4(1^-_{\etaR^c}, 2^{-}_{\chiL}, 3_\gamma^-, 4_\phi) =  \frac{2\cDipole \hv}{\Lambda^2}\langle1_k3_k\rangle\langle2_k3_k\rangle\subset\mathcal{M}_3({\bf {1}}^2_{\psi^c},{\bf {2}}^2_{\psi},{\bf {3}}^{22}_{Z}) \,.
\end{align}
This term (along with its conjugate) can now be  bolded, giving the full massive amplitude
\begin{align}\label{eqn:mnonr}
\mathcal{M}_3({\bf {1}}_{\psi^c}, {\bf {2}}_{\psi},{\bf {3}}_{Z}) = \frac{2\cbarDipole \hv}{\Lambda^2}[{\bf 13}][{\bf 23}]+\sqrt{2}g Q_\chi\frac{[{\bf{1 3 }}]\langle  {\bf{2 3}} \rangle}{m_Z}+\sqrt{2}g Q_\eta\frac{\langle{\bf{1 3 }}\rangle[  {\bf{2 3}} ]}{m_Z}+\frac{2\cDipole \hv}{\Lambda^2}\langle{\bf 13}\rangle\langle{\bf 23}\rangle\,.
\end{align}
These additional non-renormalizable interactions also contribute mass-suppressed corrections to other components of the massive amplitude.\footnote{For example, the other mass-suppressed terms generated by the dipole interactions include a $\mathcal{O}(m^2)$  $(2,1,11)$ amplitude, which can be matched to the five-point amplitude $\mathcal{A}_5(1^-_{\etaR^c}, 2^{+}_{\etaR}, 3^+_{\gamma}, 4_{h}, 5_{h})$; an $\mathcal{O}(m^2)$  $(1,1,\{12\})$ amplitude, which can matched to
$\mathcal{A}_5(1^+_{\chiL^c}, 2^{+}_{\etaR}, 3_{\theta}, 4_{h}, 5_{h})$; and
an $\mathcal{O}(m^3)$  $(1,2,\{12\})$ contribution which arises from the six-point $\mathcal{A}_6(1^+_{\chiL^c}, 2^{-}_{\chiL}, 3_{\theta}, 4_{h}, 5_{h}, 6_{h})$. The matching here proceeds as in previous examples, thus we do not discuss it further.} 
Here, we limit our discussion to one choice of little-group indices, namely $(2,2,11)$, for which the matching has some novel features.
The dipole interactions generate the following $\mathcal{O}(m^3)$ correction to this component,
\begin{align}\label{eqn:m32211}
  {\cal M}_3({\bf {1}}^{2}_{\psi^c}, {\bf {2}}^{2}_{\psi},{\bf {3}}^{11}_{Z}) &\supset \frac{2\cbarDipole \hv}{\Lambda^2}[1_q 3_k][2_q 3_k] +\frac{2\cDipole \hv}{\Lambda^2}\langle 1_k 3_q\rangle\langle 2_k 3_q\rangle \nonumber
  \\
  &= 2\cbarDipole \hv m_1m_2 \left(\frac{[1_q 3_k][2_q 3_k]}{\Lambda^2 [1_q 1_k][2_q 2_k]}\right) +2\cDipole \hv m_Z^2 \left(\frac{\langle 1_k 3_q\rangle\langle 2_k 3_q\rangle}{\Lambda^2 \langle 3_k 3_q \rangle^2}\right)\,.
\end{align}
These originate from the six-point amplitude ${\mathcal A}_6(1^-_{\etaR^c}, 2^{-}_{\chiL}, 3^+_{\gamma}, 4_{h}, 5_{h}, 6_{h})$,
given in~\autoref{eq:6pt_mmphhh},
with one soft and two frozen Higgs momenta. Due to the symmetrization of the scalar momenta, we can arbitrarily choose 6 to be soft.
The $\hv\, m_1m_2$ piece of~\autoref{eqn:m32211} is  obtained from this amplitude as,
\begin{align}
    {\cal M}_3({\bf {1}}^{2}_{\psi^c}, {\bf {2}}^{2}_{\psi},{\bf {3}}^{11}_{Z})\Big\vert_{{\cal O}(\hv\,m_1m_2)} =
    \lim_{\substack{\anb{14},\anb{25}\to0\\ 6\to0}}
    \hv^3\, {\mathcal A}_6(1^-_{\etaR^c}, 2^{-}_{\chiL}, 3^+_{\gamma}, 4_{h}, 5_{h}, 6_{h})  \,,
    \label{eq:2211_match_bstar}
\end{align}
where $\anb{14},\anb{25}\propto\hv$ as usual.
Thus, we have identified 4 as $1_q$, and 5 as $2_q$. 

Finally, the $m_Z^2$ term in~\autoref{eqn:m32211} is the first example we encounter in which the same mass appears squared, which can be thought of  as a \emph{vector} chirality flip; both Higgses are required in order to supply the vector $3_q$ in this case,
and the massive amplitude is obtained by identifying $3_q\propto 4+5$, in the limit that 4 and 5 are collinear. Specifically, starting with the expression in \autoref{eq:dipole_massless_6point}, we first take the limit $5\rangle\to c\, 4\rangle$ for some constant $c$, and then take $\anb{34}\to0$
(see \autoref{app::massless_amplitudes} for more details).
We then find that,
\begin{align}
     {\cal M}_3({\bf {1}}^{2}_{\psi^c}, {\bf {2}}^{2}_{\psi},{\bf {3}}^{11}_{Z})\Big\vert_{{\cal O}(\hv\,m_Z^2)} =
 \frac12
\lim_{\substack{\anb{34},\anb{45}\to0\\ 6\to0}}
 \hv^3 {\mathcal A}_6(1^-_{\etaR^c}, 2^{-}_{\chiL}, 3^+_{\gamma}, 4_{h}, 5_{h}, 6_{h})\,,
\end{align}
reproduces the second term of~\autoref{eqn:m32211}, independently of the value of the constant $c$. 
The additional factor of $1/2$ is introduced to account for over-counting due to the original $(4\leftrightarrow5)$ exchange symmetry of the amplitude.

%%%%%%%%%%%%%%%%%%%%%%%%%%%%%%%%%%%%%%%%%%%%%
\section{Discussion and conclusions}
We presented a method for deriving the massive contact terms of Higgsed theories and their coefficients, from massless ones which are easier to construct.
To derive the contact term for a massive $n$-point amplitude, one starts from the bases of massless $(n+n_H)$-point contact terms, where $n_H\geq0$ denotes the number of additional Higgs legs.
The $n_H$ Higgs momenta are set to zero and the remaining spinor labels are bolded.
A factor of $\hv^{n_H}$ is included to recover the dimension of an $n$-point amplitude and leads to the correct matching between massless and massive contact-term coefficients.
The list of independent spinor structures obtained for increasing numbers of Higgses is exhausted for some $n_H$.
Beyond this, one generates spinor structures that are already in the list, times powers of the Lorentz invariants or masses.
In the examples we studied, for $n=4$ and a SM particle content, only $n_H=0,1$ were required.
It would be interesting to develop a systematic understanding of  where the process truncates in general, and to automate it for general multiplicity and spin content.
We have also shown how the subleading helicity-flipped components of each massive spinor structure arise from the poles of massless factorizable amplitudes with additional soft or collinear Higgs legs.

This top-down construction yields a physical interpretation of the structure of massive contact terms, written in terms of little-group-covariant massive spinors~\cite{Arkani-Hamed:2017jhn}.
The lightlike pair  $i_k$ and $i_q$, which together describe the momentum and polarization of a massive particle $\bs i$, are mapped to the momenta of a massless particle $i$ and of a Higgs.
The components of an $n$-point contact term, which is a tensor in the little-group space,
are associated with various $(n+n_H)$-point massless amplitudes.
To construct all of these components, one can in principle  start with a massless $2n$-point amplitude with one, or more, additional Higgs legs of momentum $i_q$ for each massless leg of momentum $i_k$, with each of the Higgs legs frozen in a kinematic configuration such that $(i_k+i_q)^2=m_i^2$.
The residues of these poles give rise to the subleading components of the contact terms.
This choice is particularly intuitive, since momentum conservation holds in both the HE and LE amplitudes.
As we have seen in~\autoref{sec:three_point}, other choices are however possible, and sometimes more economical.
Note that different choices of $i_q$ correspond to different choices of the spin quantization axis for particle $i$ but, thanks to little-group covariance, any of them can be used to determine the full massive amplitude.
Our analysis mostly focused on the leading-order corrections to the contact terms, which arise from a single additional Higgs leg.
Notable exceptions were discussed in~\autoref{sec:three_point},
where three-point amplitudes were derived from massless amplitudes with up to three additional Higgses.

For vector amplitudes, we have also seen an alternative approach to bolding, which relies on the arbitrary reference momenta associated with vector polarizations in the (gauge-invariant) HE theory. 
In an on-shell bootstrap derivation of the amplitudes, these reference momenta never appear.  
However, as we saw in~\autoref{sec:three_point}, it is sometimes useful to introduce them by hand in order to rewrite the amplitude such that each term only involves a single factorization channel.
A massless contact term featuring a vector $i$ of momentum $i_k$ and a reference momentum $i_r$ can then be obtained via the identification $i_r\equiv i_q$.
As discussed in~\autoref{sec:wwhh_sm}, the relation between perturbative unitarity and gauge symmetry is clearly manifest with this identification.

Through most of our analysis, we have restricted our attention to the minimal number of Higgs legs required for describing the massive amplitude.
When multiple Higgs legs are involved, their contributions are automatically symmetrized over, since the physical Higgs $h$ can originate from any $H$ or $H^\dagger$ leg.
For the particles of spin $\leq1$ we studied, one or two Higgses connecting to any external leg are required to account for all the helicity categories contributing to each amplitude.
It would be interesting to extend our analysis to the full massive EFT amplitudes,
and in particular to develop an on-shell formulation of their field space geometry~\cite{Alonso:2015fsp,Alonso:2016oah,Cohen:2021ucp,Cheung:2021yog,Helset:2020yio,Nagai:2019tgi,Nagai:2021gmo}.
Our approach can also be applied to derive massive EFT amplitudes featuring particles of higher-spins, including both gravitons and the composite states of strongly-coupled theories.

\acknowledgments
We thank Camila S.\ Machado for valuable discussions.
Research supported in part by the Israel Science Foundation (Grant No.\,751/19), and by BSF-NSF grant 2020-785.
Part of this work was performed at the Aspen Center for Physics, which is supported by National Science Foundation grant PHY-1607611.
The research of RB is also supported by an Azrieli Fellowship of the Azrieli Foundation.
The work of TK is supported by the Japan Society for the Promotion of Science (JSPS) Grant-in-Aid for Early-Career Scientists (Grant No.\,19K14706) and the JSPS Core-to-Core Program (Grant No.\,JPJSCCA20200002).

\appendix

%%%%%%%%%%%%%%%%%%%%%%%%%%%%%%%%%%%%%%%%%%%%%%%%%%%%%
%%%%%%%%%%%%%%%%%%%%%%%%%%%%%%%%%%%%%%%%%%%%%%%%%%%%%
\section{Factorizable versus contact-term LE amplitudes}\label{app:fnf}
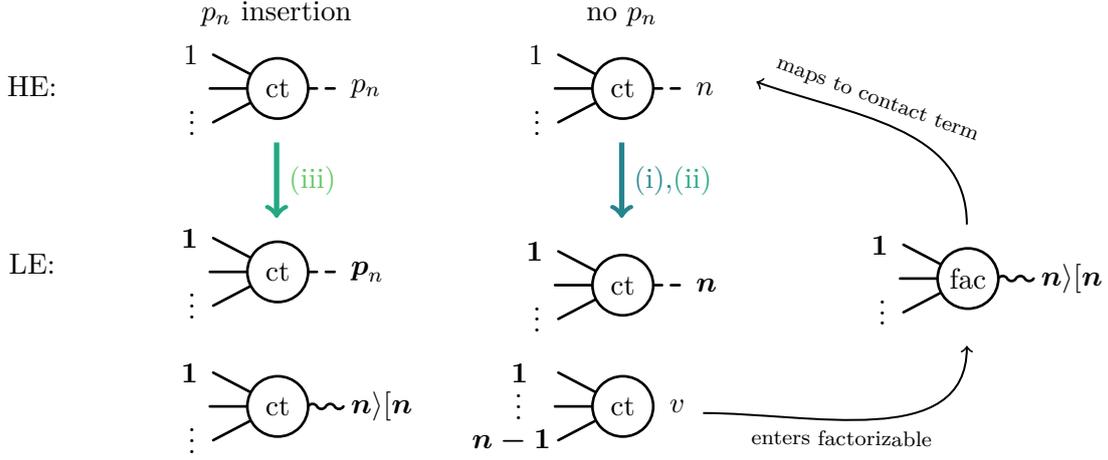
\begin{figure}
\begin{tikzpicture}
\node (uv) [inner sep=0mm] {
	\fmfframe(25,0)(25,-5){\begin{fmfgraph*}(50,50)
	\fmfleft{l0,l1,l2,l3,l4}
	\fmfright{r1}
	\fmf{vanilla}{l1,v,l3}
	\fmf{vanilla}{l2,v}
	\fmfv{lab=ct, d.shape=circle, d.size=8mm, d.fill=0, l.dist=0mm}{v}
	\fmf{dashes,tens=3}{v,r1}
	\fmfv{lab=$1$, l.dist=2mm, l.angl=180}{l3}
	\fmfv{lab=\raisebox{2mm}{$\vdots$}, l.dist=2.25mm, l.angl=180}{l1}
	\fmfv{lab=$n$, l.dist=1mm, l.angl=0}{r1}
	\end{fmfgraph*}}
};
\node (uv0) [inner sep=0mm, left=1cm of uv] {
	\fmfframe(25,0)(25,-5){\begin{fmfgraph*}(50,50)
	\fmfleft{l0,l1,l2,l3,l4}
	\fmfright{r1}
	\fmf{vanilla}{l1,v,l3}
	\fmf{vanilla}{l2,v}
	\fmfv{lab=ct, d.shape=circle, d.size=8mm, d.fill=0, l.dist=0mm}{v}
	\fmf{dashes,tens=3}{v,r1}
	\fmfv{lab=$1$, l.dist=2mm, l.angl=180}{l3}
	\fmfv{lab=\raisebox{2mm}{$\vdots$}, l.dist=2.25mm, l.angl=180}{l1}
	\fmfv{lab=$p_n$, l.dist=1mm, l.angl=0}{r1}
	\end{fmfgraph*}}
};
\node (nop) [below=1cm of uv0, xshift=0cm, inner sep=0mm]{
	\fmfframe(25,-5)(25,0){\begin{fmfgraph*}(50,50)
	\fmfleft{l0,l1,l2,l3,l4}
	\fmfright{r1}
	\fmf{vanilla}{l1,v,l3}
	\fmf{vanilla}{l2,v}
	\fmfv{lab=ct, d.shape=circle, d.size=8mm, d.fill=0, l.dist=0mm}{v}
	\fmf{dashes,tens=3}{v,r1}
	\fmfv{lab=$\bs{1}$, l.dist=2mm, l.angl=180}{l3}
	\fmfv{lab=\raisebox{2mm}{$\bs{\vdots}$}, l.dist=2.25mm, l.angl=180}{l1}
	\fmfv{lab=$\bs{p}_n$, l.dist=1mm, l.angl=0}{r1}
	\end{fmfgraph*}}
};
\node (nop2) [below=0mm of nop, inner sep=0mm]{
	\fmfframe(25,0)(25,0){\begin{fmfgraph*}(50,50)
	\fmfleft{l0,l1,l2,l3,l4}
	\fmfright{r1}
	\fmf{vanilla}{l1,v,l3}
	\fmf{vanilla}{l2,v}
	\fmfv{lab=ct, d.shape=circle, d.size=8mm, d.fill=0, l.dist=0mm}{v}
	\fmf{photon,tens=3}{v,r1}
	\fmfv{lab=$\bs{1}$, l.dist=2mm, l.angl=180}{l3}
	\fmfv{lab=\raisebox{2mm}{$\bs{\vdots}$}, l.dist=2.25mm, l.angl=180}{l1}
	\fmfv{lab=$\bs{n}\rangle[\bs{n}$, l.dist=1mm, l.angl=0}{r1}
	\end{fmfgraph*}}
};
\draw[->,c2, line width=2] (uv0.south)  to[out=-90,in=90]
	node[right]{\hyperref[item:iii]{\textcolor{c3}{(iii)}}}
	(nop.north);
\node[above=-2mm of uv0]{$p_n$ insertion};
\node (irlab) [left=1cm of nop, yshift=2mm] {LE:};
\node (uvlab) [above=1.8cm of irlab] {HE:};
\node (p2) [below=0mm of p, inner sep=0mm]{
	\fmfframe(25,-5)(25,0){\begin{fmfgraph*}(50,50)
	\fmfleft{l0,l1,l2,l3,l4}
	\fmfright{r1}
	\fmf{vanilla}{l1,v,l3}
	\fmf{vanilla}{l2,v}
	\fmfv{lab=ct, d.shape=circle, d.size=8mm, d.fill=0, l.dist=0mm}{v}
	\fmf{phantom,tens=3}{v,r1}
	\fmfv{lab=$\bs{1}$, l.dist=4mm, l.angl=180}{l3}
	\fmfv{lab=\raisebox{2mm}{$\bs{\vdots}$}, l.dist=4.25mm, l.angl=180}{l2}
	\fmfv{lab=$\bs{n-1}$, l.dist=1mm, l.angl=180}{l1}
	\fmfv{lab=$\hv$, l.dist=-2.5mm, l.angl=0}{r1}
	\end{fmfgraph*}}
};
\node (p) [below=1cm of uv, xshift=0cm, inner sep=0mm]{
	\fmfframe(25,0)(25,0){\begin{fmfgraph*}(50,50)
	\fmfleft{l0,l1,l2,l3,l4}
	\fmfright{r1}
	\fmf{vanilla}{l1,v,l3}
	\fmf{vanilla}{l2,v}
	\fmfv{lab=ct, d.shape=circle, d.size=8mm, d.fill=0, l.dist=0mm}{v}
	\fmf{dashes,tens=3}{v,r1}
	\fmfv{lab=$\bs{1}$, l.dist=2mm, l.angl=180}{l3}
	\fmfv{lab=\raisebox{2mm}{$\bs{\vdots}$}, l.dist=2.25mm, l.angl=180}{l1}
	\fmfv{lab=$\bs{n}$, l.dist=1mm, l.angl=0}{r1}
	\end{fmfgraph*}}
};
\draw[->,c1, line width=2] (uv.south) to[out=-90,in=90]
	node[right]{\hyperref[item:i]{\textcolor{c1}{(i)}},\hyperref[item:ii]{\textcolor{c2}{(ii)}}}
	(p.north);
\node[above=-2mm of uv]{no $p_n$};
\node (fac) [right=1cm of p, inner sep=0mm]{
	\fmfframe(25,-5)(25,0){\begin{fmfgraph*}(50,50)
	\fmfleft{l0,l1,l2,l3,l4}
	\fmfright{r1}
	\fmf{vanilla}{l1,v,l3}
	\fmf{vanilla}{l2,v}
	\fmfv{lab=fac, d.shape=circle, d.size=8mm, d.fill=0, l.dist=0mm}{v}
	\fmf{photon,tens=3}{v,r1}
	\fmfv{lab=$\bs{1}$, l.dist=2mm, l.angl=180}{l3}
	\fmfv{lab=\raisebox{2mm}{$\bs{\vdots}$}, l.dist=2.25mm, l.angl=180}{l1}
	\fmfv{lab=$\bs{n}\rangle[\bs{n}$, l.dist=1mm, l.angl=0}{r1}
	\end{fmfgraph*}}
};
\draw[->] ($(p2.east)+(-7mm,0cm)$) to[out=0,in=-90] node[pos=.4, sloped, below, font=\scriptsize]{enters factorizable} (fac.south);
\draw[->] (fac.north) to[out=90,in=-20] node[pos=.6, sloped, above, font=\scriptsize]{maps to contact term} (uv.east);
\end{tikzpicture}
\caption{The low-energy factorizable contributions featuring a longitudinal vector (bottom right) map to high-energy contact terms (top right) without affecting the determination of massive low-energy contact terms featuring a longitudinal vector (bottom left).}
\label{fig:factorizable}
\end{figure}

%%%%%%%%%%%%%%%%%%%%%%%%%%%%%%%%%%%%%%%%%%%%%%%%%%%%%
%%%%%%%%%%%%%%%%%%%%%%%%%%%%%%%%%%%%%%%%%%%%%%%%%%%%%
In order to  determine the LE contact terms, we start  from the HE contact terms and bold them.
This essentially means that we match the high-energy limit of each LE contact term to a HE contact term. Naively, it may seem as if we are matching just the contact-term parts of the LE and HE amplitudes, so let us now explain the matching in more detail.
Obviously, the only quantities that can be sensibly matched to each other are  the \emph{full} amplitudes, which contain both factorizable and contact-term pieces. The matching is then,
\beqa\label{eq:match}
{\cal M}^\text{ct}_n(\bs p_1^{h_1}, \ldots, \bs p_n^{h_n})\Big\vert_{{\rm high-E}} &+&
{\cal M}^\text{fac}_n(\bs p_1^{h_1}, \ldots, \bs p_n^{h_n})\Big\vert_{{\rm high-E}}\nonumber\\
&=& {\cal A}^\text{ct}_n( p_1^{h_1}, \ldots, p_n^{h_n},\cdots) +
{\cal A}^\text{fac}_n( p_1^{h_1}, \ldots, p_n^{h_n},\cdots)\,,
\eeqa
where $n\geq 4 $ and the ellipses on the right-hand side stand for additional Higgs legs which may be required by the gauge symmetry, and whose momenta are frozen.
We then need to isolate the contact terms, namely pole-free pieces, on the left- and right-hand sides of~\autoref{eq:match}, and equate their coefficients to each other. 
For longitudinal-vector amplitudes, this identification can be subtle, because the LE \emph{factorizable} amplitude ${\cal M}^\text{fac}_n$ (right of \autoref{fig:factorizable}), can sometimes give a pole-free piece in the high-energy limit.
The reason is that LE amplitudes feature a factor of $\bs{p}_i\rangle[\bs{p}_i$ for each external longitudinal vector of  momentum $p_i$.
In the high energy limit, this gives a power of the momentum, which can cancel a power of the momentum appearing in the propagator, such that one is left with a pole-free term on the left-hand side of~\autoref{eq:match} (top right of \autoref{fig:factorizable}).
This does not happen for spin-$1/2$ fermions and transverse vectors since in these cases
each factorization channel is identical in the high-energy limit to the corresponding HE contribution.

It is useful to clarify the origin of the different contributions to the LE amplitudes in this case.
Consider the matching~\autoref{eq:match} when a longitudinal is involved,
\beqa\label{eq:match1}
{\cal M}^\text{ct}_n(\bs1, \ldots, \bs{n-1}; V_L(\bs{p}))\Big\vert_{{\rm high-E}} &+&
{\cal M}^\text{fac}_n(\bs1, \ldots, \bs{n-1}; V_L(\bs{p}))\Big\vert_{{\rm high-E}}\\
&=& {\cal A}^\text{ct}_n(1,\ldots,n-1;H) + {\cal A}^\text{fac}_n(1,\ldots,n-1;H)\,.\nonumber
\eeqa
Suppose that  ${\cal A}^\text{ct}_n(1,\ldots,n-1;H)$ 
contains the contact term (top diagrams of \autoref{fig:factorizable})
\beq
c_n\, K_n\left(1,..,n-1;H(p)\right).
\eeq
If $K_n$ contains an insertion of $p$, it can be bolded into a longitudinal vector contact term, 
\beq
c_n\, \bs{K}_n\left(\bs{1},..,\bs{n-1};V_L(\bs{p})\right)\,,
\eeq
which features the factor $\bs{p}]\langle \bs{p}$, as required for a vector amplitude (bottom left of \autoref{fig:factorizable}).
This term will appear in the contact-term $n$-point amplitude ${\cal M}^\text{ct}_n(\bs1, \ldots, \bs{n-1}; V_L(\bs{p}))$.

If, however, $K_n(1,..,n-1;H(p))$ does not vanish for $p=0$, it generates in the LE
two types of contact terms. The first is a contribution to ${\cal M}^\text{ct}_n(\bs1, \ldots, \bs{n-1}; h(\bs{p}))$, which is obtained by directly bolding $K_n(1,..,n-1;H(p))$. The second is 
an $(n-1)$-point LE contact term with the Higgs leg removed, $\hv \, c_n \,\bs{K}_{n-1}(\bs{1},..,\bs{n-1})$ (bottom center of \autoref{fig:factorizable}).
This $(n-1)$-point contact term contributes to the \emph{factorizable} LE amplitude, ${\cal M}^\text{fac}_n(\bs1, \ldots, \bs{n-1}; V_L(\bs{p}))$.
In the HE limit this  contribution matches the $c K_n(1,..,n-1;H(p))$ term in the contact-term $n$-point amplitude 
as required.

To summarize, an $n$-point HE contact term with external Higgs legs 
can give rise to several contributions in the LE vector amplitudes.
First, it can generate an $n$-point LE contact term with the Higgs legs replaced by vector legs. Second, it generates lower-point LE contact terms with some Higgs legs removed. These contribute in 
${\cal M}^\text{fac}_n(\bs1, \ldots, \bs{n-1}; V_L(\bs{p}))$.

%%%%%%%%%%%%%%%%%%%%%%%%%%%%%%%%%%%%%%%%%%%%%%%%%%%%%

As a concrete example, consider the $\bar ddZZ$ amplitude.
The LE factorizable amplitude contains the contribution,
\beq\label{eq:leamp}
g \frac{\Sqb{14}\Anb{p4}}{m}\, \frac{1}{p^2-m^2}\,C^{+++}_{ddZ}\,\Sqb{23}\Sqb{3p} =
\frac{g C^{+++}_{ddZ}}{m}
\Sqb{14}\Sqb{23}\,\frac{[\bs{3p4}\rangle}{p^2-m^2}\,,
\eeq
with $p=1+4$.
Here $g$ is a gauge coupling, $m$ is the $Z$ mass, and $C^{+++}_{ddZ}$ is the LE Wilson coefficient.
This low-energy three-point amplitude coefficient is generated from the HE four-point $Q^cDWH$ contact term by setting the Higgs to its VEV, 
\beq
C^{+++}_{ddZ}= \hv \frac{ c_{Q^cDWH;6}}{\Lambda^2}\,,
\eeq
where the `$;6$' subscript refers to the dimension of the corresponding EFT operator in the HE theory.
In the high-energy limit, \autoref{eq:leamp} reduces to
\beq
 \frac{c_{Q^cDWH;6}}{\Lambda^2} [13][23]\,,
\eeq
which indeed matches the HE amplitude $Q^cDWH$. 
 
On the other hand, $c_{Q^cDWH;6}$ does not appear in the contact-term part of  $\bar d dZZ$ because it does not have the correct Lorentz structure (the only LE four-point contact term it generates is in amplitudes where 4 is a scalar line, such as $\bar d dZh$).
Instead, the only high-energy $Q^cDWH$ contact terms that generate low-energy $\bar d dZZ$ contact terms require an additional $p_4$ momentum insertion and are thus of higher dimension.
One could for instance proceed to the following bolding:
\beq
\frac{c_{Q^cDWH;8}}{\Lambda^4}\,[13][23]s_{14}\to
\frac{c_{Q^cDWH;8}}{\Lambda^4}\,\Sqb{13}\Sqb{23}\Asb{414}\,.
\eeq

\section{Subleading component of massive vector spinors}
\label{app:subleading-massive-vector-spinors}
Similarly to the fermion case detailed in \autoref{sec:subleading-fermion}, we discuss in this appendix how the subleading spinor components of transverse and longitudinal vectors arise from factorizable amplitudes with additional Higgs legs.

%%%%%%%%%%%%%%%%%%%%%%%%%%%%%%%%%%%%%%%%%%%%%%%
%%%%%%%%%%%%%%%%%%%%%%%%%%%%%%%%%%%%%%%%%%%%%%%
\paragraph{Transverse vector categories.}
%%%%%%%%%%%%%%%%%%%%%%%%%%%%%%%%%%%%%%%%%%%%%%%
%%%%%%%%%%%%%%%%%%%%%%%%%%%%%%%%%%%%%%%%%%%%%%%
An $n$-point massive spinor structure whose helicity category involves a massive transverse vector has a leading high-energy component coming from an $n$-point contact term with a massless vector of helicity $\pm1$, and a subleading component coming from an $(n+1)$-point amplitude with the vector coupled to a pair of scalars, or $H^\dagger H$ (see \autoref{fig:massive_comp_origins-c}). 
For concreteness, let us consider the positive helicity case.
The leading contribution arises from a massless contact term which is schematically of the form,  ${\cal A}_n
(k^{h=+1},2,\ldots,n)
= c_n [k\cdots k]$, where $k$ is the momentum of the vector of helicity $+1$.
It maps to the massive transverse LE amplitude ${\cal M}_n= C_n [\bs{p}^{I=1}\cdots \bs{p}^{J=1}]$ for $C_n=c_n + \mathcal{O}(m)$.

The subleading term of the form $[\bs{p}^{I=2}\cdots \bs{p}^{J=1}]$ arises from the $(n+1)$-point amplitude as
$\hv \, {\cal A}_{n+1}(H^\dagger(k),H(q);2,\ldots,n)$.
In the small mass limit, the subleading contribution comes from
the $\eta^2=(k+q)^2$ pole, whose
residue is ${\cal A}_3(H^\dagger(k), H(q), \fv (\eta))$ times the contact term ${\cal A}_n
(\eta^{h=+1},2,\ldots,n)
$, glued along the vector line. Altogether we have,
\begin{align}
\hv \, {\cal A}_{n+1}(H^\dagger(k),H(q);2,\ldots,n)
=&
\hv\, \sqrt{2} i g\, \frac{\anb{k\eta}\anb{q\eta}}{\anb{kq}}\, \frac{1}{(k+q)^2}\, c_n\, [\eta \cdots \eta] \nonumber \\
=&  \hv\, \sqrt{2} i \frac{ c_n g}{[kq]} \, \frac{\anb{k(k+q)\cdots (k+q)q}}{\anb{kq}^2}\, \nonumber \\
=& \sqrt{2} i  \frac{c_n g \hv}{[kq]} \, [q\cdots k]
\,.
\end{align}
Using $g\hv=m_V=\anb{kq}=-[kq]$, this becomes
\begin{align}\label{eqn:transkq}
\hv \, {\cal A}_{n+1}(H^\dagger(k),H(q);2,\ldots,n)
=&-\sqrt{2} i c_n\, [q\cdots k] \nonumber \\
=& \sqrt{2} i c_n\, [\bs{p}^{I=2}\cdots \bs{p}^{J=1}]\nonumber \\
=&  i c_n\, \frac{[\bs{p}^{I=1}\cdots \bs{p}^{J=2}]+ [\bs{p}^{I=2}\cdots \bs{p}^{J=1}]}{\sqrt{2}} \,,
\end{align}
which is indeed the subleading component of the LE massive amplitude ${\cal M}_n= C_n [\bs{p}\cdots \bs{p}]$ for $C_n=c_n + \mathcal{O}(m)$.
Note that the little-group symmetrization can be traced to $k\leftrightarrow q$, since $h$ can come from either Higgs leg.

%%%%%%%%%%%%%%%%%%%%%%%%%%%%%%%%%%%%%%%%%%%%%%%
%%%%%%%%%%%%%%%%%%%%%%%%%%%%%%%%%%%%%%%%%%%%%%%
\paragraph{Longitudinal vector categories.}
%%%%%%%%%%%%%%%%%%%%%%%%%%%%%%%%%%%%%%%%%%%%%%%
%%%%%%%%%%%%%%%%%%%%%%%%%%%%%%%%%%%%%%%%%%%%%%%
A massive $n$-point spinor structure whose helicity category involves a longitudinal vector arises at leading order from an $n$-point HE contact term with a scalar (Goldstone) leg.
This HE contact term is of the form  $\mathcal{A}_n
(k^{h=0},2,\ldots,n)
=c_n^\prime \sab{k\cdots k}$, where $k\rangle[k$ is the Goldstone momentum $k$.
This bolds into $\mathcal{M}_n=C_n\,\sab{\bs{p}^{I=1}\cdots \bs{p}^{J=2}}/\sqrt{2}$ with $C_n= \sqrt{2}c_n^\prime + \mathcal{O}(m)$.
The subleading component comes from an $(n+1)$-point HE amplitude with an additional vector leg (see \autoref{fig:massive_comp_origins-d}).
As in the examples above, this subleading term can be identified in the small mass limit as,
\begin{align}
\hv \, 
{\cal A}_{n+1}(k^{h=+1},H(q);2,\ldots,n)  =& -  \hv\, \sqrt{2}ig\, \frac{[k\eta][k q]}{[\eta q]}\, \frac{1}{(k+q)^2}\, c_n^\prime\, \sab{\eta \cdots \eta}\nonumber \\
=&    \sqrt{2}i g\hv\, c_n^\prime\,\frac{[k\eta]}{[\eta q]\anb{kq}}\, \sab{\eta \cdots \eta} \,.
\end{align}
Using $g\hv=m_V=\anb{kq}=-[kq]$, this becomes,
\begin{align}
\hv \, 
{\cal A}_{n+1}(k^{h=+1},H(q);2,\ldots,n) =& \sqrt{2}i c_n^\prime\, \frac{[k\eta]}{[\eta q]} \, \sab{\eta\cdots\eta} \nonumber \\
=& - \sqrt{2}i c_n^\prime\, \frac{\langle q \eta \cdots \eta k]}{\anb{q\eta}[\eta q]} 
\nonumber \\
=&  \sqrt{2}i c_n^\prime\, \sab{k\cdots q} \nonumber \\
=& \sqrt{2}i  c_n^\prime\, \sab{\bs{p}^{I=1}\cdots \bs{p}^{J=1}}\,.
\end{align}
Taking instead a negative-helicity vector, one obtains the same result with $I=J=2$.
The two subleading components of the LE amplitude $\mathcal{M}_n=C_n\sab{\bs{p}\cdots \bs{p}}$ are thus generated, and $C_n= \sqrt{2}c'_n + \mathcal{O}(m)$.
Note there remains a last sub-sub-leading $\sab{q\cdots q}$ component which we do not discuss.

%%%%%%%%%%%%%%%%%%%%%%%%%%%%%%%%%%%%%%%%%%%%%%%%%%%%%
%%%%%%%%%%%%%%%%%%%%%%%%%%%%%%%%%%%%%%%%%%%%%%%%%%%%%
\section{\texorpdfstring{$\bar{u}dWh$}{udWh} contact terms with SU(2) structures}
\label{app:udWh}
\label{sec:su2_basis}
%%%%%%%%%%%%%%%%%%%%%%%%%%%%%%%%%%%%%%%%%%%%%%%%%%%%%
%%%%%%%%%%%%%%%%%%%%%%%%%%%%%%%%%%%%%%%%%%%%%%%%%%%%%

We work with the SU(2) generators $\tilde \tau^a$, $a=+,-,3$,
\begin{equation}
\suTwoTildeGen{a}{j}{i} = 
\frac{1}{\sqrt{2}}\begin{pmatrix}0 & 1\\
0 & 0
\end{pmatrix}\,,\quad \frac{1}{\sqrt{2}}\begin{pmatrix}0 & 0\\
1 & 0
\end{pmatrix},\, \quad \begin{pmatrix}\frac{1}{2} & 0\\
0 & -\frac{1}{2}
\end{pmatrix}
\,.
\end{equation}
These satisfy,
\beq
\sqb{\suTwoTildeGen{a}{}{},\suTwoTildeGen{b}{}{}} 
=i\suTwoCabc{a}{b}{c}\suTwoTildeGen{c}{}{}\,, \quad
\tr\parn{\suTwoTildeGen{a}{}{}\suTwoTildeGen{b}{}{}} 
=\frac{1}{2}\suTwoTgab{a}{b}\,,
\eeq
where 
\begin{align}
 & g^{+-}=g^{-+}=g^{33}=1\eqncomma\\
 & \suTwoCabc+-3=-\suTwoCabc+3+=-\suTwoCabc-+3=\suTwoCabc-3-=\suTwoCabc3++=-\suTwoCabc3--=-i\eqndot
\end{align}
The  structure constants $\suTwoCabc{a}{b}{c}$ are complex, and antisymmetric
only in the first two indices.

The four- and five-point HE amplitudes contributing to the $\bar u d W h$ amplitude are listed bellow. 
Instead of using color-ordered amplitudes, we find it more convenient here to strip off the $\suN 2$ group theory factors above, writing,
\begin{subequations}\label{eqn:f1f2Wh_reduced_amplitudes}
\begin{align}
	\ampFourPt{Q^{c,i}}{D}{W^{a},h_3}{H_j}  &= \suTwoTildeGen{a}{j}{i} \redAmpFourPt{Q^{c}}{D}{W,h_3}{H} \eqncomma \\
	\ampFourPt{U^c}{Q_i}{W^{a},h_3}{H_{j}} &= \epsilon^{ik} \suTwoTildeGen{a}{j}{k}  \redAmpFourPt{U^c}{Q}{W,h}{H} \eqncomma\\
	\ampFourPt{Q^{c, i}}{Q_j}{H_k}{H^{\dagger l}} &= 
		 \suTwoNDelta{k}{l} \suTwoNDelta{j}{i} A_{1}\parn{\dots}
		+ \suTwoTgab{a}{b} \suTwoTildeGen{a}{j}{i} \suTwoTildeGen{b}{k}{l} A_{2}\parn{\dots}
		\label{eqn:f1f2Wh_reduced_amplitudes_SU2_1}\eqncomma\\
			\ampFourPt{U^{c}}{D}{H_i}{H_j}  &= \epsilon^{ij} \redAmpFourPtN{-}{U^{c}}{D}{H}{H}
			\eqncomma \\
	\ampFivePt{Q^{c, i}}{Q_j}{W^{a},h_3}{H_k}{H^{\dagger l}} &=
		\suTwoTildeGen{a}{j}{i} \suTwoNDelta{k}{l} \, A_{1}\parn{\dots}
		+\suTwoTildeGen{a}{k}{i} \suTwoNDelta{j}{l} \, A_{2}\parn{\dots}
		\nonumber\\
		&\quad +\suTwoTildeGen{a}{k}{l} \suTwoNDelta{j}{i} \, A_{3}\parn{\dots}
		\label{eqn:f1f2Wh_reduced_amplitudes_SU2_2}\eqncomma\\
			\ampFivePt{U^{c}}{D}{W^{a},h_3}{H_i}{H_j} &= 
	\parn{\epsilon^{jl} \suTwoTildeGen{a}{i}{l} + \epsilon^{il} \suTwoTildeGen{a}{j}{l}}A_+\parn{\dots}  \eqncomma \\
\ampFivePt{Q^{c,i}}{D}{H_j}{H_k}{H^{\dagger l}} &=\parn{\suTwoNDelta{j}{i}\suTwoNDelta{k}{l} + 
\suTwoNDelta{j}{l}\suTwoNDelta{k}{i}} A_{+}\parn{\dots}+\epsilon^{jk}\epsilon_{il}A_{-}\parn{\dots} \label{eqn:f1f2Wh_reduced_amplitudes_bose_1}\eqncomma\\
	\ampFivePt{U^{c}}{Q_i}{H_j}{H_k}{H^{\dagger l}} &=    
	\parn{\epsilon^{ij}\suTwoNDelta{k}{l}+\epsilon^{ik}\suTwoNDelta{j}{l}}A_{+}\parn{\dots}+\epsilon^{jk}\suTwoNDelta{i}{l}A_{-}\parn{\dots}   \eqncomma  \label{eqn:f1f2Wh_reduced_amplitudes_bose_2}
\end{align}%
\end{subequations}%
where, for brevity, we sometimes dropped the particle labels.

%%%%%%%%%%%%%%%%%%%%%%%%%%%%%%%%%%%%%%%%%%%%%%%%%%%%%%
\section{Three-point Higgsing details}
\label{app::massless_amplitudes}

%%%%%%%%%%%%%%%%%%%%%%%%%%%%%%%%%%%%%%%%%%%%%%%%%%%%%%%%%%%
\subsection*{Freezing Higgs momenta and four-point to three-point kinematics}
\label{app:4to3}
%%%%%%%%%%%%%%%%%%%%%%%%%%%%%%%%%%%%%%%%%%%%%%%%%%%%%%%%%
Here we discuss the soft or collinear Higgs limit in some detail, paying special attention to
the transition from four-point to three-point kinematics.
Consider an $n$-point amplitude in which the $n$-th leg is a Higgs leg, which we want to freeze such that
$\tilde s_{1n}^2=m_1^2$. In the massless limit, working with complex momenta, either $[1n]$ or $\anb{1n}$, or both, go to zero, while for real momenta, $[1n],\anb{1n}\sim m_1\to0$.

We can always write
$n]=a 1]+\epsilon r]$ for some arbitrary spinor $r]$ with $[1r]$ nonzero and finite.
To take the small mass limit we can either keep $a$ finite with $\epsilon \to0$, or we can take both $a$ and $\epsilon$ to zero. 
Doing the former,  $n]$ remains finite and becomes collinear with $1]$, while for the latter choice, $n$ becomes soft. 

Now let us specialize to the case $n=4$, where it would be useful to work with complex momenta.
To end up in square three-point kinematics, for example, we take $[14]\to0$.
Using momentum conservation,
\beq\label{eqn:momc}
1]\langle1+\cdots 4]\langle4=0\,,
\eeq
and dotting with either $[1$ or $[4$,  we find that to leading order,
$2\rangle\propto 3\rangle$:
\beq
3\rangle \sim-\frac{[12]}{[13]}2\rangle\,,~~ \text{or,}~~ 3\rangle \sim-\frac{[24]}{[34]}2\rangle~~\text{for}~[14]\to0\,.
\eeq
Thus for some arbitrary $\xi\rangle$,
\beq\label{eqn:arb}
\frac{\anb{2\xi}}{\anb{3\xi}}\sim -\frac{[13]}{[12]}\,,
\eeq
which is independent of $\xi\rangle$. We will use~\autoref{eqn:arb} extensively. 
Furthermore, substituting $4]=a 1]+\epsilon 4]$ into~\autoref{eqn:momc}, and dotting with $3]$, we find also
\beq
\langle 1^\prime\equiv \langle 1 + a\langle 4
\propto \langle 2 \,,
\eeq
up to ${\cal O}(\epsilon)$ terms. 
Taking also $4\rangle \to0$, we end up with square spinor kinematics.

\subsection*{Massless amplitudes}
In this appendix we present the full four-, five- and six-point massless amplitudes which match the mass-suppressed components of ${\cal M}_3({\bf {1}}_{\psi^c},{\bf {2}}_{\psi},{\bf {3}}_{Z})$ discussed in \autoref{sec:three_point}.
The amplitudes were obtained using Feynman diagrams; for convenience, we leave (when relevant) the explicit dependence on the reference momentum.
As expected, all of the amplitudes are gauge invariant under the assumption $Q_\phi = Q_\eta-Q_\chi$. Starting with the four-point amplitudes, we have
\begin{align}
    i{\mathcal A}_4(1^+_{\chiL^c}, 2^{-}_{\chiL},3_{\theta},4_{h}) 
    &= - \abs{\frac{y_\phi}{\sqrt{2}}}^2 \frac{\langle 24 \rangle}{\langle 14 \rangle} 
    -
    \abs{\frac{y_\phi}{\sqrt{2}}}^2 \frac{[14] }{[24]} 
    +2
    (g Q_\chi)(gQ_{\phi})\frac{ [142\rangle}{s_{34}} \,,
    \label{eq:massles_4pt_pmGBh}
    \\
    {\mathcal A}_4(1^-_{\etaR^c},2^{-}_{\chiL},3^+_{\gamma},4_{h})  
     & =
     -\left( \frac{y_\phi}{\sqrt{2}} \right)(\sqrt{2}g)\left[
     Q_\chi   \frac{[34]\langle 2\xi  \rangle}{[14]\langle 3\xi\rangle} 
    +
     Q_{\eta} \frac{[34] \langle 1\xi\rangle }{[24]\langle 3\xi\rangle}
    -
     Q_{\phi}   \frac{\langle12 \rangle}{ \langle 34 \rangle}\frac{\langle   4 \xi\rangle }{\langle  34 \rangle}\right] \nonumber
     \\
     &=
     -\left( \frac{y_\phi}{\sqrt{2}} \right)(\sqrt{2}g)\left[
     Q_\chi   \frac{[34]\langle 24  \rangle}{[14]\langle 34\rangle} 
    +
     Q_{\eta} \frac{[34] \langle 14\rangle }{[24]\langle 34\rangle}
\right]\,,
     \label{eq:massles_4pt_mmph}
     \\
  {\mathcal A}_4(1^-_{\etaR^c},2^{-}_{\chiL},3^-_{\gamma},4_{h}) 
  & = 
 \left( \frac{y_\phi}{\sqrt{2}} \right)\left[
   \sqrt{2} g Q_\chi   \frac{ \langle 23\rangle[4\xi]}{[14][ 3\xi]} 
  +
   \sqrt{2} gQ_{\eta} \frac{\langle 13\rangle [ 4 \xi] }{[24][ 3\xi]}
  +
   \sqrt{2}gQ_{\phi}   \frac{\langle12 \rangle}{[34]} \frac{ [ 4\xi]}{[  3\xi ] }
  \right]
    \nonumber
  \\
 & = 0
 \label{eq:massles_4pt_mmmh}\,.  
\end{align}

Next, we have the five-point amplitudes, starting with
\begin{align}\label{eq:massles_5pt_ffghh}
\mathcal{A}_5(1^-_{\etaR^c},2^{+}_{\etaR},3^+_{\gamma},4_{h},5_{h}) = \sum_{i=1}^{7} (D_i+\overline{D}_i)\,,
\end{align}
where $\overline{D}_i(1,2,3,4,5) = {D}_i(1,2,3,5,4)$ and
\begin{subequations}
  \label{eq:massles_5pt_mpphh}
\begin{align}
D_1
&=
- \left|\frac{y_\phi}{\sqrt{2}}\right|^2\left(\sqrt{2}g Q_\chi\right) \frac{ [34]\langle 5\xi\rangle}{[14]\langle25\rangle\langle 3\xi\rangle}\,,
\label{eq:massles_4pt_mmmh_match}
\\
D_2
&=
\left|\frac{y_\phi}{\sqrt{2}}\right|^2\left(\sqrt{2}gQ_{\phi}  \right)\frac{ [24]\langle 5\xi  \rangle}{[14]\langle35\rangle\langle 3\xi  \rangle} \,,
\\
D_3
&=
 \left|\frac{y_\phi}{\sqrt{2}}\right|^2
\left(\sqrt{2}gQ_{\phi}\right)
 \frac{\langle14\rangle \langle 5\xi \rangle }{\langle24\rangle \langle 35\rangle\langle 3\xi  \rangle}\,, 
\\
D_4
&=
\left| \frac{y_{\phi}}{\sqrt{2}}\right|^2
\left(\sqrt{2}gQ_{\eta}\right)\frac{ \langle15\rangle\langle1\xi\rangle}{\langle13\rangle \langle 25\rangle \langle 3\xi\rangle}\,,
\\
D_5 
&=
\left| \frac{y_{\phi}}{\sqrt{2}}\right|^2\left(\sqrt{2}gQ_{\eta}\right)
\left( \frac{[24]\langle 2\xi\rangle}{[14]\langle 23\rangle \langle 3\xi\rangle}+ \frac{[34]}{[14]\langle 23\rangle  }\right)\,,
\\
D_{6}
&=
2\sqrt{2}\left(gQ_{\eta}\right)\left(gQ_{\phi}\right)^2 \frac{\langle 142]\langle 5  \xi \rangle}{s_{12}\anb{35}\langle 3  \xi \rangle}\,,
\\
D_{7}
&=
\sqrt{2} \left(gQ_{\eta}\right)\left(g Q_\phi\right)^2
\frac{ [23 ] \langle1 \xi \rangle}{s_{12}\langle 3\xi\rangle}\,.
\end{align}
\end{subequations}
Choosing $\xi=5$, for example, we find the gauge-invariant expression
\begin{align}
    &\mathcal{A}_5(1^-_{\etaR^c},2^{+}_{\etaR},3^+_{\gamma},4_{h},5_{h}) \nonumber
    \\
    &= \left(\sqrt{2}gQ_{\eta}\right)\left[
    \left| \frac{y_{\phi}}{\sqrt{2}}\right|^2
    \left( \frac{ \langle15\rangle^2}{\langle13\rangle \langle 25\rangle \langle 35\rangle}
    -\frac{\langle 15\rangle}{\langle 23\rangle \langle 35\rangle}\right)
+
     \left(g Q_\phi\right)^2
    \frac{ [23 ] \langle1 5 \rangle}{s_{12}\langle 35\rangle}\right]
    +(4\leftrightarrow5)\,.
\end{align}

The next five-point amplitude is
\begin{align}
i {\cal A}_5(1^+_{\chiL^c},2^{+}_{\etaR},3_{\theta},4_{h},5_{h}) &= \sum_{i=1}^{7} (A_i+\overline{A}_i)\,,
\end{align}
where $\overline{A}_i(1,2,3,4,5) = {A}_i(1,2,3,5,4)$ and
\begin{subequations}
 \label{eq:massles_5pt_ppGBhh}
\begin{align}
A_1&=  \left(\frac{y_\phi^\ast}{\sqrt{2}} \right)\left(2 g^2 
Q_{\chi}Q_\phi\right)\frac{[1542] }{s_{24} s_{35}}
=\left(\frac{y_\phi^\ast}{\sqrt{2}} \right)\left(2 g^2 
Q_{\chi}Q_\phi\right)\frac{[154\rangle }{s_{35}\anb{24} } \,,
\\
A_2&= - \left(\frac{y_\phi^\ast}{\sqrt{2}} \right)\left(2 g^2 Q_{\eta} Q_\phi\right) \frac{[1452] }{s_{14} s_{35}} = 
\left(\frac{y_\phi^\ast}{\sqrt{2}} \right)\left(2 g^2 Q_{\eta} Q_\phi\right) \frac{[ 254\rangle}{s_{35}\anb{14} }\,,
\\
 A_3&= - \left|\frac{y_\phi}{\sqrt{2}} \right|^2
 \left(\frac{y_\phi^\ast}{\sqrt{2}}\right)
\frac{[1452]}{s_{14}s_{25}}
= \left|\frac{y_\phi}{\sqrt{2}} \right|^2
 \left(\frac{y_\phi^\ast}{\sqrt{2}}\right)
\frac{\anb{45}}{\anb{14}\anb{25}}\,,
 \\
  A_4&= - \left(\frac{y_\phi^\ast}{\sqrt{2}} \right)\left(gQ_\phi\right)^2 
 \frac{[12]}{ s_{35}}\,,
\\
A_5&=-\left| \frac{y_\phi}{\sqrt{2}}\right|^2 \left(\frac{y_\phi^\ast}{\sqrt{2}} \right) \left( \frac{[12]}{s_{24}}+\frac{[1352]}{s_{13}s_{24}}\right)
=-\left| \frac{y_\phi}{\sqrt{2}}\right|^2 \left(\frac{y_\phi^\ast}{\sqrt{2}} \right)\frac{\anb{34}}{\anb{13}\anb{24}}\,,
\\
A_6&=
-\left| \frac{y_\phi}{\sqrt{2}}\right|^2 \left(\frac{y_\phi^\ast}{\sqrt{2}} \right) \left(\frac{[12]}{s_{14}}+\frac{[1532]}{s_{23}s_{14}}\right)
= \left| \frac{y_\phi}{\sqrt{2}}\right|^2 \left(\frac{y_\phi^\ast}{\sqrt{2}} \right) \frac{\anb{34}}{\anb{14}\anb{23}}\,,
\\
A_7&=
 \left(\frac{y^*_\phi}{\sqrt{2}} \right)\left(gQ_{\phi}\right)^2
 \frac{s_{34}-s_{45}}{s_{12}s_{35}}[12]\,.
\end{align}
\end{subequations}

Lastly, we have the single six-point amplitude required in the text, namely
\begin{align}
{\mathcal A}_6(1^-_{\etaR^c},2^{-}_{\chiL},3^+_{\gamma},4_{h},5_{h},6_{h}) = \sum_{i=1}^{i=3}(B_i+\overline{B}_i+\overline{\overline{B}}_i)+...\,,
\label{eq:6pt_mmphhh}
\end{align}
where $\overline{B}_i(1,2,3,4,5,6) = {B}_i(1,2,3,6,5,4)$, $\overline{\overline{B}}_i(1,2,3,4,5,6) = {B}_i(1,2,3,4,6,5)$ and 
\begin{subequations}
\begin{align}
B_1 
&= \left(\frac{2\cbarDipole}{\Lambda^2} \right) \left(\frac{y_\phi}{\sqrt{2}} \right)^2\frac{  [34] [3 5] }{[14][25]}+(4\leftrightarrow5)\,,
\\
B_2 
&=-
 \left(\frac{2\cDipole}{\Lambda^2}\right)\left(gQ_{\phi}\right)^2\frac{(s_{24}-s_{14})\langle 12 \rangle+\langle  24 \rangle\langle 16\rangle [ 4 6] +\langle 14\rangle \langle 26 \rangle[4 6]}{s_{126}\anb{35}}    \frac{\langle 5 \xi\rangle}{\langle 3 \xi  \rangle}+(4\leftrightarrow5)\,,
\\
B_3 
&=
\left(\frac{2\cDipole}{\Lambda^2}\right)\left(g Q_\phi\right)^2\frac{\langle 12\rangle[3  (2-1)\xi\rangle +\langle 16\rangle [36]\langle 2\xi \rangle + \langle2 6 \rangle[36]\langle 1\xi \rangle}{s_{126}\langle 3 \xi\rangle} \,.
\end{align}
\end{subequations}
Note we are only keeping track of terms which are relevant to our discussion, namely terms which are proportional to $\cDipole$ and $\cbarDipole$. $B_1$ is trivially gauge invariant, while one can show (\eg, numerically) that $B_2+B_3$ is also gauge invariant.
In the limit $6\to 0$, we find
\begin{align}
\lim_{6\to0}\left(B_2+B_3\right) = \left(\frac{2\cDipole}{\Lambda^2}\right)\left(gQ_{\phi}\right)^2
\frac{1}{[12]\langle 3\xi \rangle}
 \left[ \left(s_{24}-s_{14}\right)    \frac{\langle 5  \xi \rangle }{\anb{35}}+ \left(s_{25}-s_{15}\right)    \frac{\langle 4  \xi \rangle}{\anb{34}}  -[3 (2-1)\xi \rangle  \right] \,,
\end{align}
which is also a gauge invariant quantity.
Setting $\xi=1$ for concreteness, we get
\begin{align}
    \lim_{6\to0}\left(B_2+B_3\right) 
 & = - \left(\frac{2\cDipole}{\Lambda^2}\right)\left(gQ_{\phi}\right)^2
\frac{1}{[2(4+5)3 \rangle}
 \left[ \left(s_{24}-s_{14}\right)    \frac{\langle 5  1 \rangle}{\anb{35}}  + \left(s_{25}-s_{15}\right)   \frac{\langle 4  1\rangle}{\anb{34}}  
 +[3 (4+5)1 \rangle  \right]\,.
 \label{eq:dipole_massless_6point}
\end{align}
For the purpose of matching, we calculate the collinear limit  $p_5 =c \,p_4$ and find
\begin{align}
\lim_{\substack{4 \parallel 5\\ 6\to0}}\left(B_2+B_3\right) 
& = - \left(\frac{2\cDipole}{\Lambda^2}\right)\left(gQ_{\phi}\right)^2
\frac{1}{\left(1+c\right)[2 4 3 \rangle}
 \left[\left(1+c\right) \left(s_{24}-s_{14}\right)  \frac{ \langle 4  1 \rangle}{\anb{34}} 
 +\left(1+c\right)[3 41 \rangle  \right]\nonumber \\
 & = - \left(\frac{2\cDipole}{\Lambda^2}\right)\left(gQ_{\phi}\right)^2
\frac{1}{[2 4] \langle 4 3 \rangle}
 \left[ \left(s_{24}-s_{14}\right)  \frac{ \langle 4  1 \rangle}{\anb{34}} 
 - s_{34} \frac{\langle 4 1 \rangle}{\anb{34}}   \right]\nonumber \\
&=  2\left(\frac{2\cDipole}{\Lambda^2}\right)\left(gQ_{\phi}\right)^2  
\frac{ \langle 14 \rangle  \langle 2 4 \rangle}{ \langle 34 \rangle ^2  }\,.
\end{align}
%%%%%%%%%%%%%%%%%%%%%%%%%%%%%%%%%%%%%%%%%%%%%%%%%%%%%%

\subsection*{Matching with reference momenta}
The identification of the $q$ vector of a massive transverse vector with its HE reference momentum $\xi$, facilitates (and in some cases, enables) the identification of several mass-suppressed components of ${\cal M}_3({\bf {1}}_{\psi^c},{\bf {2}}_{\psi},{\bf {3}}_{Z})$ with higher-point amplitudes. For example, let us consider the little-group indices $(2,2,\{22\})$, for which the HE amplitude in \autoref{eq:massles_4pt_mmmh} vanishes. However, by taking the first line of  \autoref{eq:massles_4pt_mmmh} as a starting point, one can easily show that
\begin{subequations}
\begin{align}
\mathcal{M}_3({\bf {1}}^{2}_{\psi^c},{\bf {2}}^{2}_{\psi},{\bf {3}}^{22}_{Z})\Big\vert_{{\cal O}(m_1)}& = \lim_{\substack{[14]\to0 \\ \xi\to 3_q}} \hv\, {\mathcal A}_4(1^-_{\etaR^c}, 2^{-}_{\chiL}, 3^-_{\gamma}(\xi), 4_{h}) \,, 
 \\
\mathcal{M}_3({\bf {1}}^{2}_{\psi^c},{\bf {2}}^{2}_{\psi},{\bf {3}}^{22}_{Z})\Big\vert_{{\cal O}(m_2)} &= \lim_{\substack{[24]\to0 \\ \xi\to 3_q}}\hv\, {\mathcal A}_4(1^-_{\etaR^c}, 2^{-}_{\chiL}, 3^-_{\gamma}(\xi), 4_{h}) \,,
\end{align}
\end{subequations}
where the frozen Higgs momentum $4$ is identified either $1_q$ and $2_q$ in the first and second line, respectively. For notational clarity, we added the explicit dependence on $\xi$ on the right-hand side.

One can match, in a similar fashion, the $\mathcal{O}(m^2)$ correction to the (would-be) gauge interaction $(2,1,11)$
\begin{align}
{\mathcal M}_3({\bf {1}}^{2}_{\psi^c},{\bf {2}}^{1}_{\psi},{\bf {3}}^{11}_{Z})
&=
\sqrt{2}g Q_\eta\frac{\langle1_k 3_q \rangle[  2_k 3_k ]}{m_Z}-\sqrt{2}g Q_\chi\frac{[1_q 3_k ]\langle  2_q 3_q \rangle}{m_Z}\nonumber
\\
&=
\sqrt{2}g Q_\eta\frac{\langle1_k 3_q \rangle[  2_k 3_k ]}{\langle 3_k 3_q \rangle}+\sqrt{2}g Q_\chi m_1 m_2 \left(\frac{[1_q 3_k ]\langle 2_q 3_q \rangle}{[1_q 1_k]\langle 2_q 2_k  \rangle \langle 3_k 3_q \rangle}\right)\,.
\end{align}
By starting from the $\xi$-dependent expression for $\mathcal{A}_5(1^-_{\etaR^c},2^{+}_{\etaR},3^+_{\gamma},4_{h},5_{h})$ given in \autoref{eq:massles_5pt_ffghh} (and in particular, the term in \autoref{eq:massles_4pt_mmmh_match}), we find that
\begin{align}
\mathcal{M}_3({\bf {1}}^{2}_{\psi^c},{\bf {2}}^{1}_{\psi},{\bf {3}}^{11}_{Z})\Big\vert_{{\cal O}(m_1 m_2)}& = \lim_{\substack{[14],\anb{25}
\to0 \\
\xi \to 3_q}}
\, \hv^2
    {\mathcal A}_5(1^-_{\etaR^c}2^{+}_{\etaR}3^+_{\gamma}(\xi)4_{h}5_{h})\,,
 \end{align}
 where $4$ and $5$ are identified with $1_q$ and $2_q$, respectively. Note that although $\mathcal{A}_5(1^-_{\etaR^c},2^{+}_{\etaR},3^+_{\gamma},4_{h},5_{h})$ contains additional terms, in the limit stated above all the terms vanish except the one in \autoref{eq:massles_4pt_mmmh_match}.
%%%%%%%%%%%%%%%%%%%%%%%%%%%%%%%%%%%%%%%%%%%%%%%%%%%%%%%%%%%%%
%%%%%%%%%%%%%%%%%%%%%%%%%%%%%%%%%%%%%%%%%%%%%%%%%%%%%%

\end{fmffile}
\bibliographystyle{apsrev4-1_title}
\bibliography{Higgsing}

%merlin.mbs apsrev4-1.bst 2010-07-25 4.21a (PWD, AO, DPC) hacked
%Control: key (0)
%Control: author (72) initials jnrlst
%Control: editor formatted (1) identically to author
%Control: production of article title (1) required
%Control: page (0) single
%Control: year (1) truncated
%Control: production of eprint (0) enabled
\begin{thebibliography}{44}%
\makeatletter
\providecommand \@ifxundefined [1]{%
 \@ifx{#1\undefined}
}%
\providecommand \@ifnum [1]{%
 \ifnum #1\expandafter \@firstoftwo
 \else \expandafter \@secondoftwo
 \fi
}%
\providecommand \@ifx [1]{%
 \ifx #1\expandafter \@firstoftwo
 \else \expandafter \@secondoftwo
 \fi
}%
\providecommand \natexlab [1]{#1}%
\providecommand \enquote  [1]{``#1''}%
\providecommand \bibnamefont  [1]{#1}%
\providecommand \bibfnamefont [1]{#1}%
\providecommand \citenamefont [1]{#1}%
\providecommand \href@noop [0]{\@secondoftwo}%
\providecommand \href [0]{\begingroup \@sanitize@url \@href}%
\providecommand \@href[1]{\@@startlink{#1}\@@href}%
\providecommand \@@href[1]{\endgroup#1\@@endlink}%
\providecommand \@sanitize@url [0]{\catcode `\\12\catcode `\$12\catcode
  `\&12\catcode `\#12\catcode `\^12\catcode `\_12\catcode `\%12\relax}%
\providecommand \@@startlink[1]{}%
\providecommand \@@endlink[0]{}%
\providecommand \url  [0]{\begingroup\@sanitize@url \@url }%
\providecommand \@url [1]{\endgroup\@href {#1}{\urlprefix }}%
\providecommand \urlprefix  [0]{URL }%
\providecommand \Eprint [0]{\href }%
\providecommand \doibase [0]{http://dx.doi.org/}%
\providecommand \selectlanguage [0]{\@gobble}%
\providecommand \bibinfo [0]{\@secondoftwo}%
\providecommand \bibfield [0]{\@secondoftwo}%
\providecommand \translation [1]{[#1]}%
\providecommand \BibitemOpen [0]{}%
\providecommand \bibitemStop [0]{}%
\providecommand \bibitemNoStop [0]{.\EOS\space}%
\providecommand \EOS [0]{\spacefactor3000\relax}%
\providecommand \BibitemShut  [1]{\csname bibitem#1\endcsname}%
\let\auto@bib@innerbib\@empty
%</preamble>
\bibitem [{\citenamefont{Cohen} \emph {et\,al.}(2011)\citenamefont{Cohen},
  \citenamefont{Elvang}, and \citenamefont{Kiermaier}}]{Cohen:2010mi}%
  \BibitemOpen
  \bibfield{author}{\bibinfo{author}{\bibfnamefont{T.}\,\bibnamefont{Cohen}},
  \bibinfo{author}{\bibfnamefont{H.}\,\bibnamefont{Elvang}},  and
  \bibinfo{author}{\bibfnamefont{M.}\,\bibnamefont{Kiermaier}},
  }\bibfield{title}{\emph {\bibinfo{title}{{On-shell constructibility of tree
  amplitudes in general field theories}}}, }\href {\doibase
  10.1007/JHEP04(2011)053}
  {\bibfield{journal}{\bibinfo{journal}{JHEP}\,}\textbf{\bibinfo{volume}{04}}\,(\bibinfo{year}{2011})\,\bibinfo{pages}{053}},
  \Eprint {http://arxiv.org/abs/1010.0257}{arXiv:1010.0257
  [hep-th]}\BibitemShut {NoStop}%
%%CITATION = ARXIV:1010.0257;%%
\bibitem [{\citenamefont{Cheung} and
  \citenamefont{Shen}(2015)}]{Cheung:2015aba}%
  \BibitemOpen
  \bibfield{author}{\bibinfo{author}{\bibfnamefont{C.}\,\bibnamefont{Cheung}}
  and \bibinfo{author}{\bibfnamefont{C.-H.} \bibnamefont{Shen}},
  }\bibfield{title}{\emph {\bibinfo{title}{{Nonrenormalization Theorems without
  Supersymmetry}}}, }\href {\doibase 10.1103/PhysRevLett.115.071601}
  {\bibfield{journal}{\bibinfo{journal}{Phys. Rev.
  Lett.}\,}\textbf{\bibinfo{volume}{115}}\,(\bibinfo{year}{2015})\,\bibinfo{pages}{071601}},
  \Eprint {http://arxiv.org/abs/1505.01844}{arXiv:1505.01844
  [hep-ph]}\BibitemShut {NoStop}%
%%CITATION = ARXIV:1505.01844;%%
\bibitem [{\citenamefont{Azatov} \emph {et\,al.}(2017)\citenamefont{Azatov},
  \citenamefont{Contino}, \citenamefont{Machado}, and
  \citenamefont{Riva}}]{Azatov:2016sqh}%
  \BibitemOpen
  \bibfield{author}{\bibinfo{author}{\bibfnamefont{A.}\,\bibnamefont{Azatov}},
  \bibinfo{author}{\bibfnamefont{R.}\,\bibnamefont{Contino}},
  \bibinfo{author}{\bibfnamefont{C.~S.} \bibnamefont{Machado}},  and
  \bibinfo{author}{\bibfnamefont{F.}\,\bibnamefont{Riva}},
  }\bibfield{title}{\emph {\bibinfo{title}{{Helicity selection rules and
  noninterference for BSM amplitudes}}}, }\href {\doibase
  10.1103/PhysRevD.95.065014} {\bibfield{journal}{\bibinfo{journal}{Phys.
  Rev.}\,}\textbf{\bibinfo{volume}{D95}}\,(\bibinfo{year}{2017})\,\bibinfo{pages}{065014}},
  \Eprint {http://arxiv.org/abs/1607.05236}{arXiv:1607.05236
  [hep-ph]}\BibitemShut {NoStop}%
%%CITATION = ARXIV:1607.05236;%%
\bibitem [{\citenamefont{Elias~Mir\'o} \emph
  {et\,al.}(2020)\citenamefont{Elias~Mir\'o}, \citenamefont{Ingoldby}, and
  \citenamefont{Riembau}}]{EliasMiro:2020tdv}%
  \BibitemOpen
  \bibfield{author}{\bibinfo{author}{\bibfnamefont{J.}\,\bibnamefont{Elias~Mir\'o}},
  \bibinfo{author}{\bibfnamefont{J.}\,\bibnamefont{Ingoldby}},  and
  \bibinfo{author}{\bibfnamefont{M.}\,\bibnamefont{Riembau}},
  }\bibfield{title}{\emph {\bibinfo{title}{{EFT anomalous dimensions from the
  S-matrix}}}, }\href {\doibase 10.1007/JHEP09(2020)163}
  {\bibfield{journal}{\bibinfo{journal}{JHEP}\,}\textbf{\bibinfo{volume}{09}}\,(\bibinfo{year}{2020})\,\bibinfo{pages}{163}},
  \Eprint {http://arxiv.org/abs/2005.06983}{arXiv:2005.06983
  [hep-ph]}\BibitemShut {NoStop}%
\bibitem [{\citenamefont{Baratella} \emph
  {et\,al.}(2020)\citenamefont{Baratella}, \citenamefont{Fernandez}, and
  \citenamefont{Pomarol}}]{Baratella:2020lzz}%
  \BibitemOpen
  \bibfield{author}{\bibinfo{author}{\bibfnamefont{P.}\,\bibnamefont{Baratella}},
  \bibinfo{author}{\bibfnamefont{C.}\,\bibnamefont{Fernandez}},  and
  \bibinfo{author}{\bibfnamefont{A.}\,\bibnamefont{Pomarol}},
  }\bibfield{title}{\emph {\bibinfo{title}{{Renormalization of
  Higher-Dimensional Operators from On-shell Amplitudes}}}, }\href {\doibase
  10.1016/j.nuclphysb.2020.115155} {\bibfield{journal}{\bibinfo{journal}{Nucl.
  Phys.
  B}\,}\textbf{\bibinfo{volume}{959}}\,(\bibinfo{year}{2020})\,\bibinfo{pages}{115155}},
  \Eprint {http://arxiv.org/abs/2005.07129}{arXiv:2005.07129
  [hep-ph]}\BibitemShut {NoStop}%
\bibitem [{\citenamefont{Jiang} \emph
  {et\,al.}(2021{\natexlab{a}})\citenamefont{Jiang}, \citenamefont{Ma}, and
  \citenamefont{Shu}}]{Jiang:2020mhe}%
  \BibitemOpen
  \bibfield{author}{\bibinfo{author}{\bibfnamefont{M.}\,\bibnamefont{Jiang}},
  \bibinfo{author}{\bibfnamefont{T.}\,\bibnamefont{Ma}},  and
  \bibinfo{author}{\bibfnamefont{J.}\,\bibnamefont{Shu}},
  }\bibfield{title}{\emph {\bibinfo{title}{{Renormalization Group Evolution
  from On-shell SMEFT}}}, }\href {\doibase 10.1007/JHEP01(2021)101}
  {\bibfield{journal}{\bibinfo{journal}{JHEP}\,}\textbf{\bibinfo{volume}{01}}\,(\bibinfo{year}{2021}{\natexlab{a}})\,\bibinfo{pages}{101}},
  \Eprint {http://arxiv.org/abs/2005.10261}{arXiv:2005.10261
  [hep-ph]}\BibitemShut {NoStop}%
\bibitem [{\citenamefont{Accettulli~Huber} and
  \citenamefont{De~Angelis}(2021)}]{AccettulliHuber:2021uoa}%
  \BibitemOpen
  \bibfield{author}{\bibinfo{author}{\bibfnamefont{M.}\,\bibnamefont{Accettulli~Huber}}
  and \bibinfo{author}{\bibfnamefont{S.}\,\bibnamefont{De~Angelis}},
  }\bibfield{title}{\emph {\bibinfo{title}{{Standard Model EFTs via on-shell
  methods}}}, }\href {\doibase 10.1007/JHEP11(2021)221}
  {\bibfield{journal}{\bibinfo{journal}{JHEP}\,}\textbf{\bibinfo{volume}{11}}\,(\bibinfo{year}{2021})\,\bibinfo{pages}{221}},
  \Eprint {http://arxiv.org/abs/2108.03669}{arXiv:2108.03669
  [hep-th]}\BibitemShut {NoStop}%
\bibitem [{\citenamefont{Baratella} \emph {et\,al.}()\citenamefont{Baratella},
  \citenamefont{Haslehner}, \citenamefont{Ruhdorfer}, \citenamefont{Serra}, and
  \citenamefont{Weiler}}]{Baratella:2021guc}%
  \BibitemOpen
  \bibfield{author}{\bibinfo{author}{\bibfnamefont{P.}\,\bibnamefont{Baratella}},
  \bibinfo{author}{\bibfnamefont{D.}\,\bibnamefont{Haslehner}},
  \bibinfo{author}{\bibfnamefont{M.}\,\bibnamefont{Ruhdorfer}},
  \bibinfo{author}{\bibfnamefont{J.}\,\bibnamefont{Serra}},  and
  \bibinfo{author}{\bibfnamefont{A.}\,\bibnamefont{Weiler}},
  }\bibfield{title}{\emph {\bibinfo{title}{{RG of GR from On-shell
  Amplitudes}}}, }\href@noop {} {}\,\Eprint
  {http://arxiv.org/abs/2109.06191}{arXiv:2109.06191 [hep-th]}\BibitemShut
  {NoStop}%
\bibitem [{\citenamefont{Jiang} \emph
  {et\,al.}(2021{\natexlab{b}})\citenamefont{Jiang}, \citenamefont{Shu},
  \citenamefont{Xiao}, and \citenamefont{Zheng}}]{Jiang:2020rwz}%
  \BibitemOpen
  \bibfield{author}{\bibinfo{author}{\bibfnamefont{M.}\,\bibnamefont{Jiang}},
  \bibinfo{author}{\bibfnamefont{J.}\,\bibnamefont{Shu}},
  \bibinfo{author}{\bibfnamefont{M.-L.} \bibnamefont{Xiao}},  and
  \bibinfo{author}{\bibfnamefont{Y.-H.} \bibnamefont{Zheng}},
  }\bibfield{title}{\emph {\bibinfo{title}{{Partial Wave Amplitude Basis and
  Selection Rules in Effective Field Theories}}}, }\href {\doibase
  10.1103/PhysRevLett.126.011601} {\bibfield{journal}{\bibinfo{journal}{Phys.
  Rev.
  Lett.}\,}\textbf{\bibinfo{volume}{126}}\,(\bibinfo{year}{2021}{\natexlab{b}})\,\bibinfo{pages}{011601}},
  \Eprint {http://arxiv.org/abs/2001.04481}{arXiv:2001.04481
  [hep-ph]}\BibitemShut {NoStop}%
\bibitem [{\citenamefont{Baratella} \emph
  {et\,al.}(2021)\citenamefont{Baratella}, \citenamefont{Fernandez},
  \citenamefont{von\,Harling}, and \citenamefont{Pomarol}}]{Baratella:2020dvw}%
  \BibitemOpen
  \bibfield{author}{\bibinfo{author}{\bibfnamefont{P.}\,\bibnamefont{Baratella}},
  \bibinfo{author}{\bibfnamefont{C.}\,\bibnamefont{Fernandez}},
  \bibinfo{author}{\bibfnamefont{B.}\,\bibnamefont{von\,Harling}},  and
  \bibinfo{author}{\bibfnamefont{A.}\,\bibnamefont{Pomarol}},
  }\bibfield{title}{\emph {\bibinfo{title}{{Anomalous Dimensions of Effective
  Theories from Partial Waves}}}, }\href {\doibase 10.1007/JHEP03(2021)287}
  {\bibfield{journal}{\bibinfo{journal}{JHEP}\,}\textbf{\bibinfo{volume}{03}}\,(\bibinfo{year}{2021})\,\bibinfo{pages}{287}},
  \Eprint {http://arxiv.org/abs/2010.13809}{arXiv:2010.13809
  [hep-ph]}\BibitemShut {NoStop}%
\bibitem [{\citenamefont{Shu} \emph {et\,al.}()\citenamefont{Shu},
  \citenamefont{Xiao}, and \citenamefont{Zheng}}]{Shu:2021qlr}%
  \BibitemOpen
  \bibfield{author}{\bibinfo{author}{\bibfnamefont{J.}\,\bibnamefont{Shu}},
  \bibinfo{author}{\bibfnamefont{M.-L.} \bibnamefont{Xiao}},  and
  \bibinfo{author}{\bibfnamefont{Y.-H.} \bibnamefont{Zheng}},
  }\bibfield{title}{\emph {\bibinfo{title}{{Constructing general partial waves
  and renormalization in EFT}}}, }\href@noop {} {}\,\Eprint
  {http://arxiv.org/abs/2111.08019}{arXiv:2111.08019 [hep-th]}\BibitemShut
  {NoStop}%
\bibitem [{\citenamefont{Bern} \emph
  {et\,al.}(2020{\natexlab{a}})\citenamefont{Bern},
  \citenamefont{Parra-Martinez}, and \citenamefont{Sawyer}}]{Bern:2019wie}%
  \BibitemOpen
  \bibfield{author}{\bibinfo{author}{\bibfnamefont{Z.}\,\bibnamefont{Bern}},
  \bibinfo{author}{\bibfnamefont{J.}\,\bibnamefont{Parra-Martinez}},  and
  \bibinfo{author}{\bibfnamefont{E.}\,\bibnamefont{Sawyer}},
  }\bibfield{title}{\emph {\bibinfo{title}{{Nonrenormalization and Operator
  Mixing via On-Shell Methods}}}, }\href {\doibase
  10.1103/PhysRevLett.124.051601} {\bibfield{journal}{\bibinfo{journal}{Phys.
  Rev.
  Lett.}\,}\textbf{\bibinfo{volume}{124}}\,(\bibinfo{year}{2020}{\natexlab{a}})\,\bibinfo{pages}{051601}},
  \Eprint {http://arxiv.org/abs/1910.05831}{arXiv:1910.05831
  [hep-ph]}\BibitemShut {NoStop}%
%%CITATION = ARXIV:1910.05831;%%
\bibitem [{\citenamefont{Bern} \emph
  {et\,al.}(2020{\natexlab{b}})\citenamefont{Bern},
  \citenamefont{Parra-Martinez}, and \citenamefont{Sawyer}}]{Bern:2020ikv}%
  \BibitemOpen
  \bibfield{author}{\bibinfo{author}{\bibfnamefont{Z.}\,\bibnamefont{Bern}},
  \bibinfo{author}{\bibfnamefont{J.}\,\bibnamefont{Parra-Martinez}},  and
  \bibinfo{author}{\bibfnamefont{E.}\,\bibnamefont{Sawyer}},
  }\bibfield{title}{\emph {\bibinfo{title}{{Structure of two-loop SMEFT
  anomalous dimensions via on-shell methods}}}, }\href {\doibase
  10.1007/JHEP10(2020)211}
  {\bibfield{journal}{\bibinfo{journal}{JHEP}\,}\textbf{\bibinfo{volume}{10}}\,(\bibinfo{year}{2020}{\natexlab{b}})\,\bibinfo{pages}{211}},
  \Eprint {http://arxiv.org/abs/2005.12917}{arXiv:2005.12917
  [hep-ph]}\BibitemShut {NoStop}%
\bibitem [{\citenamefont{Jin} \emph {et\,al.}(2021)\citenamefont{Jin},
  \citenamefont{Ren}, and \citenamefont{Yang}}]{Jin:2020pwh}%
  \BibitemOpen
  \bibfield{author}{\bibinfo{author}{\bibfnamefont{Q.}\,\bibnamefont{Jin}},
  \bibinfo{author}{\bibfnamefont{K.}\,\bibnamefont{Ren}},  and
  \bibinfo{author}{\bibfnamefont{G.}\,\bibnamefont{Yang}},
  }\bibfield{title}{\emph {\bibinfo{title}{{Two-Loop anomalous dimensions of
  QCD operators up to dimension-sixteen and Higgs EFT amplitudes}}}, }\href
  {\doibase 10.1007/JHEP04(2021)180}
  {\bibfield{journal}{\bibinfo{journal}{JHEP}\,}\textbf{\bibinfo{volume}{04}}\,(\bibinfo{year}{2021})\,\bibinfo{pages}{180}},
  \Eprint {http://arxiv.org/abs/2011.02494}{arXiv:2011.02494
  [hep-ph]}\BibitemShut {NoStop}%
\bibitem [{\citenamefont{Shadmi} and
  \citenamefont{Weiss}(2019)}]{Shadmi:2018xan}%
  \BibitemOpen
  \bibfield{author}{\bibinfo{author}{\bibfnamefont{Y.}\,\bibnamefont{Shadmi}}
  and \bibinfo{author}{\bibfnamefont{Y.}\,\bibnamefont{Weiss}},
  }\bibfield{title}{\emph {\bibinfo{title}{{Effective Field Theory Amplitudes
  the On-Shell Way: Scalar and Vector Couplings to Gluons}}}, }\href {\doibase
  10.1007/JHEP02(2019)165}
  {\bibfield{journal}{\bibinfo{journal}{JHEP}\,}\textbf{\bibinfo{volume}{02}}\,(\bibinfo{year}{2019})\,\bibinfo{pages}{165}},
  \Eprint {http://arxiv.org/abs/1809.09644}{arXiv:1809.09644
  [hep-ph]}\BibitemShut {NoStop}%
%%CITATION = ARXIV:1809.09644;%%
\bibitem [{\citenamefont{Ma} \emph {et\,al.}()\citenamefont{Ma},
  \citenamefont{Shu}, and \citenamefont{Xiao}}]{Ma:2019gtx}%
  \BibitemOpen
  \bibfield{author}{\bibinfo{author}{\bibfnamefont{T.}\,\bibnamefont{Ma}},
  \bibinfo{author}{\bibfnamefont{J.}\,\bibnamefont{Shu}},  and
  \bibinfo{author}{\bibfnamefont{M.-L.} \bibnamefont{Xiao}},
  }\bibfield{title}{\emph {\bibinfo{title}{{Standard Model Effective Field
  Theory from On-shell Amplitudes}}}, }\href@noop {} {}\,\Eprint
  {http://arxiv.org/abs/1902.06752}{arXiv:1902.06752 [hep-ph]}\BibitemShut
  {NoStop}%
%%CITATION = ARXIV:1902.06752;%%
\bibitem [{\citenamefont{Li} \emph
  {et\,al.}(2021{\natexlab{a}})\citenamefont{Li}, \citenamefont{Ren},
  \citenamefont{Shu}, \citenamefont{Xiao}, \citenamefont{Yu}, and
  \citenamefont{Zheng}}]{Li:2020gnx}%
  \BibitemOpen
  \bibfield{author}{\bibinfo{author}{\bibfnamefont{H.-L.} \bibnamefont{Li}},
  \bibinfo{author}{\bibfnamefont{Z.}\,\bibnamefont{Ren}},
  \bibinfo{author}{\bibfnamefont{J.}\,\bibnamefont{Shu}},
  \bibinfo{author}{\bibfnamefont{M.-L.} \bibnamefont{Xiao}},
  \bibinfo{author}{\bibfnamefont{J.-H.} \bibnamefont{Yu}},  and
  \bibinfo{author}{\bibfnamefont{Y.-H.} \bibnamefont{Zheng}},
  }\bibfield{title}{\emph {\bibinfo{title}{{Complete set of dimension-eight
  operators in the standard model effective field theory}}}, }\href {\doibase
  10.1103/PhysRevD.104.015026} {\bibfield{journal}{\bibinfo{journal}{Phys. Rev.
  D}\,}\textbf{\bibinfo{volume}{104}}\,(\bibinfo{year}{2021}{\natexlab{a}})\,\bibinfo{pages}{015026}},
  \Eprint {http://arxiv.org/abs/2005.00008}{arXiv:2005.00008
  [hep-ph]}\BibitemShut {NoStop}%
\bibitem [{\citenamefont{Li} \emph
  {et\,al.}(2021{\natexlab{b}})\citenamefont{Li}, \citenamefont{Ren},
  \citenamefont{Xiao}, \citenamefont{Yu}, and
  \citenamefont{Zheng}}]{Li:2020xlh}%
  \BibitemOpen
  \bibfield{author}{\bibinfo{author}{\bibfnamefont{H.-L.} \bibnamefont{Li}},
  \bibinfo{author}{\bibfnamefont{Z.}\,\bibnamefont{Ren}},
  \bibinfo{author}{\bibfnamefont{M.-L.} \bibnamefont{Xiao}},
  \bibinfo{author}{\bibfnamefont{J.-H.} \bibnamefont{Yu}},  and
  \bibinfo{author}{\bibfnamefont{Y.-H.} \bibnamefont{Zheng}},
  }\bibfield{title}{\emph {\bibinfo{title}{{Complete set of dimension-nine
  operators in the standard model effective field theory}}}, }\href {\doibase
  10.1103/PhysRevD.104.015025} {\bibfield{journal}{\bibinfo{journal}{Phys. Rev.
  D}\,}\textbf{\bibinfo{volume}{104}}\,(\bibinfo{year}{2021}{\natexlab{b}})\,\bibinfo{pages}{015025}},
  \Eprint {http://arxiv.org/abs/2007.07899}{arXiv:2007.07899
  [hep-ph]}\BibitemShut {NoStop}%
\bibitem [{\citenamefont{Li} \emph
  {et\,al.}(2021{\natexlab{c}})\citenamefont{Li}, \citenamefont{Ren},
  \citenamefont{Xiao}, \citenamefont{Yu}, and
  \citenamefont{Zheng}}]{Li:2020tsi}%
  \BibitemOpen
  \bibfield{author}{\bibinfo{author}{\bibfnamefont{H.-L.} \bibnamefont{Li}},
  \bibinfo{author}{\bibfnamefont{Z.}\,\bibnamefont{Ren}},
  \bibinfo{author}{\bibfnamefont{M.-L.} \bibnamefont{Xiao}},
  \bibinfo{author}{\bibfnamefont{J.-H.} \bibnamefont{Yu}},  and
  \bibinfo{author}{\bibfnamefont{Y.-H.} \bibnamefont{Zheng}},
  }\bibfield{title}{\emph {\bibinfo{title}{{Low energy effective field theory
  operator basis at d \ensuremath{\leq} 9}}}, }\href {\doibase
  10.1007/JHEP06(2021)138}
  {\bibfield{journal}{\bibinfo{journal}{JHEP}\,}\textbf{\bibinfo{volume}{06}}\,(\bibinfo{year}{2021}{\natexlab{c}})\,\bibinfo{pages}{138}},
  \Eprint {http://arxiv.org/abs/2012.09188}{arXiv:2012.09188
  [hep-ph]}\BibitemShut {NoStop}%
\bibitem [{\citenamefont{Ruhdorfer} \emph
  {et\,al.}(2020)\citenamefont{Ruhdorfer}, \citenamefont{Serra}, and
  \citenamefont{Weiler}}]{Ruhdorfer:2019qmk}%
  \BibitemOpen
  \bibfield{author}{\bibinfo{author}{\bibfnamefont{M.}\,\bibnamefont{Ruhdorfer}},
  \bibinfo{author}{\bibfnamefont{J.}\,\bibnamefont{Serra}},  and
  \bibinfo{author}{\bibfnamefont{A.}\,\bibnamefont{Weiler}},
  }\bibfield{title}{\emph {\bibinfo{title}{{Effective Field Theory of Gravity
  to All Orders}}}, }\href {\doibase 10.1007/JHEP05(2020)083}
  {\bibfield{journal}{\bibinfo{journal}{JHEP}\,}\textbf{\bibinfo{volume}{05}}\,(\bibinfo{year}{2020})\,\bibinfo{pages}{083}},
  \Eprint {http://arxiv.org/abs/1908.08050}{arXiv:1908.08050
  [hep-ph]}\BibitemShut {NoStop}%
\bibitem [{\citenamefont{Durieux} and
  \citenamefont{Machado}(2020)}]{Durieux:2019siw}%
  \BibitemOpen
  \bibfield{author}{\bibinfo{author}{\bibfnamefont{G.}\,\bibnamefont{Durieux}}
  and \bibinfo{author}{\bibfnamefont{C.~S.} \bibnamefont{Machado}},
  }\bibfield{title}{\emph {\bibinfo{title}{{Enumerating higher-dimensional
  operators with on-shell amplitudes}}}, }\href {\doibase
  10.1103/PhysRevD.101.095021} {\bibfield{journal}{\bibinfo{journal}{Phys.
  Rev.}\,}\textbf{\bibinfo{volume}{D101}}\,(\bibinfo{year}{2020})\,\bibinfo{pages}{095021}},
  \Eprint {http://arxiv.org/abs/1912.08827}{arXiv:1912.08827
  [hep-ph]}\BibitemShut {NoStop}%
%%CITATION = ARXIV:1912.08827;%%
\bibitem [{\citenamefont{Henning} and
  \citenamefont{Melia}(2019)}]{Henning:2019enq}%
  \BibitemOpen
  \bibfield{author}{\bibinfo{author}{\bibfnamefont{B.}\,\bibnamefont{Henning}}
  and \bibinfo{author}{\bibfnamefont{T.}\,\bibnamefont{Melia}},
  }\bibfield{title}{\emph {\bibinfo{title}{{Constructing effective field
  theories via their harmonics}}}, }\href {\doibase
  10.1103/PhysRevD.100.016015} {\bibfield{journal}{\bibinfo{journal}{Phys.
  Rev.}\,}\textbf{\bibinfo{volume}{D100}}\,(\bibinfo{year}{2019})\,\bibinfo{pages}{016015}},
  \Eprint {http://arxiv.org/abs/1902.06754}{arXiv:1902.06754
  [hep-ph]}\BibitemShut {NoStop}%
%%CITATION = ARXIV:1902.06754;%%
\bibitem [{\citenamefont{Falkowski}()}]{Falkowski:2019zdo}%
  \BibitemOpen
  \bibfield{author}{\bibinfo{author}{\bibfnamefont{A.}\,\bibnamefont{Falkowski}},
  }\bibfield{title}{\emph {\bibinfo{title}{{Bases of massless EFTs via momentum
  twistors}}}, }\href@noop {} {}\,\Eprint
  {http://arxiv.org/abs/1912.07865}{arXiv:1912.07865 [hep-ph]}\BibitemShut
  {NoStop}%
%%CITATION = ARXIV:1912.07865;%%
\bibitem [{\citenamefont{Christensen} and
  \citenamefont{Field}(2018)}]{Christensen:2018zcq}%
  \BibitemOpen
  \bibfield{author}{\bibinfo{author}{\bibfnamefont{N.}\,\bibnamefont{Christensen}}
  and \bibinfo{author}{\bibfnamefont{B.}\,\bibnamefont{Field}},
  }\bibfield{title}{\emph {\bibinfo{title}{{Constructive standard model}}},
  }\href {\doibase 10.1103/PhysRevD.98.016014}
  {\bibfield{journal}{\bibinfo{journal}{Phys.
  Rev.}\,}\textbf{\bibinfo{volume}{D98}}\,(\bibinfo{year}{2018})\,\bibinfo{pages}{016014}},
  \Eprint {http://arxiv.org/abs/1802.00448}{arXiv:1802.00448
  [hep-ph]}\BibitemShut {NoStop}%
%%CITATION = ARXIV:1802.00448;%%
\bibitem [{\citenamefont{Herderschee} \emph
  {et\,al.}(2019{\natexlab{a}})\citenamefont{Herderschee},
  \citenamefont{Koren}, and \citenamefont{Trott}}]{Herderschee:2019ofc}%
  \BibitemOpen
  \bibfield{author}{\bibinfo{author}{\bibfnamefont{A.}\,\bibnamefont{Herderschee}},
  \bibinfo{author}{\bibfnamefont{S.}\,\bibnamefont{Koren}},  and
  \bibinfo{author}{\bibfnamefont{T.}\,\bibnamefont{Trott}},
  }\bibfield{title}{\emph {\bibinfo{title}{{Massive On-Shell Supersymmetric
  Scattering Amplitudes}}}, }\href {\doibase 10.1007/JHEP10(2019)092}
  {\bibfield{journal}{\bibinfo{journal}{JHEP}\,}\textbf{\bibinfo{volume}{10}}\,(\bibinfo{year}{2019}{\natexlab{a}})\,\bibinfo{pages}{092}},
  \Eprint {http://arxiv.org/abs/1902.07204}{arXiv:1902.07204
  [hep-th]}\BibitemShut {NoStop}%
\bibitem [{\citenamefont{Herderschee} \emph
  {et\,al.}(2019{\natexlab{b}})\citenamefont{Herderschee},
  \citenamefont{Koren}, and \citenamefont{Trott}}]{Herderschee:2019dmc}%
  \BibitemOpen
  \bibfield{author}{\bibinfo{author}{\bibfnamefont{A.}\,\bibnamefont{Herderschee}},
  \bibinfo{author}{\bibfnamefont{S.}\,\bibnamefont{Koren}},  and
  \bibinfo{author}{\bibfnamefont{T.}\,\bibnamefont{Trott}},
  }\bibfield{title}{\emph {\bibinfo{title}{{Constructing $ \mathcal{N} $ = 4
  Coulomb branch superamplitudes}}}, }\href {\doibase 10.1007/JHEP08(2019)107}
  {\bibfield{journal}{\bibinfo{journal}{JHEP}\,}\textbf{\bibinfo{volume}{08}}\,(\bibinfo{year}{2019}{\natexlab{b}})\,\bibinfo{pages}{107}},
  \Eprint {http://arxiv.org/abs/1902.07205}{arXiv:1902.07205
  [hep-th]}\BibitemShut {NoStop}%
\bibitem [{\citenamefont{Aoude} and
  \citenamefont{Machado}(2019)}]{Aoude:2019tzn}%
  \BibitemOpen
  \bibfield{author}{\bibinfo{author}{\bibfnamefont{R.}\,\bibnamefont{Aoude}}
  and \bibinfo{author}{\bibfnamefont{C.~S.} \bibnamefont{Machado}},
  }\bibfield{title}{\emph {\bibinfo{title}{{The Rise of SMEFT On-shell
  Amplitudes}}}, }\href {\doibase 10.1007/JHEP12(2019)058}
  {\bibfield{journal}{\bibinfo{journal}{JHEP}\,}\textbf{\bibinfo{volume}{12}}\,(\bibinfo{year}{2019})\,\bibinfo{pages}{058}},
  \Eprint {http://arxiv.org/abs/1905.11433}{arXiv:1905.11433
  [hep-ph]}\BibitemShut {NoStop}%
%%CITATION = ARXIV:1905.11433;%%
\bibitem [{\citenamefont{Christensen} \emph
  {et\,al.}(2020)\citenamefont{Christensen}, \citenamefont{Field},
  \citenamefont{Moore}, and \citenamefont{Pinto}}]{Christensen:2019mch}%
  \BibitemOpen
  \bibfield{author}{\bibinfo{author}{\bibfnamefont{N.}\,\bibnamefont{Christensen}},
  \bibinfo{author}{\bibfnamefont{B.}\,\bibnamefont{Field}},
  \bibinfo{author}{\bibfnamefont{A.}\,\bibnamefont{Moore}},  and
  \bibinfo{author}{\bibfnamefont{S.}\,\bibnamefont{Pinto}},
  }\bibfield{title}{\emph {\bibinfo{title}{{Two-, three-, and four-body decays
  in the constructive standard model}}}, }\href {\doibase
  10.1103/PhysRevD.101.065019} {\bibfield{journal}{\bibinfo{journal}{Phys.
  Rev.}\,}\textbf{\bibinfo{volume}{D101}}\,(\bibinfo{year}{2020})\,\bibinfo{pages}{065019}},
  \Eprint {http://arxiv.org/abs/1909.09164}{arXiv:1909.09164
  [hep-ph]}\BibitemShut {NoStop}%
%%CITATION = ARXIV:1909.09164;%%
\bibitem [{\citenamefont{Durieux} \emph
  {et\,al.}(2020{\natexlab{a}})\citenamefont{Durieux}, \citenamefont{Kitahara},
  \citenamefont{Shadmi}, and \citenamefont{Weiss}}]{Durieux:2019eor}%
  \BibitemOpen
  \bibfield{author}{\bibinfo{author}{\bibfnamefont{G.}\,\bibnamefont{Durieux}},
  \bibinfo{author}{\bibfnamefont{T.}\,\bibnamefont{Kitahara}},
  \bibinfo{author}{\bibfnamefont{Y.}\,\bibnamefont{Shadmi}},  and
  \bibinfo{author}{\bibfnamefont{Y.}\,\bibnamefont{Weiss}},
  }\bibfield{title}{\emph {\bibinfo{title}{{The electroweak effective field
  theory from on-shell amplitudes}}}, }\href {\doibase 10.1007/JHEP01(2020)119}
  {\bibfield{journal}{\bibinfo{journal}{JHEP}\,}\textbf{\bibinfo{volume}{01}}\,(\bibinfo{year}{2020}{\natexlab{a}})\,\bibinfo{pages}{119}},
  \Eprint {http://arxiv.org/abs/1909.10551}{arXiv:1909.10551
  [hep-ph]}\BibitemShut {NoStop}%
%%CITATION = ARXIV:1909.10551;%%
\bibitem [{\citenamefont{Durieux} \emph
  {et\,al.}(2020{\natexlab{b}})\citenamefont{Durieux}, \citenamefont{Kitahara},
  \citenamefont{Machado}, \citenamefont{Shadmi}, and
  \citenamefont{Weiss}}]{Durieux:2020gip}%
  \BibitemOpen
  \bibfield{author}{\bibinfo{author}{\bibfnamefont{G.}\,\bibnamefont{Durieux}},
  \bibinfo{author}{\bibfnamefont{T.}\,\bibnamefont{Kitahara}},
  \bibinfo{author}{\bibfnamefont{C.~S.} \bibnamefont{Machado}},
  \bibinfo{author}{\bibfnamefont{Y.}\,\bibnamefont{Shadmi}},  and
  \bibinfo{author}{\bibfnamefont{Y.}\,\bibnamefont{Weiss}},
  }\bibfield{title}{\emph {\bibinfo{title}{{Constructing massive on-shell
  contact terms}}}, }\href {\doibase 10.1007/JHEP12(2020)175}
  {\bibfield{journal}{\bibinfo{journal}{JHEP}\,}\textbf{\bibinfo{volume}{12}}\,(\bibinfo{year}{2020}{\natexlab{b}})\,\bibinfo{pages}{175}},
  \Eprint {http://arxiv.org/abs/2008.09652}{arXiv:2008.09652
  [hep-ph]}\BibitemShut {NoStop}%
\bibitem [{\citenamefont{Dong} \emph {et\,al.}()\citenamefont{Dong},
  \citenamefont{Ma}, and \citenamefont{Shu}}]{Dong:2021yak}%
  \BibitemOpen
  \bibfield{author}{\bibinfo{author}{\bibfnamefont{Z.-Y.} \bibnamefont{Dong}},
  \bibinfo{author}{\bibfnamefont{T.}\,\bibnamefont{Ma}},  and
  \bibinfo{author}{\bibfnamefont{J.}\,\bibnamefont{Shu}},
  }\bibfield{title}{\emph {\bibinfo{title}{{Constructing on-shell operator
  basis for all masses and spins}}}, }\href@noop {} {}\,\Eprint
  {http://arxiv.org/abs/2103.15837}{arXiv:2103.15837 [hep-ph]}\BibitemShut
  {NoStop}%
\bibitem [{\citenamefont{Arkani-Hamed} \emph
  {et\,al.}(2021)\citenamefont{Arkani-Hamed}, \citenamefont{Huang}, and
  \citenamefont{Huang}}]{Arkani-Hamed:2017jhn}%
  \BibitemOpen
  \bibfield{author}{\bibinfo{author}{\bibfnamefont{N.}\,\bibnamefont{Arkani-Hamed}},
  \bibinfo{author}{\bibfnamefont{T.-C.} \bibnamefont{Huang}},  and
  \bibinfo{author}{\bibfnamefont{Y.-t.} \bibnamefont{Huang}},
  }\bibfield{title}{\emph {\bibinfo{title}{{Scattering amplitudes for all
  masses and spins}}}, }\href {\doibase 10.1007/JHEP11(2021)070}
  {\bibfield{journal}{\bibinfo{journal}{JHEP}\,}\textbf{\bibinfo{volume}{11}}\,(\bibinfo{year}{2021})\,\bibinfo{pages}{070}},
  \Eprint {http://arxiv.org/abs/1709.04891}{arXiv:1709.04891
  [hep-th]}\BibitemShut {NoStop}%
\bibitem [{\citenamefont{Craig} \emph {et\,al.}(2011)\citenamefont{Craig},
  \citenamefont{Elvang}, \citenamefont{Kiermaier}, and
  \citenamefont{Slatyer}}]{Craig:2011ws}%
  \BibitemOpen
  \bibfield{author}{\bibinfo{author}{\bibfnamefont{N.}\,\bibnamefont{Craig}},
  \bibinfo{author}{\bibfnamefont{H.}\,\bibnamefont{Elvang}},
  \bibinfo{author}{\bibfnamefont{M.}\,\bibnamefont{Kiermaier}},  and
  \bibinfo{author}{\bibfnamefont{T.}\,\bibnamefont{Slatyer}},
  }\bibfield{title}{\emph {\bibinfo{title}{{Massive amplitudes on the Coulomb
  branch of N=4 SYM}}}, }\href {\doibase 10.1007/JHEP12(2011)097}
  {\bibfield{journal}{\bibinfo{journal}{JHEP}\,}\textbf{\bibinfo{volume}{12}}\,(\bibinfo{year}{2011})\,\bibinfo{pages}{097}},
  \Eprint {http://arxiv.org/abs/1104.2050}{arXiv:1104.2050
  [hep-th]}\BibitemShut {NoStop}%
%%CITATION = ARXIV:1104.2050;%%
\bibitem [{\citenamefont{Dittmaier}(1998)}]{Dittmaier:1998nn}%
  \BibitemOpen
  \bibfield{author}{\bibinfo{author}{\bibfnamefont{S.}\,\bibnamefont{Dittmaier}},
  }\bibfield{title}{\emph {\bibinfo{title}{{Weyl-van der Waerden formalism for
  helicity amplitudes of massive particles}}}, }\href {\doibase
  10.1103/PhysRevD.59.016007} {\bibfield{journal}{\bibinfo{journal}{Phys.
  Rev.}\,}\textbf{\bibinfo{volume}{D59}}\,(\bibinfo{year}{1998})\,\bibinfo{pages}{016007}},
  \Eprint
  {http://arxiv.org/abs/hep-ph/9805445}{arXiv:hep-ph/9805445}\BibitemShut
  {NoStop}%
%%CITATION = HEP-PH/9805445;%%
\bibitem [{\citenamefont{Ochirov}(2018)}]{Ochirov:2018uyq}%
  \BibitemOpen
  \bibfield{author}{\bibinfo{author}{\bibfnamefont{A.}\,\bibnamefont{Ochirov}},
  }\bibfield{title}{\emph {\bibinfo{title}{{Helicity amplitudes for QCD with
  massive quarks}}}, }\href {\doibase 10.1007/JHEP04(2018)089}
  {\bibfield{journal}{\bibinfo{journal}{JHEP}\,}\textbf{\bibinfo{volume}{04}}\,(\bibinfo{year}{2018})\,\bibinfo{pages}{089}},
  \Eprint {http://arxiv.org/abs/1802.06730}{arXiv:1802.06730
  [hep-ph]}\BibitemShut {NoStop}%
\bibitem [{\citenamefont{Kiermaier}()}]{Kiermaier:2011cr}%
  \BibitemOpen
  \bibfield{author}{\bibinfo{author}{\bibfnamefont{M.}\,\bibnamefont{Kiermaier}},
  }\bibfield{title}{\emph {\bibinfo{title}{{The Coulomb-branch S-matrix from
  massless amplitudes}}}, }\href@noop {} {}\,\Eprint
  {http://arxiv.org/abs/1105.5385}{arXiv:1105.5385 [hep-th]}\BibitemShut
  {NoStop}%
%%CITATION = ARXIV:1105.5385;%%
\bibitem [{\citenamefont{Bachu} and
  \citenamefont{Yelleshpur}(2020)}]{Bachu:2019ehv}%
  \BibitemOpen
  \bibfield{author}{\bibinfo{author}{\bibfnamefont{B.}\,\bibnamefont{Bachu}}
  and \bibinfo{author}{\bibfnamefont{A.}\,\bibnamefont{Yelleshpur}},
  }\bibfield{title}{\emph {\bibinfo{title}{{On-Shell Electroweak Sector and the
  Higgs Mechanism}}}, }\href {\doibase 10.1007/JHEP08(2020)039}
  {\bibfield{journal}{\bibinfo{journal}{JHEP}\,}\textbf{\bibinfo{volume}{08}}\,(\bibinfo{year}{2020})\,\bibinfo{pages}{039}},
  \Eprint {http://arxiv.org/abs/1912.04334}{arXiv:1912.04334
  [hep-th]}\BibitemShut {NoStop}%
%%CITATION = ARXIV:1912.04334;%%
\bibitem [{\citenamefont{Alonso} \emph
  {et\,al.}(2016{\natexlab{a}})\citenamefont{Alonso}, \citenamefont{Jenkins},
  and \citenamefont{Manohar}}]{Alonso:2015fsp}%
  \BibitemOpen
  \bibfield{author}{\bibinfo{author}{\bibfnamefont{R.}\,\bibnamefont{Alonso}},
  \bibinfo{author}{\bibfnamefont{E.~E.} \bibnamefont{Jenkins}},  and
  \bibinfo{author}{\bibfnamefont{A.~V.} \bibnamefont{Manohar}},
  }\bibfield{title}{\emph {\bibinfo{title}{{A Geometric Formulation of Higgs
  Effective Field Theory: Measuring the Curvature of Scalar Field Space}}},
  }\href {\doibase 10.1016/j.physletb.2016.01.041}
  {\bibfield{journal}{\bibinfo{journal}{Phys.
  Lett.}\,}\textbf{\bibinfo{volume}{B754}}\,(\bibinfo{year}{2016}{\natexlab{a}})\,\bibinfo{pages}{335}},
  \Eprint {http://arxiv.org/abs/1511.00724}{arXiv:1511.00724
  [hep-ph]}\BibitemShut {NoStop}%
%%CITATION = ARXIV:1511.00724;%%
\bibitem [{\citenamefont{Alonso} \emph
  {et\,al.}(2016{\natexlab{b}})\citenamefont{Alonso}, \citenamefont{Jenkins},
  and \citenamefont{Manohar}}]{Alonso:2016oah}%
  \BibitemOpen
  \bibfield{author}{\bibinfo{author}{\bibfnamefont{R.}\,\bibnamefont{Alonso}},
  \bibinfo{author}{\bibfnamefont{E.~E.} \bibnamefont{Jenkins}},  and
  \bibinfo{author}{\bibfnamefont{A.~V.} \bibnamefont{Manohar}},
  }\bibfield{title}{\emph {\bibinfo{title}{{Geometry of the Scalar Sector}}},
  }\href {\doibase 10.1007/JHEP08(2016)101}
  {\bibfield{journal}{\bibinfo{journal}{JHEP}\,}\textbf{\bibinfo{volume}{08}}\,(\bibinfo{year}{2016}{\natexlab{b}})\,\bibinfo{pages}{101}},
  \Eprint {http://arxiv.org/abs/1605.03602}{arXiv:1605.03602
  [hep-ph]}\BibitemShut {NoStop}%
\bibitem [{\citenamefont{Cohen} \emph {et\,al.}(2021)\citenamefont{Cohen},
  \citenamefont{Craig}, \citenamefont{Lu}, and
  \citenamefont{Sutherland}}]{Cohen:2021ucp}%
  \BibitemOpen
  \bibfield{author}{\bibinfo{author}{\bibfnamefont{T.}\,\bibnamefont{Cohen}},
  \bibinfo{author}{\bibfnamefont{N.}\,\bibnamefont{Craig}},
  \bibinfo{author}{\bibfnamefont{X.}\,\bibnamefont{Lu}},  and
  \bibinfo{author}{\bibfnamefont{D.}\,\bibnamefont{Sutherland}},
  }\bibfield{title}{\emph {\bibinfo{title}{{Unitarity violation and the
  geometry of Higgs EFTs}}}, }\href {\doibase 10.1007/JHEP12(2021)003}
  {\bibfield{journal}{\bibinfo{journal}{JHEP}\,}\textbf{\bibinfo{volume}{12}}\,(\bibinfo{year}{2021})\,\bibinfo{pages}{003}},
  \Eprint {http://arxiv.org/abs/2108.03240}{arXiv:2108.03240
  [hep-ph]}\BibitemShut {NoStop}%
\bibitem [{\citenamefont{Cheung} \emph {et\,al.}()\citenamefont{Cheung},
  \citenamefont{Helset}, and \citenamefont{Parra-Martinez}}]{Cheung:2021yog}%
  \BibitemOpen
  \bibfield{author}{\bibinfo{author}{\bibfnamefont{C.}\,\bibnamefont{Cheung}},
  \bibinfo{author}{\bibfnamefont{A.}\,\bibnamefont{Helset}},  and
  \bibinfo{author}{\bibfnamefont{J.}\,\bibnamefont{Parra-Martinez}},
  }\bibfield{title}{\emph {\bibinfo{title}{{Geometric Soft Theorems}}},
  }\href@noop {} {}\,\Eprint {http://arxiv.org/abs/2111.03045}{arXiv:2111.03045
  [hep-th]}\BibitemShut {NoStop}%
\bibitem [{\citenamefont{Helset} \emph {et\,al.}(2020)\citenamefont{Helset},
  \citenamefont{Martin}, and \citenamefont{Trott}}]{Helset:2020yio}%
  \BibitemOpen
  \bibfield{author}{\bibinfo{author}{\bibfnamefont{A.}\,\bibnamefont{Helset}},
  \bibinfo{author}{\bibfnamefont{A.}\,\bibnamefont{Martin}},  and
  \bibinfo{author}{\bibfnamefont{M.}\,\bibnamefont{Trott}},
  }\bibfield{title}{\emph {\bibinfo{title}{{The Geometric Standard Model
  Effective Field Theory}}}, }\href {\doibase 10.1007/JHEP03(2020)163}
  {\bibfield{journal}{\bibinfo{journal}{JHEP}\,}\textbf{\bibinfo{volume}{03}}\,(\bibinfo{year}{2020})\,\bibinfo{pages}{163}},
  \Eprint {http://arxiv.org/abs/2001.01453}{arXiv:2001.01453
  [hep-ph]}\BibitemShut {NoStop}%
\bibitem [{\citenamefont{Nagai} \emph {et\,al.}(2019)\citenamefont{Nagai},
  \citenamefont{Tanabashi}, \citenamefont{Tsumura}, and
  \citenamefont{Uchida}}]{Nagai:2019tgi}%
  \BibitemOpen
  \bibfield{author}{\bibinfo{author}{\bibfnamefont{R.}\,\bibnamefont{Nagai}},
  \bibinfo{author}{\bibfnamefont{M.}\,\bibnamefont{Tanabashi}},
  \bibinfo{author}{\bibfnamefont{K.}\,\bibnamefont{Tsumura}},  and
  \bibinfo{author}{\bibfnamefont{Y.}\,\bibnamefont{Uchida}},
  }\bibfield{title}{\emph {\bibinfo{title}{{Symmetry and geometry in a
  generalized Higgs effective field theory: Finiteness of oblique corrections
  versus perturbative unitarity}}}, }\href {\doibase
  10.1103/PhysRevD.100.075020} {\bibfield{journal}{\bibinfo{journal}{Phys. Rev.
  D}\,}\textbf{\bibinfo{volume}{100}}\,(\bibinfo{year}{2019})\,\bibinfo{pages}{075020}},
  \Eprint {http://arxiv.org/abs/1904.07618}{arXiv:1904.07618
  [hep-ph]}\BibitemShut {NoStop}%
\bibitem [{\citenamefont{Nagai} \emph {et\,al.}(2021)\citenamefont{Nagai},
  \citenamefont{Tanabashi}, \citenamefont{Tsumura}, and
  \citenamefont{Uchida}}]{Nagai:2021gmo}%
  \BibitemOpen
  \bibfield{author}{\bibinfo{author}{\bibfnamefont{R.}\,\bibnamefont{Nagai}},
  \bibinfo{author}{\bibfnamefont{M.}\,\bibnamefont{Tanabashi}},
  \bibinfo{author}{\bibfnamefont{K.}\,\bibnamefont{Tsumura}},  and
  \bibinfo{author}{\bibfnamefont{Y.}\,\bibnamefont{Uchida}},
  }\bibfield{title}{\emph {\bibinfo{title}{{Scalar and fermion on-shell
  amplitudes in generalized Higgs effective field theory}}}, }\href {\doibase
  10.1103/PhysRevD.104.015001} {\bibfield{journal}{\bibinfo{journal}{Phys. Rev.
  D}\,}\textbf{\bibinfo{volume}{104}}\,(\bibinfo{year}{2021})\,\bibinfo{pages}{015001}},
  \Eprint {http://arxiv.org/abs/2102.08519}{arXiv:2102.08519
  [hep-ph]}\BibitemShut {NoStop}%
\end{thebibliography}%
\end{document}